%% file: ChruscielCostagrqc3.tex
\documentclass[a4,english]{smfart}
\usepackage{amsmath,amssymb}%

\usepackage{graphicx}  % standard LaTe\changedX  graphics tool

\usepackage{color}

\usepackage{psfrag}
\usepackage{epsfig}
\usepackage{epstopdf}
\usepackage{wrapfig}
\usepackage{a4}
\usepackage{amssymb,latexsym}%,amsmath}
\usepackage[mathscr]{eucal}
\usepackage{cite}
\usepackage{url}

\makeindex

\DeclareFontFamily{OT1}{rsfs}{} \DeclareFontShape{OT1}{rsfs}{m}{n}{
<-7> rsfs5 <7-10> rsfs7 <10-> rsfs10}{}
\DeclareMathAlphabet{\mycal}{OT1}{rsfs}{m}{n}
\def\scri{{\mycal I}}%
\def\scrip{\scri^{+}}%

\begin {document}

\frontmatter

\author [P.T.~Chru\'sciel]{Piotr T.~Chru\'sciel}
\address {LMPT,
F\'ed\'eration Denis Poisson, Tours; Mathematical Institute and
Hertford College, Oxford}
 \email {chrusciel@maths.ox.ac.uk}
\urladdr {www.phys.univ-tours.fr/$\sim$piotr}%url address

\author [J.L.~Costa]{Jo\~ao~Lopes~Costa}
\address {Lisbon University Institute (ISCTE); Mathematical Institute and Magdalen College, Oxford}
\email {jlca@iscte.pt}
%\urladdr {}
\input{bhmacros}

%\thanks {}%sponsors

\title{On uniqueness of stationary vacuum black holes}
\alttitle {Sur l'unicit\'e de trous noirs stationnaires dans le vide}%translated title
\begin {altabstract}
On d\'emontre l'unicit\'e de trous noirs de Kerr dans la classe de trous
noirs connexes, analytiques, r\'eguliers, non-d\'eg\'en\'er\'es, solutions des
\'equations d'Einstein du vide.
\end {altabstract}
\begin {abstract}
 We prove uniqueness of the Kerr black holes
 within the connected, non-degenerate,
 analytic class of regular
 vacuum black holes.
\end {abstract}
\keywords {Stationary black holes, no-hair theorems}
\altkeywords {Trous noirs stationnaires, th\'eor\^emes d'unicit\'e}%translated keywords
\subjclass {83C57}%subject classification
\dedicatory {It is a pleasure to dedicate this work to J.-P.~Bourguignon on the occasion of his 60th birthday.}
\maketitle

\tableofcontents

\mainmatter

\section{Introduction}
 \label{Sintro}

It is widely expected that   the Kerr metrics provide the only
stationary, asymptotically flat,  sufficiently well-behaved, vacuum,
four-dimensional black holes. Arguments to this effect have been given in
the literature~\cite{RobinsonKerr,CarterlesHouches} (see
also~\cite{Heusler:book,Weinstein1,Neugebauer:2003qe}), with  the
hypotheses needed not always spelled out, and with some notable
technical gaps. The aim of this work is to prove a precise version of one
such uniqueness result for analytic space-times, with detailed filling of the
gaps alluded to above.

The results presented here can be used to obtain a similar result for
electro-vacuum black holes (compare~\cite{Mazur,CarterCMP}), or for
five-dimensional black holes with three commuting Killing vectors (see
also~\cite{HY2,HY});  this will be discussed elsewhere~\cite{CostaPhD}.

We start with some terminology. The reader is referred to
Section~\ref{Saf} for a precise definition of asymptotic flatness, to
Section~\ref{sSdoc} for that of a domain of outer communications $\doc$,
and to Section~\ref{ssZKV} for the definition of mean-non-degenerate
horizons. A Killing vector $\changedX $%
\index{$\changedX $}
is said to be complete if its orbits
are complete, i.e., for every $p\in \mcM$ the orbit $\phi_t[\changedX ](p)$%
\index{$\phi_t[X]$}
of $\changedX $ is defined for all $t\in \R$; in an asymptotically flat context,
$\changedX $ is called \emph{stationary} if it is  timelike at large distances.

\begin{figure}[ht]
\begin{center} { \psfrag{Mext}{$\phantom{x,}\Mext$}
\psfrag{H}{ } \psfrag{B}{ }
\psfrag{H}{ }
 \psfrag{pSigma}{$\!\!\pohyp\qquad\phantom{xxxxxx}$}
\psfrag{Sigma}{ $\hyp$ }
 \psfrag{toto}{$\!\!\!\!\!\!\!\!\!\!\doc$}
 \psfrag{S}{}
\psfrag{H'}{ } \psfrag{W}{$\mathcal{W}$}
\psfrag{scriplus} {} %{ $\mathcal{I}^+$}
\psfrag{scriminus} {} %{ $\mathcal{I}^-$}
 \psfrag{i0}{}%{ $i^0$}
\psfrag{i-}{ } \psfrag{i+}{}
 \psfrag{E+}{ $\phantom{.}{\mycal E}^+$}
%\resizebox{3in}{!}
{\includegraphics{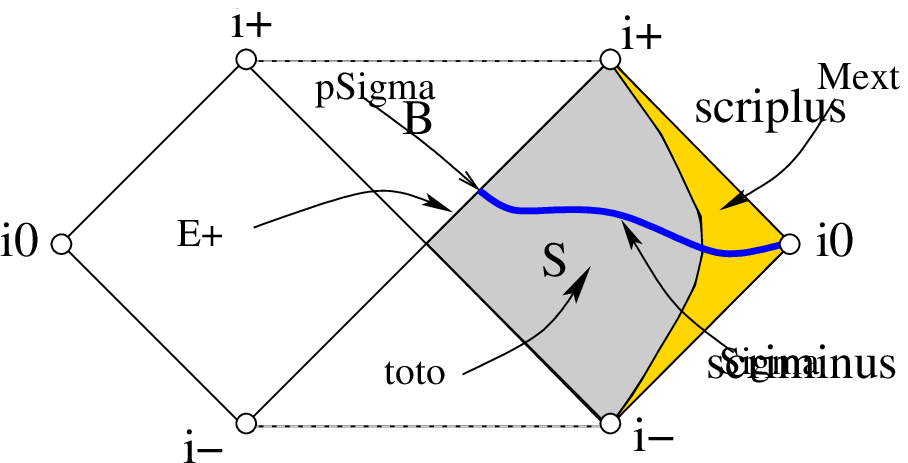}}
}
\caption{The hypersurface $\hyp$ from the definition of \regular ity.
\protect\label{fregu}}
\end{center}
\end{figure}

 A key definition for our work is the following:

\begin{Definition}
 \label{Dmain}
Let $(\mcM,\fourg)$ be a space-time containing an asymptotically
flat end $\Sext$, and let  $\changedX $ be stationary Killing vector field  on $\mcM$.
We will say that $(\mcM,\fourg,\changedX)$ is $\mbox{\rm {\regular}}$%
\index{$\mbox{\rm {\regular}}$}
if $\changedX $ is complete, if the domain of outer communications
$\doc$ is globally hyperbolic, and if $\doc$ contains a spacelike,
connected, acausal hypersurface $\hyp\supset\Sext $,%
\index{$\hyp$}
the  closure $\ohyp $ of which is a topological manifold with boundary,
consisting of  the union of a compact set and of a finite number of
asymptotic ends, such that the boundary $ \pohyp:= \ohyp \setminus \hyp$
is a  topological manifold satisfying
\bel{subs}
\pohyp \subset \mcE^+:= \partial \doc \cap I^+(\Mext)
 \;,
 \ee
with $\pohyp$
meeting every generator of $\mcE^+$ precisely once. (See Figure~\ref{fregu}.)
\end{Definition}

%An equivalent set of conditions, simpler to
%verify in practice, can be found in Remark~\ref{Raltcond} below.

In Definition~\ref{Dmain}, the hypothesis of asymptotic
flatness is made for definiteness, and is not needed for
several of the results presented below. Thus, this definition
appears to be convenient in a wider context, e.g. if asymptotic
flatness is replaced by Kaluza-Klein asymptotics, as
in~\cite{CGS,ChHighDim}.

Some comments about the definition  are in order. First we
require completeness of the orbits of the stationary Killing
vector because we need an action of $\R$ on $\mcM$ by
isometries. Next, we require global hyperbolicity of the domain
of outer communications to guarantee its simple connectedness,
to make sure that the area theorem holds, and to avoid
causality violations as well as certain kinds of naked
singularities in $\doc$. Further, the existence of a
well-behaved spacelike hypersurface gives us reasonable control
of the geometry of $\doc$, and is a prerequisite to any
elliptic PDEs analysis, as is extensively needed for the
problem at hand. The existence of compact cross-sections of the
future event horizon prevents singularities on the future part
of the boundary of the domain of outer communications, and
eventually guarantees the smoothness of that boundary.
(Obviously $I^+$ could have been replaced by $I^-$ throughout
the definition, whence $\mcE^+$ would have become $\mcE^-$.) We
find the requirement \eq{subs} somewhat unnatural, as there are
perfectly well-behaved hypersurfaces in, e.g.,  the
Schwarzschild space-time which do not satisfy this condition,
but we have not been able
to develop a coherent theory without assuming some version of \eq{subs}.
Its main point is to avoid certain zeros of the stationary Killing vector
$\changedX$ at the boundary
of $\hyp$, which otherwise create various difficulties; e.g., it is not clear how to guarantee then smoothness of $\mcEp$,
or the static-or-axisymmetric alternative.%
\footnote{In fact, this condition is not needed for \emph{static} metric if, e.g.,  one assumes at the outset that
all horizons are non-degenerate, as we do in Theorem~\ref{Tubh} below, see the discussion in the Corrigendum to~\cite{Chstatic}.}
Needless to say, all those conditions are satisfied
by the Schwarzschild, Kerr, or Majumdar-Papapetrou solutions.

We have the following,  long-standing conjecture, it being
understood that both the Minkowski and the Schwarzschild
space-times are  members of the Kerr family:

\begin{Conjecture}
 \label{Cubh} Let $(\mcM,\fourg)$ be a stationary, vacuum,
four-dimensional space-time containing a spacelike, connected, acausal
hypersurface $\hyp $, such that $\ohyp $ is a topological manifold with
boundary, consisting of  the union of a compact set and of a finite number
of asymptotically flat ends. Suppose that there exists on $\mcM$ a
complete stationary Killing vector $\changedX$, that $\doc$ is globally
hyperbolic, and that $ \pohyp\subset\mcM\setminus \doc$. Then   $\doc$
is isometric  to the domain of outer communications of a Kerr space-time.
\end{Conjecture}

In this work we establish the following special case thereof:

\begin{theorem}
 \label{Tubh} Let $(\mcM,\fourg)$ be a stationary,
asymptotically flat, {\regular}, vacuum,  four-dimensional
analytic   space-time. If each component of the event horizon
is mean non-degenerate, then $\doc$ is isometric  to the domain
of outer communications of one of the Weinstein solutions of
Section~\ref{ssCs}.  In particular, if $\mcE^+$ is connected
and mean non-degenerate, then $\doc$ is isometric  to the
domain of outer communications of a Kerr space-time.
\end{theorem}

In addition to the references already cited, some key steps of
the proof are due to Hawking~\cite{Ha1}, and to Sudarsky and
Wald~\cite{Sudarsky:wald}, with the construction of the
candidate solutions with several non-degenerate horizons  due
to Weinstein~\cite{Weinstein4,Weinstein:Hadamard}.   It should
be emphasized that the hypotheses of analyticity and
non-degeneracy are highly unsatisfactory,  and one believes
that they are not needed for the conclusion.

One also believes that no candidate solutions  with more than one
component of $\mcE^+$ are singularity-free, but no proof is available except for
some special cases~\cite{LiTian,Weinstein:trans}.

A few words comparing our work with the existing literature are
in order. First, the event horizon in a smooth or analytic
black hole space-time is a priori only a Lipschitz surface,
which is way insufficient to prove the usual
static-or-axisymmetric alternative. Here we use the  results
of~\cite{ChDGH} to show that event horizons in regular
stationary\kk{} black hole space-times are as differentiable as
the differentiability of the metric allows. Next, no paper that
we are aware of adequately shows that the ``area function" is
non-negative within the domain of outer communications; this is
due both to a potential lack of regularity of the intersection
of the rotation axis with the  zero-level-set of the area
function, and to the fact that the gradient of the area
function could  vanish on its zero level set \emph{regardless
of whether or not the event horizon itself is degenerate}. The
second new result of this paper is Theorem~\ref{Tdoc2}, which
proves this result. The difficulty here is to exclude
\emph{non-embedded Killing prehorizons} (for terminology, see
below), and we have not been able to do it without assuming
analyticity or axisymmetry, \emph{even for static solutions}.
Finally, no previous work  known to us  establishes
the  behavior, as needed for the proof of uniqueness,  of the relevant
harmonic map at  points where the horizon meets the rotation axis. The
third new result of this paper is Theorem~\ref{Tbdist}, settling this question
for non-degenerate black-holes. (This last result requires, in turn, the
Structure Theorem~\ref{Tgt} and the Ergoset Theorem~\ref{TdocA}, and
relies heavily on the analysis in~\cite{ChUone}.)  Last but not least, we
provide a coherent set of conditions under which all  pieces of the proof
can be combined to obtain the uniqueness result.

We note that various intermediate results are established under conditions
weaker than previously cited, or are generalized to higher dimensions; this
is of potential interest for further work on the subject.

\subsection{Static case}
Assuming  \emph{staticity}, i.e., stationarity and hypersurface-orthogonality
of the stationary Killing vector, a more satisfactory result is available
in space dimensions less than or equal to seven, and in higher dimensions
on manifolds on which the Riemannian rigid positive energy theorem
holds: non-connected
configurations are excluded, without any \emph{a priori} restrictions on the
gradient $\nabla (\fourg(\changedX ,\changedX ))$ at event horizons.

More precisely, we shall say that  a manifold $\hahyp$ is of \emph{positive
energy type}
\index{positive energy type}%
if there are no asymptotically flat complete Riemannian
metrics on $\hahyp$ with positive scalar curvature and vanishing mass
except perhaps for a flat one. This property has been proved so far for all
$n$--dimensional manifolds $\hahyp$ obtained by removing a finite
number of points from a compact manifold of dimension  $3\le n\le
7$~\cite{SchoenCatini}, or under the hypothesis that $\hahyp$ is a spin
manifold of any dimension $n\ge 3$, and is expected to be true in
general~\cite{ChristLohkamp,Lohkamp}.

We have the following result, which finds its
roots in the work of Israel~\cite{Israel:vacuum}, with further
simplifications by Robinson~\cite{RobinsonSP}, and with a
significant strengthening by Bunting and
Masood-ul-Alam~\cite{bunting:masood}:

\begin{theorem}
\label{Tubhs} Under the hypotheses of Conjecture~\ref{Cubh},
 suppose moreover that $(\doc,\fourg)$ is analytic and $\changedX$ is
 hypersurface-orthogonal.
%Let $(\mcM,\fourg)$ be an asymptotically flat, vacuum,
%$(n+1)$--dimensional space-time, $n\ge 3$, with a hypersurface
%$\hyp$ satisfying the hypotheses of Conjecture~\ref{Cubh}, with
%a complete static Killing vector field $\changedX$, and with a
%globally hyperbolic $\doc$.
Let
 $\,\,\widehat{\!\! \hyp}$ denote the manifold obtained by
doubling $\hyp$ across the non-degenerate components of its boundary
and compactifying,  in the doubled manifold, all  asymptotically flat regions
but one to a point. If $\,\,\widehat{\!\! \hyp}$ is of positive energy type, then
$\doc$ is isometric  to the domain of outer communications of a
Schwarzschild space-time.
\end{theorem}

\begin{Remark}
\label{RCRT} {\rm As a corollary of Theorem~\ref{Tubhs} one obtains
 non-existence of black holes as above with some
 components of the horizon degenerate. In space-time dimension four an
 elementary proof of this fact has been given in~\cite{CRT}, but the simple
 argument there does not seem to generalize to higher dimensions in any
 obvious way.
}
\end{Remark}
\begin{Remark}
\label{Rst2} {\rm   Analyticity is only  needed to exclude
non-embedded degenerate prehorizons within $\doc$. In
space-time dimension four it can be replaced by the condition
of axisymmetry and \regular ity, compare Theorem~\ref{Tdoc3}.}
\end{Remark}

\proof We want to invoke~\cite{Chstatic}, where $n=3$ has been
assumed; the argument given there generalizes immediately to those
higher dimensional manifolds on which the positive energy theorem holds.
However, the proof in~\cite{Chstatic} contains one mistake, and one gap,
both of which need to be addressed.

First, in the case of  degenerate horizons $\mcH$, the analysis
of~\cite{Chstatic} assumes that the static Killing vector has
no zeros on $\mcH$; this is used in the key Proposition~3.2
there, which could be wrong without this assumption. The
non-vanishing of the static Killing vector  is justified
in~\cite{Chstatic} by an incorrectly quoted version of Boyer's
theorem~\cite{Boyer}, see~\cite[Theorem~3.1]{Chstatic}. Under a
supplementary  assumption of  \regular ity,  the zeros of a
Killing vector which could arise in the closure of a degenerate
Killing horizon can be excluded using Corollary~\ref{Cnov}. In
general, the problem is dealt with in  the addendum to the arXiv
versions v$N$, $N\ge 3$, of~\cite{Chstatic} in space-dimension
three, and in~\cite{ChHighDim} in higher dimensions.

Next, neither the original proof, nor that given
in~\cite{Chstatic}, of the Vishveshwara-Carter Lemma, takes
properly into account the possibility that the hypersurface
$\mcN$ of~\cite[Lemma~4.1]{Chstatic} could fail to be embedded.%
\footnote{This problem affects points 4c,d,e and f of
\cite[Theorem~1.3]{Chstatic}, which require the supplementary
hypothesis of existence of an embedded closed hypersurface
within $\mcN$; the remaining claims
of~\cite[Theorem~1.3]{Chstatic} are justified by the arguments
described here.}
This problem is taken care of by Theorem~\ref{Tdoc2} below with
$s=1$, which shows that $\doc$ cannot intersect the  set where
$W:=-\fourg(\changedX ,\changedX )$ vanishes.  This implies that
$\changedX $ is timelike on $\doc\supset \hyp$, and null on
$\pohyp$. The remaining details are as
in~\cite{Chstatic}.
\qed

\section{Preliminaries}
 \label{SPrelim}
\subsection{Asymptotically flat stationary  metrics}
 \label{Saf}
  \renewcommand{\changedX}{X}
A space-time $(\mcM,\fourg)$ will  be said to possess an
\emph{asymptotically flat end }%
\index{asymptotic flatness}%
if $\mcM$ contains a spacelike
hypersurface $\Mtext$  diffeomorphic to $\R^n\setminus B(R)$, where
$B(R)$ is an open coordinate ball of radius $R$, with the following properties:
there exists a constant $\alpha>0$ such that, in local coordinates on
$\Mtext$ obtained from $\R^n\setminus B(R)$, the metric $\threeg$
induced by $\fourg$ on $\Mtext$, and the extrinsic curvature tensor $K_{ij}$ of
$\Mtext$,  satisfy the fall-off conditions
\beal{falloff1}
 & \threeg_{ij}-\delta_{ij}=O_k(r^{-\alpha})\;,  \qquad
  K_{ij}=O_{k-1}(r^{-1-\alpha})\;,
 \eeal{falloff2}
for some $k > 1$, where we write $f=O_k(r^{\alpha})$ if $f$ satisfies
\bel{okdef}
  \partial_{k_1}\ldots\partial_{k_\ell}
f=O(r^{\alpha-\ell})\;, \quad 0\le \ell \le k
 \;.
\ee
For simplicity we assume that the space-time is vacuum, though similar
results hold in general under appropriate conditions on matter fields,
see~\cite{ChMaerten,ChBeigKIDs} and references therein. Along any
spacelike hypersurface $\hyp$, a Killing vector field $\changedX$ of
$(\mcM,\fourg)$ can be decomposed as
$$
 \changedX = N n + Y \;,
$$
where $Y$ is tangent to $\hyp$, and $n$ is the unit future-directed normal
to $\Mtext$.   The vacuum field equations, together with the Killing
equations imply  the following set of equations on $\hyp$,
where $
 R_{ij}(\threeg) $ is the Ricci tensor of $\threeg$:
\begin{eqnarray}
\label{K.13}
&
 D_i Y_j + D_j Y_i = 2 N K_{ij}\;,
 &
 \\
 &
 R_{ij}(\threeg) +
K^k{_k}
K_{ij} - 2 K_{ik} K^k{}_j    - N^{-1}(\mcL_Y K_{ij} + D_i
 D_j N  )=0 \;.
 & \label{K.15}
\end{eqnarray}

Under the boundary conditions \eq{falloff2} with $k\ge 2$, an
analysis of \eq{K.13}-\eq{K.15} provides detailed information
about the asymptotic behavior of $(N,Y)$. In particular, one
can prove that if the asymptotic region $\Sext$ is contained in
a hypersurface $\hyp$ satisfying the requirements of the
positive energy theorem, and if $\changedX$ is timelike along
$\Sext$, then $(N,Y^i)\to_{r\to\infty} (A^0,A^i)$, where the
$A^\mu$'s are constants satisfying $(A^0)^2>\sum_i( A^i)^2$.
One can then choose adapted coordinates so that the metric can,
locally, be written as
\beal{gme1}
 &\fourg  =
 -V^2(dt+\underbrace{\theta_i dx^i}_{=\theta})^2 +
 \underbrace{\threeg_{ij}dx^i dx^j}_{=\threeg}\;,
\eeal{gme2}
with
\beal{foff} &
 \partial_t V = \partial_t \theta = \partial_t \threeg=0
 &
 \\
 & \threeg_{ij}-\delta_{ij}=O_k(r^{-\alpha})\;, \quad
  \theta_{i } =O_k(r^{-\alpha})\;, \quad V-1= O_k(r^{-\alpha})
 \;,
\eeal{foff2}
for any $k\in \N$. As discussed in more detail
in~\cite{ChBeig3}, in $\threeg$-harmonic coordinates, and in
e.g. a maximal time-slicing, the vacuum equations for $\fourg$
form a quasi-linear elliptic system with diagonal principal
part, with principal symbol identical to that of the scalar
Laplace operator. Methods known in principle show that, in this
``gauge", all metric functions have a full
asymptotic expansion%
\footnote{One can  use the results in, e.g.,~\cite{ChAFT} together
with a simple iterative argument to obtain the expansion. This
analysis holds in any dimension.}
 in terms of powers of $\ln r$ and inverse
powers of $r$. In the new coordinates we can in fact take
\bel{decrate}
 \alpha= {n-2}
 \;.
 \ee
By inspection of the equations one can further infer that the
leading order corrections in the metric can be written in a
Schwarzschild form, which in ``isotropic" coordinates reads
\bean
 \fourg_m&=&  - \left(\frac{1-\frac{m}{2|x|^{n-2}}}{1+\frac{m}{2|x|^{n-2}}}\right)^2
dt^2+\left(1 + \frac{m}{2|x|^{n-2}}\right)^{\frac4{n-2}}
\left(\sum_{i=1}^n dx_i^2\right)
% \\
% & \approx &
%- \Big(1-\frac {m}{|x|^{n-2}}\Big)^2 dt^2+\Big(1+\frac
%{m}{|x|^{n-2}}\Big)^{\frac 2 {n-2}}\left(\sum_{i=1}^n dx_i^2\right)
 \;,
 \eeal{stschw}
where $m\in \R$.%
\renewcommand{\changedX}{K}%
\kk{add a subsection on KK-asymptotic flatness}

\subsection{Domains of outer communications, event horizons}
 \label{sSdoc}

\begin{figure}[t]
\begin{center} { \psfrag{Mext}{\Large$\,\Mext$}
\psfrag{H}{ } \psfrag{B}{ }
\psfrag{pasthorizon}{\Large $\!\!\!\!\!{\partial I^+(\Mext)}$ }
 \psfrag{pSigma}{$\!\!\pohyp\qquad\phantom{xxxxxx}$}
\psfrag{Sigma}{\Large $\!\Sext$}
 \psfrag{toto}{\Large$\!\!\!\!\!\!\!\!\!\!\!\!\!\!\!\!\! I^-(\Mext)$}
 \psfrag{S}{}
 \psfrag{future}{\Large $\!\!\!\!\!{I^+(\Mext)}$}
\psfrag{H'}{ } \psfrag{W}{$\mathcal{W}$}
\psfrag{scriplus} {} %{ $\mathcal{I}^+$}
\psfrag{scriminus} {} %{ $\mathcal{I}^-$}
 \psfrag{i0}{}%{ $i^0$}
\psfrag{i-}{ } \psfrag{i+}{}
 \psfrag{E+}{\Large $\!\!\!\!\!{\partial I^-(\Mext)}$}
\resizebox{2.3in}{!}{\includegraphics{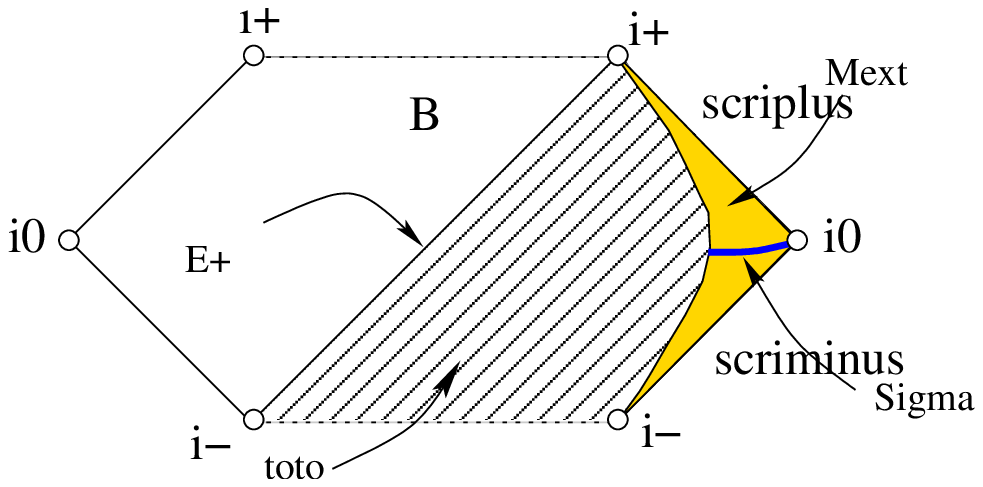}}
\resizebox{2.3in}{!}{\includegraphics{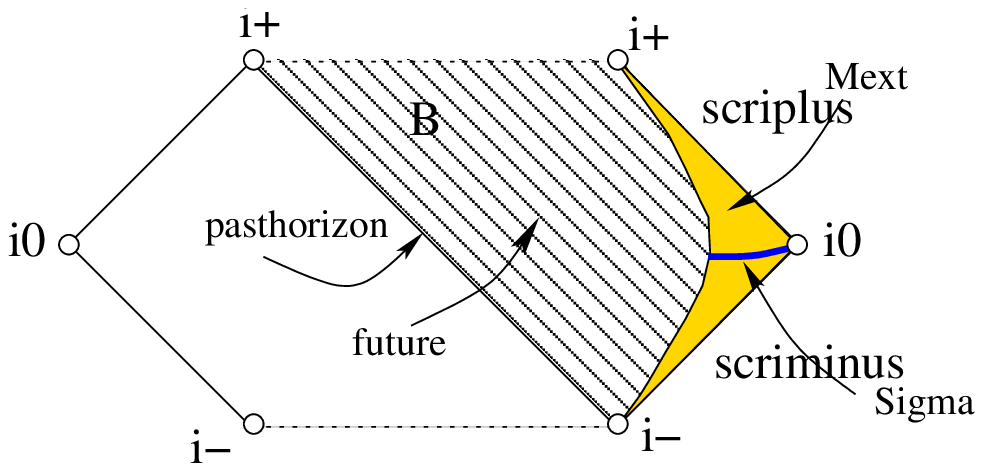}}
}
\caption{$\Sext$, $\Mext$, together with the future and  the past of $\Mext$. One has $\Mext\subset I^\pm(\Mext)$, even
though this is not immediately apparent from the figure.
The domain of outer communications is the intersection $ I^+(\Mext)\cap I^-(\Mext)$, compare Figure~\ref{fregu}.
\label{FPast}}
\end{center}
\end{figure}

A key notion in the theory of black holes is that of the \emph{domain
of outer communications}:
A  space-time $(\mcM,\fourg)$ will be
called  stationary\kk{} if there exists on $\mcM$ a complete Killing
vector field $\changedX $ which is  \emph{timelike} in the asymptotically flat\kk{} region
$\Sext$.%
\index{$\Sext$}%
\footnote{In fact,  in the literature it is always implicitly assumed that
$\changedX $ is  \emph{uniformly timelike} in the asymptotic region
$\Sext$, by this we mean that $\fourg(\changedX ,\changedX
)<-\epsilon<0$ for some $\epsilon$ and for all $r$ large enough. This
uniformity condition excludes the possibility of a timelike vector which
asymptotes to a null one. This involves no loss of generality in well-behaved space-times: indeed, uniformity always holds for Killing vectors
which are timelike for all large distances if the conditions of the positive
energy theorem are met~\cite{ChBeig1,ChMaerten}.}
\index{$I^{\pm}$, $\dot I^\pm$}%
\index{$J^{\pm}$, $\dot J^\pm$}%
For $t\in {\R }$ let $\phi_t[\changedX ]:\mcM\to \mcM$ denote the
one-parameter group of diffeomorphisms generated by $\changedX $; we will
write $\phi_t$ for $\phi_t[\changedX ]$
\index{$\phi_t[X]$}
whenever ambiguities are unlikely to
occur. The exterior region $\Mext$
\index{$\Mext$}
and the \emph{domain of outer communications}%
\index{domain of outer communications}
$\doc$ are  then defined as%
\footnote{Recall that $I^-(\Omega)$, respectively $J^-(\Omega)$, is
the set covered by past-directed timelike, respectively causal,
curves originating from $\Omega$, while $\dot I^- $ denotes the
boundary of $I^-$, etc. The sets $I^+$, etc., are defined as $I^-$,
etc., after changing time-orientation.}
(compare Figure~\ref{FPast})
\bel{docdef}
 \doc = I^+(\underbrace{\cup_t \phi_t(
 \Sext )}_{=:\Mext})\cap  I^-(\cup_t \phi_t
 (\Sext ))
 \;.
\ee
\index{$\doc$}%
The \emph{black hole region} $\mcB$%
\index{black hole!region}
and the \emph{black hole event horizon} $\mcH^+$
\index{black hole!event horizon}
are defined  as
$$
 \mcB= \mcM\setminus I^-(\Mext)\;,\quad \mcH^+=\partial \mcB
 \;.
$$
The \emph{white hole region} $\mcW$ and the \emph{white hole event horizon} $\mcH^-$%
\index{white hole event horizon}
are defined  as above after changing time orientation:
$$
 \mcW= \mcM\setminus I^+(\Mext)\;,\quad \mcH^-=\partial \mcW
 \;, \quad \mcH=\mcH^+\cup \mcH^-\;.
$$
\index{$\mcH^\pm$}
It follows that the boundaries of  $\doc $ are included in the {event horizons}. We set
\bel{epm}
  \mcE^\pm = \partial \doc \cap I^\pm (\Mext)
 \;, \qquad \mcE=\mcE^+\cup \mcE^- \;.
\ee
\index{$\mcE^\pm$}

There is considerable freedom in choosing the asymptotic region $\Sext$.
However, it is not too difficult to show, using Lemma~\ref{Lvis} below, that
$I^\pm (\Mext)$, and hence $\doc$, $\mcH^\pm$ and $\mcE^\pm$, are
independent of the choice of $\Sext$ whenever the associated $\Mext$'s
overlap.

Several results below hold without assuming asymptotic flatness: for
example, one could assume that we have a region $\Sext$ on which
$\changedX$ is timelike, and carry on with the definitions above. An
example of interest is provided by Kaluza-Klein metrics with an
asymptotic region of the form $(\R^n\setminus B(R))\times \T^p$,
with the space metric asymptotic to a flat\kk{discussion to be
moved/removed} metric there. However, for definiteness, and to avoid
unnecessary discussions, we have chosen to assume asymptotic
flatness in the definition of \regular ity.

\subsection{Killing horizons, bifurcate horizons}
 \label{ssKH}

A null hypersurface, invariant  under the flow of a Killing vector
$\changedX $, which coincides with a connected component of the set
$$
\mcN(\changedX ):= \{\fourg(\changedX ,\changedX )=0\;,\ \changedX \ne 0\}
 \;,
$$
is called a \emph{Killing horizon} associated to $\changedX $.%
\index{Killing horizon}%

A
set will be called a  \emph{bifurcate Killing horizon}%
\index{Killing horizon!bifurcate}
if
it is the
union of four Killing horizons, the intersection of the closure
of which forms a smooth submanifold $S$ of co-dimension two,
called  the
\emph{bifurcation surface}.%
\index{bifurcation surface}
The four Killing horizons consist then of the four
null hypersurfaces obtained by shooting null geodesics in the four distinct null directions
normal to $S$.
For example, the Killing vector $x\partial_t+t\partial_x$ in
Minkowski space-time has a bifurcate Killing horizon, with the bifurcation
surface $\{t=x=0\}$.

The
\emph{surface gravity} $\kappa$ of a Killing horizon $\mcN$  is defined by
the formula
\begin{equation}
  \label{kdef0}
  d\left(\fourg(\changedX ,\changedX )\right)|_\mcN =   -2\kappa \changedX ^\flat  \ ,
\end{equation}
where $\changedX ^\flat=\fourg_{\mu\nu}\,\changedX ^\nu dx^\mu$. A fundamental property is
that {the surface gravity} $\kappa$
\index{$\kappa$}%
\index{surface gravity}%
is constant over each horizon in vacuum, or in electro-vacuum, see
e.g.~\cite[Theorem~7.1]{Heusler:book}. The proof given
in~\cite{Wald:book} generalizes to all space-time dimensions $n+1\ge 4$;
the result also follows in all dimensions from the analysis in~\cite{HIW}
when the horizon has compact spacelike sections.  (The constancy of $\kappa$
can also be established without assuming any field equations in some
cases, see~\cite{KayWald,RaczWald2}.) A Killing horizon is called
\emph{degenerate}
\index{Killing horizon!degenerate}%
if $\kappa$ vanishes, and \emph{non-degenerate}
\index{Killing horizon!non-degenerate}%
otherwise.

\subsubsection{Near-horizon geometry}
 \label{ssnhg}

Following~\cite{VinceJimcompactCauchyCMP}, near a smooth event horizon
one can
introduce \emph{Gaussian null coordinates}, in which the metric
takes the form
\bel{GNC1} \fourg=r \varphi dv^2 + 2dv dr + 2r h_a dx^a dv + h_{ab}dx^a
dx^b\;. \ee (These coordinates can be introduced for any null hypersurface,
not necessarily an event horizon, in any number of dimensions).  The
horizon is given by the equation $\{r=0\}$, replacing $r$ by $-r$ if
necessary we can without loss of generality assume that $r>0$ in the
domain of outer communications. Assuming that the horizon admits a smooth
compact cross-section $S$, the \emph{average surface gravity}
$\langle \kappa\rangle_S$
\index{$\langle \kappa\rangle_S$}
 is defined as
\bel{asg}
 \langle \kappa\rangle_S=-\frac 1 {|S|}\int_S \varphi d\mu_h
 \;,
\ee
where $d\mu_h$ is the measure induced by the metric $h$ on $S$,
and $|S|$ is the volume of $S$.  We emphasize that this is
defined regardless of whether or not some Killing vector
$\changedX$ is tangent to the horizon generators; but if
$\changedX$ is, and if the surface gravity $\kappa$ of
$\changedX$ is constant on $S$, then $\langle \kappa\rangle_S$
equals $\kappa$.

On a degenerate Killing horizon the surface gravity vanishes by definition, so that the
function $\varphi$ in \eq{GNC1} can itself be written as $rA$, for some
smooth function $A$. The vacuum Einstein equations imply
(see~\cite[eq.~(2.9)]{VinceJimcompactCauchyCMP} in dimension four
and~\cite[eq.~(5.9)]{LP2} in higher dimensions)
\bel{vEe} \zR_{ab} = \frac 12 \mzh_{a}\mzh_{b} -  \zD_{(a
}\mzh_{b)}\;,\ee where $\zR_{ab}$ is the Ricci tensor of $\mzh_{ab}:=h_{ab}|_{r=0}$,
and $\zD$ is the covariant derivative thereof, while $\mzh_a:=h_a|_{r=0}$. The Einstein equations also determine
$\zA:=A|_{r=0}$ uniquely in terms of $\mzh_a$ and $\mzh_{ab}$:
\bel{Asol}
\zA = \frac{1}{2} \mzh^{ab} \left( \mzh_a \mzh_b -  \zD_a \mzh_b  \right)
\ee
(this equation
follows again e.g. from~\cite[eq.~(2.9)]{VinceJimcompactCauchyCMP} in
dimension four, and can be checked by a calculation in all higher
dimensions). We have the following:%
\footnote{Some partial results  with a non-zero cosmological constant have also been proved in~\cite{CRT}.}

\begin{Theorem}[\cite{CRT}]
 \label{TCRT} Let  the space-time dimension be $n+1$, $n\ge 3$,
 suppose that a degenerate Killing horizon $\mcN$ has a compact
 cross-section, and   that $\mzh_a=\partial_a \lambda$ for some function
$\lambda$ (which is necessarily the case in vacuum static space-times).
Then  \eq{vEe} implies $\mzh_a\equiv0$, so that $\mzh_{ab}$ is Ricci-flat.
\end{Theorem}

\begin{Theorem}[\cite{Hajicek3Remarks,LP2}]
 \label{TLP}
In space-time dimension four
 and in vacuum, suppose that  a degenerate Killing horizon
$\mcN$ has a spherical cross-section, and that $(\mcM,\fourg)$ admits a second Killing vector field with periodic orbits. For every connected component $\mcN_0$ of $\mcN$ there exists an embedding  of $\mcN_0$ into a Kerr space-time which preserves $\mzh_a$, $\mzh_{ab}$ and $\zA$.
\end{Theorem}

It would be of interest to understand fully \eq{vEe}, in all dimensions, without  restrictive conditions.

In the four-dimensional static case, Theorem~\ref{TCRT}  enforces toroidal topology of cross-sections of $\mcN$, with a flat $\mzh_{ab}$.
On the other hand, in the four-dimensional axisymmetric case, Theorem~\ref{TLP} guarantees that the geometry tends
 to a Kerr one, up to errors made clear in the statement of the theorem, when the horizon is approached. (Somewhat more detailed information can be found in~\cite{Hajicek3Remarks}.) So, in the degenerate case, the vacuum equations
impose strong restrictions on the near-horizon geometry.

It seems that this is not the case any more for non-degenerate horizons, at
least in the analytic setting. Indeed, we claim that for any triple
$(N,\mzh_a,\mzh_{ab})$, where $N$ is a two-dimensional analytic
manifold (compact or not), $\mzh_a$ is an analytic one-form on $N$, and
$\mzh_{ab}$ is an analytic Riemannian metric on $N$, there exists a
vacuum space-time $(\mcM,\fourg)$ with a bifurcate (and thus
non-degenerate) Killing horizon, so that the metric $\fourg$ takes the form
\eq{GNC1} near each Killing horizon branching out of the bifurcation
surface $S\approx N$, with $\mzh_{ab}=h_{ab}|_{r=0}$ and
$\mzh_{a}=h_{a}|_{r=0}$;  in fact $\mzh_{ab}$  is the metric induced by
$\fourg$ on $S$. When $N$ is  the two-dimensional torus  $\T^2$ this can
be inferred from~\cite{MoncriefCauchyANOP} as follows:
using~\cite[Theorem~(2)]{MoncriefCauchyANOP} with
$(\phi,\beta_a,g_{ab})|_{t=0}=(0,2\mzh_a,\mzh_{ab})$ one obtains a
vacuum space-time $(\mcM'=S^1\times \T^2\times
(-\epsilon,\epsilon),\fourg')$ with a compact Cauchy horizon $S^1\times
\T^2$ and Killing vector $\changedX $ tangent to the $S^1$ factor of
$\mcM'$. One can then  pass to a covering space where  $S^1$  is
replaced by  $\R$, and use a construction of R\'acz and
Wald~\cite[Theorem~4.2]{RaczWald2} to obtain the desired $\mcM$
containing the bifurcate horizon. This argument generalizes to any analytic
$(N,\mzh_a,\mzh_{ab})$ without difficulties.

\subsection{Globally hyperbolic asymptotically flat domains of outer communications are simply  connected}
 \label{sghsc}
Simple connectedness of the domain of outer communication is an essential ingredient in several steps of the uniqueness argument below. It was first noted in
\cite{ChWald} that this stringent topological restriction is a consequence of
the ``topological censorship theorem" of Friedman, Schleich and Witt
\cite{FriedmanSchleichWitt}   for asymptotically flat, stationary and globally
hyperbolic domains of outer communications satisfying the null energy
condition:
\bel{NEC}
 R_{\mu\nu}Y^\mu Y^\nu\geq 0 \,\,\text{ for null }\,\, Y^\mu
  \;.
\ee

In fact, stationarity is not needed. To make things precise,
consider a space-time $(\mcM,\fourg)$ with several asymptotically flat
regions $\Mext^i$, $i=1,\ldots,N$, each generating its own  domain of
outer communications. It turns out~\cite{galloway-topology} (compare~\cite{Galloway:fitopology}) that the null
energy condition prohibits causal interactions between distinct such ends:

\begin{Theorem}
\label{topocensor} If $(\mcM,\fourg)$ is a globally hyperbolic and
asymptotically flat space-time satisfying the null energy condition
\eq{NEC}, then
\bel{asg2}
 \langle\langle \mcMext^i\rangle\rangle \cap J^{\pm}(\langle\langle \mcMext^j\rangle\rangle)=\emptyset
 \text{ for } i\neq j\,.
\ee
\end{Theorem}

A clever covering/connectedness argument%
\footnote{Under more general asymptotic conditions it was proved in
\cite{GSWW} that inclusion induces a surjective homeomorphism between
the  fundamental groups of the exterior region and the domain of outer
communications. In particular,
 $ \pi_1(\Mext)=0 \Rightarrow \pi_1(\doc)=0\;.$}
~\cite{galloway-topology} shows then:%
\footnote{Strictly speaking, our applications below of~\cite{galloway-topology} require
checking that the conditions of
asymptotic flatness in~\cite{galloway-topology} coincide with ours; this, however, can be
avoided by invoking directly~\cite{ChWald}.}

\begin{Corollary}
 \label{Cdsc}
A  globally hyperbolic and asymptotically flat domain of outer
communications satisfying the null energy condition is simply connected.
\end{Corollary}

In space-time dimension four this, together with standard topological
results~\cite{Munkres}, leads to a spherical topology of horizons (see~\cite{ChWald} together with Proposition~\ref{PCtmco} below):

\begin{Corollary}
 \label{Cst}
In {\regular}, stationary, asymptotically flat space-times satisfying the null
energy condition, cross-sections of $\mcEp$ have spherical topology.
\end{Corollary}

\section{Zeros of Killing vectors}
 \label{ssZKV}

Let $\hyp$ be a spacelike hypersurface in $\doc$; in the proof of
Theorem~\ref{Tubh} it will be essential to have no zeros  of the stationary
Killing vector $\changedX$ on $\ohyp$. Furthermore, in the
axisymmetric scenario, we need to exclude zeros of Killing vectors of the form
$\Kz +\alpha K_\kl 1$ on $\doc$, where $\Kz=\changedX$ and $K_\kl 1$
is a generator of the axial symmetry. The aim of this section is to present
conditions which guarantee that; for future reference, this is done in
arbitrary space-time dimension.

We start with the following:

\begin{Lemma}
 \label{Ponce} Let $\Sext\subset\hyp\subset \doc$, and suppose
that $\hyp$ is achronal in $\doc$. Then  for any $p\in \Mext$
there exists $t_0\in \R$ such that
$$
\ohyp\cap I^+(\phi_{t_0}(p))=\emptyset
\;.
$$
\end{Lemma}

\begin{proof}
Let $p\in \Mext$. There exists  $t_0$  such that $r:=\phi_{t_0}(p)\in \Sext$.
Suppose that $\ohyp\cap I^+(\phi_{t_0}(p))\ne \emptyset$. Then there exists
a timelike future directed curve $\gamma$ from $r$ to $q\in \ohyp$. Let
$q_i \in \hyp$ converge to $q$; then $q_i\in I^+(r)$ for $i$ large enough,
which contradicts achronality of $\hyp$ within $\doc$.
\end{proof}

\begin{Lemma}
\label{Loncew} Let $S\subset   I^+(\Mext)$ be  compact.
\begin{enumerate}
\item
 \label{PLonce1}
There exists $p\in \Mext$ such that $S$ is contained in $I^+(p)$.
 \item
   \label{PLonce2}
    If $S\subset \partial\doc\cap I^+(\Mext)$ and if $(\overline \doc,\fourg)$ is strongly causal at $S$,%
\footnote{In a sense made clear in the last sentence of the proof below.}
     then for any $p\in \Mext$      there exists $t_0\in
\R$ such that $S\cap I^+(\phi_{t_0}(p))=\emptyset$.
\end{enumerate}
\end{Lemma}

\proof
 1:
Let $q\in S$; there exists $p_q\in \Mext$ such that  $q \in
I^+(p_q)$, and since $I^+(p_q)$ is open there exists an open
neighborhood $\mcO_q\subset S$ of $q$  such that $\mcO_q\subset
I^+(p_q)$. By compactness there exists a finite collection
$\mcO_{q_i}$, $i=1,\ldots,I$, covering $S$, thus $S\subset
\cup_i I^+(p_{q_i})$. Letting $p\in \Mext$ be any point such that
$p_{q_i}\in I^+(p)$ for  $i=1,\ldots,I$, the result follows.

\medskip

2: Suppose not. Then $\phi_i(p)\in I^-(S)$ for all $i\in \N$,
hence there exists $q_i\in S$ such that $q_i \in
I^+(\phi_i(p))$. By compactness there exits $q\in S$ such that
$q_i\to q$. Let $\mcO$ be an arbitrary neighborhood of $q$;
since $q\in \mcE^+$;  there exists $r\in \mcO\cap \doc$,
$p_+\in \Mext$, and a future directed causal curve $\gamma$
from $r$ to $p_+$. For all $i$ large, this can be continued by
a future directed causal curve from $p_+$ to $\phi_i(p)$, which
can then be continued by a future directed causal curve to
$q_i$. But $q_i\in \mcO$ for $i$ large enough. This implies
that every small neighborhood of $q$ meets a future directed
causal curve entirely contained within $\doc$ which leaves the
neighborhood and returns, contradicting strong causality of
$\overline
\doc$.
\qed

 \medskip

It follows from Lemma~\ref{Ponce}, together with point 1 of
Lemma~\ref{Loncew}  with $S=\{r\}$, that

\begin{Corollary}
 \label{Cnov} If $r\in\ohyp \cap I^+(\Mext)$, then  the stationary\kk{} Killing
 vector $\changedX$ does not vanish at $r$. In particular if
 $(\mcM,\fourg)$ is \regular, then  $\changedX$ has no zeros on $\ohyp$.
\qed
\end{Corollary}

To continue, we  assume the existence of a commutative group of
isometries $\R\times \T^{s-1}$, $s\ge 1$. We denote by $\Kz $%
\index{$K_\kl 0$}%
\index{$K_\kl i$}
the Killing vector tangent to the orbits $\R$ factor, and we assume that
$\Kz$ is timelike in $\Mext$. We denote by  $K_\kl i$, $i=1,\ldots,s-1$ the
Killing vector tangent to the orbits of the $i$'th $S^1$ factor of $ \T^{s-1}$.
We assume that each $K_\kl i$ is spacelike in $\doc$ wherever
non-vanishing, which will necessarily be the case if $\doc$ is
chronological. Note that asymptotic flatness imposes $s-1\le n/2$,
\kk{$s$ is not restricted any more}
 though most of the results of this
section remain true without this hypothesis, when properly formulated.

We say that a Killing orbit $\gamma:\R\to \mcM$ is future-oriented
\index{future-oriented orbit}%
if there exist numbers $\tau_1>\tau_0$ such that $\gamma(\tau_1)\in
I^+(\gamma(\tau_0))$. Clearly all orbits of a Killing vector $\changedX $
are future-oriented in the region where $\changedX $ is timelike. A
less-trivial example is given by orbits of the Killing vector
$\partial_t+\Omega \partial_\varphi$ in Minkowski space-time. Similarly, in
stationary\kk{} axisymmetric space-times, those orbits of this last Killing vector
on which $\partial_t$ is timelike are future-oriented (let $\tau_0=0$ and
$\tau_1=2\pi/\Omega$).

We have:

\begin{Lemma}
 \label{LtorMext}
Orbits through $\Mext$ of  Killing vector fields $\changedX $ of the form $
\Kz +\sum \alpha_\kl i K_\kl i$  are future-oriented.
 \label{LTo}
\end{Lemma}
\begin{proof}
Recall that for any Killing vector field $Z$ we denote by
$\phi_t[Z]$ the flow of $Z$. Let
$$
 Y:=\sum \alpha_\kl i K_\kl i
 \;.
$$
Suppose, first, that there exists $\tau>0$ such that
$\phi_\tau[Y]$
is the identity. Since $\Kz $ and $Y$ commute we have
$$\phi_\tau[\changedX ]=\phi_\tau[\Kz +Y]=\phi_\tau[\Kz ]\circ
\phi_\tau[Y] = \phi_\tau[\Kz ]
\;.
$$
Setting $\tau_0=0$ and $\tau_1=\tau$, the result follows.

Otherwise, there exists a sequence $t_i\to\infty$ such that
$\phi_{t_i} [Y](p)$ converges to $p$. Since $I^+(p)$ is open
there exists a neighborhood $\mcU^+\subset   I^+(p)$ of {$\phi_{
1}[\Kz]\nic (p)$}. Let $\mcV^+=\phi_{-1}[\Kz](\mcU^+)$,  then
every point in $\mcU^+$ lies on a future directed timelike path
starting in $\mcV^+$, namely  an integral curve of $\Kz $.
There exists $i_0\ge 1 $ so that $t_i\ge 1$ and  $\phi_{t_i}
[Y](p) \in \mcV^+ $  for $i\ge i_0$. We then have
$$
 \phi_{t_i} [\changedX ](p)= \phi_{t_i} [\Kz +Y](p)=\phi_{t_i-1}
 [K_\kl
 0]\big(\underbrace{
 \phi_{1} \nic (\underbrace{\phi_{t_i} [Y](p)}_{\in
 \mcV^+})}_{\in \mcU^+\subset
 I^+( p)}\big)\in I^+(p)
 \;.
$$
The numbers $\tau_0=0$ and $\tau_1=t_{i_0}$ satisfy then the
requirements of the definition.
\end{proof}

For future reference we note the following:

\begin{Lemma}
The orbits   through $\doc$ of any Killing vector $K$ of the form $
\Kz +\sum \alpha_\kl i K_\kl i$ are future-oriented.
\end{Lemma}

\proof Let $p\in \doc$, thus there exist points $p_\pm \in
\Mext$
 such that $p_\pm \in I^\pm(p)$, with associated future
directed timelike curves $\gamma_\pm$. It follows from
Lemma~\ref{LtorMext} together with asymptotic flatness
that
there exists $\tau$ such that $\phi_\tau[K](p_-)\in I^+(p_+)$
for some $\tau$, as well as an associated future directed curve
$\gamma$ from $p_+$ to $\phi_\tau[K](p_-)$.
 Then the curve $\gamma_+\cdot\gamma \cdot
\phi_\tau[K](\gamma_-)$, where $\cdot$ denotes concatenation
of curves, is a timelike curve from $p$ to $\phi_\tau(p)$.
\qed

\medskip

The following result, essentially due to~\cite{ChWald1}, turns
out
to be very useful:

\begin{Lemma}
\label{Lvis} Let $\alpha_i \in \R$. For any set $C $ invariant
under
the flow of $\changedX = \Kz +\sum_i\alpha_i K_i$, the set $I^\pm(C)\cap
\Mext$ coincides with $\Mext$, if non-empty.
\end{Lemma}
\begin{proof}
The null achronal boundaries $\dot I^\mp(C)\cap \Mext$ are
invariant
under the flow of $\changedX $. This is compatible with Lemma~\ref{LTo}
if
and only if $\dot I^\mp(C)\cap \Mext=\emptyset$. If $C$
intersects $
I^+(\Mext) $ then $I^-(C)\cap \Mext $ is non-empty, hence
$I^-(C)\supset \Mext$ since $\Mext$ is connected. A similar argument applies if $C$
intersects $
I^-(\Mext)$.
\end{proof}

We have the following strengthening of Lemma~\ref{Loncew}:

\begin{Lemma}
\label{Lnozeros} Let $\alpha_i \in \R$. If  {$(\doc,\fourg)$}
is
chronological,
then there exists no nonempty set $N$ which is  invariant
under the flow of $\Kz +\sum_i\alpha_i K_i$ and which is
included in
a compact set
 $C\subset\doc$.
\end{Lemma}

\begin{proof}
Assume that $N \subset \doc$ is not empty. {}From
Lemma~\ref{Lvis} we
obtain $\Mext \subset I^+ (N)$, hence $I^+(\Mext)\subset
I^+(N)$.
Arguing similarly with $I^-$ we infer that
$$
 \doc \subset I^+(N)\cap I^-(N)
 \;.
$$
Hence every point  $q$ in $\doc$ is in $I^+(p)$ for some $p\in
N$.
We conclude  that  $\{I^+( p)\cap C\}_{p\in N}$ is an open
cover of
$C$. Assuming compactness, we may then choose a finite
subcover
$\{I^+(p_i)\cap C\}_{i=1}^I$. This implies that each $p_i$ must
be
in the future of at least one $p_j$, and since there is a
finite
number of them one eventually gets a closed timelike curve,
which is not
possible in chronological space-times.
\end{proof}

Since each zero of a Killing vector provides a  compact
invariant set, from
Lemma~\ref{Lnozeros} we conclude

\begin{Corollary}
\label{CLnozeros} Let $\alpha_i \in \R$. If  {$(\doc,\fourg)$}
is
chronological,
then  Killing vectors of the form  $\Kz +\sum_i\alpha_i K_i$
have no zeros in $\doc$
\end{Corollary}

\bigskip

\section{Horizons and domains of outer communications in regular space-times}

In this section we analyze the structure of a class of horizons, and of domains of outer communications.

\subsection{Sections of horizons}
 \label{ssNzK1}
The aim of this section is to establish the existence of cross-sections of the
event horizon with good properties.

By standard causality theory the future event horizon
$
 \mcH^+ =\dot
 I^-(\mcMext)
$
(recall that $\dot I^\pm$ denotes the boundary of $I^\pm$) is the union of
Lipschitz topological hypersurfaces. Furthermore, through every point $p\in
\mcHp $ there is a future inextendible null geodesic entirely contained in
$\mcHp $ (though it may leave $\mcHp $ when followed to the past of
$p$). Such geodesics are called \emph{generators}.
\index{generator}%
A topological
submanifold $S$ of $\mcHp $ will be called a \emph{local section}, or
simply \emph{section}, if $S$ meets the generators of $\mcHp $
transversally; it will be called a \emph{cross-section}
\index{cross-section}%
if it meets all the
generators precisely once. Similar definitions apply to any null achronal
hypersurfaces, such as $\mcH^-$ or $\mcE^\pm$.

We start with the proof of existence of  sections of the event horizon which
are moved to their future by the isometry group. The existence of such
sections has been claimed in Lemma~5.2 of~\cite{ChAscona}; here we give the proof of a somewhat more general result:

\begin{Proposition}
\label{PWald3} Let $\mcH_0\subset \mcH:=\mcH^+\cup \mcH^-\equiv \dot
I^-(\Mext)\cup \dot I^+(\Mext)$ be a connected component of the  event
horizon $\mcH$ in a  space-time $(\mcM,\fourg)$ with
stationary\kk{} Killing vector $\Kz $, and suppose that there exists a compact
cross-section $S$ of $\mcH_0$ satisfying
$$
 S \subset \mcE_0:=\mcH_0\cap I^+(\Mext)
  \;.
$$
Assume that
\begin{enumerate}
\item either
$$
 \odoc \cap I^+(\Mext) \ \textrm{is strongly causal},
$$
\item or there exists in $\doc$ a spacelike  hypersurface
    $\hyp\supset\Sext$, achronal in $\doc$, so that $S$ above coincides
    with the boundary of $\ohyp$:
    $$S= \pohyp \subset \mcEp\;.
$$
\end{enumerate}
Then there
exists a compact Lipschitz  hypersurface $\Sz$  of $\mcE_0$ which is
transverse to both the stationary\kk{} Killing vector field $\Kz $ and to the
generators of $\mcE_0$, and which meets every generator of $\mcE_0$
precisely once; in particular
$$ \mcE_0=\cup_t \phi_t(\Sz)
\;.
$$
\end{Proposition}

\proof
Changing time orientation if necessary, and replacing $\mcM$ by
$I^+(\Mext)\setminus (\mcH\setminus \mcH_0)$, we can without loss of
generality assume that $\mcE= \mcE_0=\mcH_0=\mcH=\mcH^+$.
Choose a point $p\in \Mext$, where the Killing vector $\Kz $ is timelike, and let
$$
 \gamma_p=\cup_{t\in  \R}\phi_t(p)
$$
be the orbit of $\Kz $ through $p$. Then $I^-(S)$ must intersect
$\gamma_p$ (since $\mcE_0$ is contained in the future of $\Mext$).
Further, $I^-(S)$ cannot contain all of $\gamma_p$, by
Lemma~\ref{Ponce} or by part 2 of Lemma \ref{Loncew}.  Let $q \in
\gamma_p$ lie on the boundary of $I^-(S)$, then $I^+(q)$ cannot contain
any point of $S$, so it does not contain any complete null generator of
$\mcE_0$. On the other hand, if $I^+(q)$ failed to intersect some
generator of $\mcE_0$, then (by invariance under the flow of $\Kz $) each
point
of $\gamma_p$ would also fail to intersect some generator. By
considering a sequence, $\{q_n=\phi_{t_n}(q)\}$, along $\gamma_p$ with
$ t_n \to -\infty$, one would obtain a corresponding sequence of horizon
generators lying entirely outside the future of $\{q_n\}$. Using
compactness, one would get an ``accumulation generator" that lies
outside the future of all $\{q_n\}$ and thus lies outside of
$I^+(\gamma_p)=I^+(\Mext)$, contradicting the fact that $S$ lies to the
future of $\Mext$.

Set
$$
 \Sz:= \dot I^+(q)\cap \mcE_0
 %\;,
  %\qquad \Stp:= \phi_t(\Sonep) =\dot I^+(\phi_t(q))\cap \mcE
 \;,
$$
and we have just proved that every generator of $\mcE_0$ intersects
$\Sonep$ at least once.

The fact that the only null geodesics tangent to $\mcE_0$ are the
generators of $\mcE_0$ shows that the generators of $\dot I^+(q)$
intersect $\mcE_0$ transversally. (Otherwise a generator of $\dot I^+(q)$
would become a generator, say $\Gamma$, of $\mcE_0$. Thus
$\Gamma$ would leave $\mcE_0$ when followed to the past at the
intersection point of $\dot I^+(q)$ and $\mcE_0$, reaching $q$, which
contradicts the fact that $\mcE_0$ lies at the boundary of $I^-(\Mext)$.) As
in~\cite{ChDGH}, Clarke's Lipschitz implicit function
theorem~\cite{Clarke:optimization} shows now that $\Sone $ is a Lipschitz
submanifold intersecting each horizon generator; while the argument just
given shows that it intersects each generator at most one point. Thus,
$\Sone $ is a cross-section with respect to the null generators. However,
$\Sone $ also is a cross-section with respect to the flow of $\Kz $,
because for all $t$ we have
$$
 \phi_t(\Sone)=  \dot I^+(\phi_t(q))\cap \mcE
 \;,
$$
%  $ with respect to the flow of $\Kz $ by parameter $t>0$ is
%equivalent to moving $ q$ "forward in time" by $t$ to $\phi_t(q)$,
and for $t>0$ the boundary of $I^+(\phi_t(q))$ is contained within $I^+(q)$.
In other words, $\phi_t(\Sone  )$ cannot intersect $\Sone $, which is
equivalent to saying that each orbit of the flow of $\Kz $ on the horizon
cannot intersect $\Sone $ at more than one point.
%For further reference we
%note that we have just proved the following result, which ends the proof of
%
%\qed
%
%\begin{Lemma}
%$
%\forall p \in \Mext \quad
% \mcE_0 \approx \R\times \Sonep
%\;,
%$
%
%with the flow of $\changedX$ acting by translations along the $\R$ factor.
%
%\qed
%
%\end{Lemma}
On the other hand, each orbit must intersect $\Sone $ at least once by the
type of argument already given --- one will run into a contradiction if
complete Killing orbits on the horizon are either contained within $I^+(q)$
or lie entirely outside of $I^+(q)$.
\qed

\medskip

Now, both $S$ and $\Sz$ are compact cross-sections of $\mcE_0$.
Flowing along the generators of the horizon, one obtains:

\begin{Proposition}
 \label{SSzdiff}
 $S$ is homeomorphic to $\Sz$.
\end{Proposition}

\medskip

We note that so far we only have a $C^{0,1}$ cross-section of the horizon,
and in fact this is the best one can get at this stage, since this is the natural
differentiability of $\mcE_0$. However, if $\mcE_0$ is smooth, we claim:

\begin{Proposition}
\label{PWald4} Under the hypotheses of Proposition~\ref{PWald3},
assume moreover that $\mcE_0$ is smooth, and that $\doc$ is globally
hyperbolic. Then $\Sz $ can be chosen to be smooth.
\end{Proposition}

\proof The result is obtained  with the following regularization argument:
Choose a point  $p\in \Mext$, such that the section $S$ of
Proposition~\ref{PWald3} does \emph{not} intersect the future of $p$.   Let the
function $u$ be the retarded time associated with the orbit $\gamma_p$
through $p$ parameterized by the Killing time from $p$; this is defined as
follows: For any $q\in \mcM$ we consider the intersection $J^-{(q)}\cap
\gamma_p$. If that intersection is empty we set $u(q)=\infty$. If $J^-(q)$
contains $\gamma_p$ we set $u(q)=-\infty$. Otherwise, as $\dot J^-(q)$ is
achronal, the set $\dot J^-{(q)}\cap \gamma_p$ contains precisely one
point $\phi_\tau(p)$ for some $\tau$. We then set $u(q)=\tau$. Note that,
with appropriate conventions, this is the same as setting
\bel{udef}
 u(q)=\inf\{t:\phi_t(p)\in J^-{(q)}\}
 \;.
\ee
 It follows from the definition of $u$ that we have, for all $r$,
\bel{flowp}
 u(\phi_t(r))= u(r)+t \;.
\ee
In particular, $u$ is differentiable in the direction tangent
to the orbits of $\changedX_\kl 0$, with
\bel{maindef}
 \Kz  (u) = \fourg(\Kz , \nabla u) = 1
 \;,
\ee
everywhere.

The proof  of Proposition~\ref{PWald3} shows that $u$ is finite
in a neighborhood of $\mcE_0$; let
$$
 \Sz=u^{-1}(0)\cap \mcE_0\;,
$$
and let $\mcO$ denote a conditionally compact
neighborhood of $\Sz$ on which $u$ is finite; note that $\Sz$ here is a $\phi_t[\Kz]$--translate of the section $\Sz $ of Proposition~\ref{PWald3}.

Let $n$ be the field of future directed tangents to the generators of
$\mcE_0$, normalized to unit length with some auxiliary smooth Riemannian
metric on $\mcM$.  For $q\in \Sz $ let $\mcN_q\subset T_q\mcM$ denote the
collection of all similarly normalized null vectors that are tangent to an
achronal past directed null geodesic $\gamma$ from $q$ to
$\phi_{u(q)}(p)$, with $\gamma$ contained in $\doc$ except for its initial
point. (If $u$ is differentiable at $q$ then $\mcN_q$ contains one single
element, proportional to $\nabla u$, but $\mcN_q$ can contain more than one
null vector in general.) We claim that there exists $c>0$ such that
\bel{lowb} \inf_{q\in \Sz , l_q\in \mcN_q} \fourg(l_q,n_q)\ge c>0
  \;.
 \ee
Indeed, suppose that this is not the case; then there exists a sequence
$q_i\in \Sz $ and a sequence of past directed null achronal geodesic
segments $\gamma_i$ from $q_i$ to $p$, with tangents $l_i$
at $q_i$, such that $\fourg(l_i,n)\to 0$. Compactness of $\Sz $ implies that
there exists $q\in \Sz $ such that $q_i\to q$.

Let $\gamma$ be an accumulation curve of the $\gamma_i$'s
passing through $q$.   By hypothesis, $\mcE_0$ is a smooth null
hypersurface  {contained in} the boundary of $\doc$, with $q\in
\mcE_0$. This implies that either $\gamma$ immediately enters
$\doc$, or $\gamma$ is a subsegment of a generator of $\mcE_0$
through $q$. In the latter case $\gamma$ intersects $S$ when
followed from $q$ towards the past, and therefore the
$\gamma_i$'s intersect $\dot J^-(S)\cap \doc$ for all $i$ large
enough. But this is not possible since $S\cap
J^+(p)=\emptyset$. We conclude that there exists $s_0>0$ such
that $\gamma(s_0)\in\doc$. Thus a subsequence, still denoted by
$\gamma_i(s_0)$, converges to $\gamma(s_0)$, and global
hyperbolicity of $\doc$ implies that  the $\gamma_i$'s converge
to an achronal null geodesic segment $\gamma$ through $p$,
with tangent $l$ at $\Sz$ satisfying $\fourg(l,n)=0$. Since
both $l$ and $n$ are null we conclude that $l$ is proportional
to $n$, which is not possible as the intersection must be
transverse, providing a contradiction, and establishing
\eq{lowb}.

Let $\mcO_i$, $i=1,\ldots,N$, be a family of coordinate balls of radii
$3r_i$ such that the balls of radius $r_i$ cover $\overline{\mcO}$, and let
$\varphi_i$ be an associated partition of unity; by this we mean that the
$\varphi_i$'s are supported in $\mcO_i$, and they sum to one on $\mcO$.
For $\epsilon\le r:=\min r_i$ let
$\varphi_\epsilon(x)=\epsilon^{-n-1}\varphi(x/\epsilon)$ (recall that the
dimension of $\mcM$ is $n+1$), where $\varphi$ is a positive smooth
function supported in the ball of radius one, with integral one.
Set
\bel{uepsdef}
  \ue := \sum_{i=1}^N \varphi_i \;\varphi_\epsilon * u
  \;,
\ee
where $*$ denotes a  convolution in local coordinates. Strictly speaking,
$\varphi_\epsilon$ should be denoted by $\varphi_{\epsilon,i}$, as it
depends explicitly on the local coordinates on
$\mcO_i$, but we will not overburden the notation with yet another index.%
\footnote{This is admittedly somewhat confusing since, e.g.,
$\sum_{i=1}^N \varphi_i \;\varphi_\epsilon * u\ne (\sum_{i=1}^N \varphi_i)
\;\varphi_\epsilon * u$.} Then $\ue $ tends uniformly to $u$. Further, using
the Stokes theorem for Lipschitz functions~\cite{Morrey},
\beal{eq12}
  d\ue  &=&   \sum_{i=1}^N \Big\{\varphi_\epsilon * u \;d \varphi_i  + \varphi_i \;\varphi_\epsilon * du \Big\}
  \\
  &=&  \sum_{i=1}^N \Big\{\underbrace{(\varphi_\epsilon * u -u)}_{I}\;d \varphi_i  + \varphi_i \;\underbrace{\varphi_\epsilon * du} _{II}\Big\}
  \;,
 \eean
where we have also used $\sum_id \varphi_i=d\sum_i \varphi_i=d1=0$.
It immediately follows that the term $I$ uniformly tends to zero as $\epsilon$
goes to zero. Now, the term $II$, when contracted with $\Kz $, gives a
contribution
\beal{Ktrv}
 i_{\Kz }(\varphi_\epsilon*du)(x) & = &  \int_{|y-x|\le \epsilon} K^i_\kl 0 (x)\,\partial_i u(y) \varphi_\epsilon(x-y) d^{n+1}y
\\& = &  \int_{|y-x|\le \epsilon} \Big[\underbrace{(K^i_\kl 0 (x)- K^i_\kl 0 (y))}_{=O(\epsilon)}\partial_i u(y)
 \nonumber
 \\
 \nonumber
 && \phantom{xxxxxxxx}
+\underbrace{K^i_\kl 0 (y)\partial_i u(y)}_{=1 \  \mbox{\scriptsize by \eq{maindef}}}\Big]
 \varphi_\epsilon(x-y) d^{n+1}y
 \nonumber
 \\
  & = & 1+O(\epsilon)
  \;.
\eean
It follows that, for all $\epsilon$ small enough, the differential $d\ue $ is
nowhere vanishing, and that $\Kz $  is transverse to the level sets of $\ue
$.

To conclude, let $n$ denote any future directed causal smooth vector field
on $\mcO$ which coincides with the field of tangents to the null generators
of $\mcE_0$ as defined above. By \eq{lowb} the terms $II$ in the formula
for $d\ue $, when contracted with $n$, will give a contribution
\beal{trnsveps}
 &&
 \\
 \nonumber
  i_n(\varphi_\epsilon*du)(x) & = &     \int_{|y-x|\le \epsilon}
[\underbrace{(n^i(x)- n^i (y))}_{=O(\epsilon)}\partial_i u(y)+\underbrace{n^i
(y)\partial_i u(y)}_{\ge c}] \varphi_\epsilon(x-y) d^{n+1}y
 \\
  & \ge  & c+O(\epsilon)
  \;,
\eean
and transversality of the generators of $\mcE_0$ to the level sets of $\ue
$, for $\epsilon$ small enough, follows.
\qed

\bigskip

\subsection{The structure of the domain of outer communications}
 \label{ssNzK}
\kk{this subsection uses asymp flatness}
The aim of this section is to establish   the product structure of {\regular}
domains of outer communication, Theorem~\ref{Tgt} below. The analysis
here is closely related to that of~\cite{ChWald1}.

As in Section~\ref{ssZKV}, we  assume the existence of a commutative group of
isometries $\R\times \T^{s-1}$ with $s\ge 1$. We use the notation there, with $\Kz$
timelike in $\Mext$, and  each $K_\kl i$  spacelike in $\doc$.

Let $r=\sqrt{\sum_i(x^i)^2}$
be the radius function in $\Mext$. By the
asymptotic analysis of~\cite{ChMaerten} there exists $R$ so that for $r\ge
R$ the orbits of the $K_\kl i$'s are entirely contained in $\Mext$, so that
the function
$$
 \hat r(p) = \int_{g\in \T^{s-1}} r( g(p)) d\mu_g
 \;,
$$
is well defined, and invariant under $\T^{s-1}$. Here $d\mu_g$ is the
translation invariant measure on $\T^{s-1}$ normalized to total volume one,
and $g(p)$ denotes the action on $\mcM$ of the isometry group
generated by the $K_\kl i$'s. Similarly, let $t$ be any time function on
$\doc$, the level sets of which are asymptotically flat\kk{} Cauchy surfaces.
Averaging over $\T^{s-1}$ as above, we obtain a new time function $\hat
t$, with asymptotically flat\kk{} level sets, which is invariant under $\T^{s-1}$.
(The interesting question, whether or not the level sets of $\hat t$ are
Cauchy, is irrelevant for our further considerations here.) It is then easily
seen that, for $\sigma$ large enough, the level sets
$$
 \hat S_{\tau ,\sigma}:=\{\hat t=\tau,\hat r = \sigma\}
$$
are smooth embedded spheres included in $\Mext$.

Throughout this section we assume that $(\mcM,\fourg)$ is {\regular}.
Let $\hyp$ be as in the definition of regularity, thus $\hyp$ is an
asymptotically flat\kk{}
\kk{}
spacelike acausal hypersurface in $\doc$ with compact
boundary, the latter coinciding with a compact cross-section of $\mcE^+$.
Deforming $\hyp$ if necessary, without loss of generality we may assume
that $\hyp\cap \Mext$ is a level set of $\hat t$. We choose $R$ large
enough so that $\hSzR$ is a smooth sphere, and so that the slopes of light
cones on the $\hSts$'s, for $\sigma\ge R$, are bounded from above by two, and from below
by one half, and redefine $\Sext$ so that $\partial \Sext= \hSzR$.

Consider
$$
 \mcC^+:=(\dot J^+(\hSzR)\setminus \Mext)\cap \doc
 \;.
$$
Then $\mcC^+$ is a null, achronal, Lipschitz hypersurface generated by null
geodesics initially orthogonal to $\hSzR$. Let us  write $\phi_t$ for
$\phi_t[\Kz] $, and set
$$
  \qquad \Ct:= \phi_t(\mcC^+)\;;
$$
we then have
$$
 \Ct:=(\dot J^+(\hStR)\setminus \Mext)\cap \doc
 \;,
$$
(recall that the flow of $\Kz$ consists of  translations in $t$ in $\Mext$) which implies that every orbit of $\Kz $ intersects $\mcC^+$ at most once.

Since $\hyp$ is achronal it partitions $\doc$ as
\bel{partition}
\doc= \hyp \cup
 I^+(\hyp;\doc)\cup I^-(\hyp;\doc) \ \text{(disjoint union)}
 \;.
\ee
Indeed, as $\doc$ is globally hyperbolic, the boundaries
$(\dot I^\pm(\hyp)\setminus \hyp)\cap \doc$
  are generated by null
geodesics with end points on edge($\hyp$)$\cap \doc
=\emptyset$.

We claim that every orbit of $\Kz$ intersects $\hyp$. For this, recall that for
any $q$ in $\doc$ there exist points $p_\pm\in \Mext$ such that $q \in
I^\mp (p_\pm)$. Since the flow of $\Kz$ in $\Mext$ is by time translations
there exist $t_\pm \in \R$ so that $  \phi_{t_\pm}(p_\pm )\in \Sext$. Hence
$\phi_{ t_\pm}(q)\in I^\mp(\Sext)$, which shows that every orbit of $\Kz$
meets both the future and the past of $\hyp$. By continuity and
\eq{partition} every orbit  meets $\hyp$ (perhaps more than once). Hence
\bel{conclu}
 \doc = \cup_t \phi_t(\hyp)
 \;, \quad \odoc\cap I^+(\Mext) =  \cup_t \phi_t({\ohyp})
\ee
(for the second equality  Proposition~\ref{PWald3} has been used).
Setting $\Mint=\doc\setminus \Mext$, one similarly obtains
\beal{partition2}
 &
\Mint= \Cp \cup
 I^+(\Cp;\Mint)\cup I^-(\Cp;\Mint) \ \text{(disjoint union)}\;,
 &
 \\
 &
 \Mint= \cup_t \phi_t(\Cp)
 \;.
 &
\eeal{conclu2}
By hypothesis $\overline{\hyp\setminus \Sext}$ is compact and so, by the
first part of Lemma~\ref{Loncew}, there exists $p_-\in \Mext$ such that
\bel{goodpm}
 \overline{\hyp\setminus \Sext}\subset I^+(p_-)
 \;.
 \ee
%.
Choose $t_-<0$ so that $p_-\in I^+(\hStmR)$; we obtain that
$\overline{\hyp\setminus \Sext}\subset I^+(\hStmR)$, hence
$$
 \overline{\hyp\setminus \Sext} \subset I^+(\Ctm)
 \;.
$$

Since $\hSzR\subset \hyp$  we have $\Cp\subset I^+
(\hyp)$. By acausality of $\hyp$ and \eq{partition} we infer that $\overline{\hyp\setminus \Sext} \subset I^-(
\mcC^+)$, and hence $\phi_{t_-}(\overline{\hyp\setminus \Sext}) \subset I^-(\Ctm)$.

So,  for $p\in \overline{\hyp\setminus \Sext}$ the orbit segment
$$
 [t_-,0]\ni t\mapsto
 \phi_t\nic (p)
$$
starts in the past of $\Ctm$ and finishes to its future. {}From \eq{conclu} we
conclude that
\bel{contained}
 \overline{\Ctm} \subset \cup_{t\in[t_-,0]} \phi_t\nic (\overline{\hyp\setminus \Sext})
 \;;
\ee
equivalently,
$$
 \overline{\Cp} \subset \cup_{t\in[0,-t_-]} \phi_t\nic (\overline{\hyp\setminus \Sext})
 %\overline{ \phi_t\nic (\hyp\setminus \Sext)}
 \;.
$$
As the set at the right-hand-side is compact, we have established:

\begin{Proposition}
 \label{PCtmco} Suppose that $(\mcM,\fourg)$ is {\regular}, then
$\overline{\Cp}$ is compact.
\end{Proposition}

We are ready to prove now the following version of point 2 of Lemma~5.1 of~\cite{ChAscona}:

\begin{Theorem}[Structure theorem]%
\index{structure theorem}%
 \label{Tgt}
Suppose that $(\mcM,\fourg)$ is {\aregular} stationary\kk{} space-time invariant
under a commutative group of isometries $\R\times \T^{s-1}$, $s\ge 1$,
with the stationary\kk{} Killing vector $\Kz $ tangent to the orbits of the $\R$
factor. There exists on $\doc $ a smooth time function $t$, invariant under
$ \T^{s-1}$,  which together with the flow of $\Kz$ induces the
diffeomorphisms
\bel{gooddecomp}
 \doc \approx \R \times \hypo
 \;,\qquad
\overline{ \doc}{\cap I^+(\Mext)} \approx \R \times \ohypo
 \;,
\ee
where $\hypo:=t^{-1}(0)$ is asymptotically flat\kk{}, (invariant under $ \T^{s-1}$),
with the boundary $\partial \ohypo$ being a compact cross-section of
$\mcE^+$.  The smooth hypersurface with boundary $\ohypo$ is acausal,
spacelike up-to-boundary, and the flow of $\Kz $ is a translation along the
$\R$ factor in \eq{gooddecomp}.
\end{Theorem}

\proof
{}From what has been said,
every orbit of $\Kz $ through $\doc \setminus \Mext$ intersects $\Cp$
precisely once.
For $p\in \doc \setminus \Mext$ we let $u(p)$ be the unique real number
such that $\phi_{u(p)}(p)\in \Cp$, while for $p\in \Mext$ we let $u(p)$ be the
unique real number such that $\phi_{u(p)}(p)\in \Sext$. The function
$u:\doc\to\R$ is Lipschitz, smooth in $\Mext$, with achronal level sets
transverse to the flow of $\Kz $, and provides a homeomorphism
$$
 \doc \setminus \Mext \approx \R \times \Cp
 \;, \qquad
 \doc \approx \R \times (\Cp\cup \Sext)
 \;.
$$
The desired hypersurface $\hypo$ will be a small spacelike smoothing of
$u^{-1}{(0)}$, obtained by first deforming the metric $\fourg$ to a metric
$\fourge$, the null vectors of which  are spacelike for $\fourg$. The
associated corresponding function $\ue $ will have Lipschitz level sets
which are uniformly spacelike for $\fourg$. A smoothing of $\ue $ will
provide the desired function $t$. The details are as follows:

We start by finding a smooth hypersurface, not necessarily spacelike,
transverse to the flow of $\changedX$. We shall use the following general
result, pointed out to us by R.~Wald (private communication):

\begin{Proposition}
 \label{PWald}
Let $S_0$ be a two-sided, smooth,
hypersurface in a manifold $
M$ with an open neighborhood $\mcO$ such that $M\setminus \mcO$
consists of two disconnected components $M_-$ and $M_+$. Let $X$ be
a complete vector field on $M$ and suppose that there exists $T>0$ such
that for every orbit $\phi_t(p)$ of $X$, $t\in \R$, $p\in M$, there is an
interval $ [t_0, t_1]$ with $(t_1 - t_0) < T$ such that $\phi_t(p)$ lies in
$M_-$ for all $t<t_0$, and $\phi_t(p)$ lies in $M_+$ for all $t>t_1$. If $M$
has a boundary, assume moreover that $\partial S_0 \subset \partial M$,
and that $X$ is tangent to $\partial M$. Then there exists a smooth
hypersurface $S_1 \subset M$ such that every orbit of $X$ intersects
$S_1$ once and only once.
\end{Proposition}

\proof
Let $f$ be a smooth function with the property that $f=0$ in $M_-$, $0\le
f\le 1$ in $\mcO$, and $f=1$ in $M_+$; such a function is easily
constructed by introducing Gauss coordinates, with respect to some
auxiliary Riemannian metric, near $S_0$. For $t\in \R$ and $p\in M$ let
$\phi_t(p)$ denote the flow generated by $X$. Define $F:M\to\R$ by
$$
 F(p)= \int_{-\infty}^0f\circ \phi_s(p) ds
 \;.
$$
Then $F$ is a smooth function on $M$ increasing monotonically from zero
to infinity along every orbit of $X$. Furthermore $F$ is strictly increasing
along the orbits at points at which $F \geq T$ (since such points must lie in
$M_+$, where $f=1$). In particular, the gradient of $F $ is non-vanishing at
all points where $F \geq T$. Setting $S_1=\{F=T\}$, the result follows.
\qed

\medskip

Returning to the proof of Theorem~\ref{Tgt}, we use
Proposition~\ref{PWald} with $X=\Kz$,
$$
 M=\overline{ \overline{\doc}{\cap I^+(\Mext)} \setminus\Mext}
\;,
$$
and $S_0=\hyp \cap M$. Letting $t_-$ be as in \eq{contained} we set
$$
 \mcO:= \cup_{t\in(t_-,-t_-)}\phi_t(\hyp)
 \;;
$$
%,
by what has been said, $\mcO$ is an open neighborhood of $\hyp$. Finally
$$
 M_-:= \cup_{t\in(-\infty,t_-]}\phi_t(\hyp)\;, \quad
 M_+:= \cup_{t\in[-t_-,\infty)}\phi_t(\hyp)\;.
$$
It follows now from Proposition~\ref{PWald}  that there exists
a hypersurface $S_1\subset M$ which is transverse to the flow
of $\Kz$.

Let $\hat T$ be any  smooth, timelike vector field defined
along $S_1$, and define the smooth timelike vector field $T$ on
$M$ as the unique solution
of the Cauchy problem%
\bel{hTdef}
 \mcL_{\Kz} T = 0\;, \quad T= \hat T \ \mbox{ on $S_1$}
 \;.
\ee
Since the flow of $\Kz$ acts by time translations on $\Mext$,
it is straightforward to extend $T$ to a smooth vector field
defined on $\mcM$, timelike wherever non vanishing, still
denoted by $T$, which is invariant under the flow of $\Kz$,
the support of which on $\hyp$ is compact.
Replacing $T$ by its average over $\T^{s-1}$, we can
assume that $T$ is invariant under the action of $\T^{s-1}$.

For all $\epsilon \ge0 $ sufficiently small, the formula
\bel{fourgedef}
 \fourge(Z_1,Z_2) = \fourg(Z_1,Z_2)- \epsilon \fourg(T,Z_1)\fourg(T,Z_2)
 \;.
\ee
defines a Lorentzian,  $\R\times \T^{s-1}$ invariant metric on the manifold
with ($\fourge$--timelike) boundary $\overline{\doc}{\cap I^+(\Mext)}$. By
definition of $\fourge$, vectors which are causal for $\fourg$ are timelike
for $\fourge$. Wherever $T\ne 0$ the light cones of $\fourge$ are
spacelike for $\fourg$, provided $\epsilon \ne 0$.

Since $\fourg$-causal curves are also $\fourge$-causal,
$(\doc,\fourge)$ is also a domain of outer communications with
respect to $\fourge$.

Set
$$
 \Cpe=(\dot J^+_\epsilon(\hSzR)\setminus \Mext)\cap \doc
 \;,
$$
where we denote by $  J^+_\epsilon (\Omega)$ the  future of a set $\Omega$
with respect to the metric $\fourge$. Then the $\Cpe$'s are Lipschitz,
$\fourg$-spacelike wherever differentiable, $\T^{s-1}$ invariant,
hypersurfaces.
Continuous dependence of geodesics upon the metric together with Proposition~\ref{PCtmco} shows that the $\Cpe$'s accumulate at $\Cp$ as $\epsilon$ tends to zero.

Let $\ue :M\to \R$ be defined as in \eq{udef} using the metric $\fourge$
instead of $\fourg$. As before we have
\bel{uepsco}
 \ue (\phi_t(p))= \ue (p)+t\;, \ \mbox{so that} \
 \Kz(\ue )=1
 \;.
\ee
We perform a smoothing procedure as in the proof of
Proposition~\ref{PWald4}, with $\mcO$ there replaced by a conditionally
compact neighborhood of $\Cp$. The vector field $\hat T$ in \eq{hTdef} is
chosen to be timelike on $\overline \mcO$; the same will then be true
of $T$. Analogously to \eq{uepsdef} we set
\bel{uepsdef2}
  u_{\epsilon,\eta}:= \sum_{i=1}^N \varphi_i \;\varphi_\eta* \ue
  \;,
\ee
so that the $u_{\epsilon,\eta}$'s converge uniformly on $\mcO$ to $\ue $ as $\eta$
tends to zero. The calculation in \eq{Ktrv} shows that
$$
 \Kz(u_{\epsilon,\eta})\ge \frac 12
 %\;.
$$
for $\eta$ small enough, so that the level sets of $u_{\epsilon,\eta}$ near
$\Cp$ are transverse to the flow of $\Kz$.

It remains to show that the level sets of $\uee$ are spacelike. For this we
start with some lemmata:

\begin{Lemma}
 \label{Lsuee}
Let  $\fourg$ be a Lipschitz-continuous metric on  a coordinate ball
$B(p,3r_i)\equiv \mcO_i$ of coordinate radius $3r_i$. There exists a constant $C$
such that for any $q \in B(p,r_i)$ and for any timelike, respectively causal,
vector $N_q=N^\mu_q
\partial_\mu\in T_q \mcM$ satisfying
\bel{Ncond}
  \sum_\mu (N^\mu_q)^2=
  1
\ee
there exists a timelike, respectively causal, vector field $N=N^\mu
\partial_\mu$ on $B(p,2r_i)$ such that for all points $y,z\in B(p,2r_i)$ we
have
\bel{Ncontr}
  |N^\mu_y-N^\mu_z| \le C |y-z| \;, \quad
  C^{-1}\le \sum_\mu (N^\mu_y)^2
  \le C
  \;.
\ee
\end{Lemma}

\begin{proof}
We will write both $N_q^\mu$ and $N^\mu(q)$ for the coordinate components of a
vector field at $q$. For $\nu=0,\ldots,n$, let $e_\kl \nu=e^\mu_\kl \nu
\partial_\mu $ be any Lipschitz-continuous $ON$ basis for $\fourg$ on $\mcO_i$.
there exists a constant $c$ such that on $B(p,2r_i)$ we have
$$
|e^\mu_\kl \nu(y)-e^\mu_\kl \nu(z)| \le c |y-z|
\;.
$$
Decompose $N_q$ as $N_q=N^\kl \nu _qe_\kl \nu(q)$, and for $y \in
\mcO_i$ set $N_y = N^\kl \nu _qe_\kl \nu(y)$; \eq{Ncontr}
easily follows.
\end{proof}

\begin{Lemma}
 \label{Lsuee2}
Under the hypotheses of Lemma~\ref{Lsuee}, let $f$ be differentiable on
$\mcO_i$. Then $\nabla f$ is timelike past directed on $B(p,2r_i)$ if and
only if $N^\mu
\partial_\mu f<0$ on $\mcO_i$ for all causal past directed vector fields
satisfying \eq{Ncond} and \eq{Ncontr}.
\end{Lemma}

\begin{proof}
The condition is clearly necessary. For sufficiency, suppose that there
exists $q\in B(p,2r_i)$ such that $\nabla f$ is null, let $N_q=\lambda \nabla
f(q)$, where $\lambda$ is chosen so that \eq{Ncond} holds, and let $N$ be
as in Lemma~\ref{Lsuee}; then $N^\mu
\partial_\mu f$ vanishes at $q$. If $\nabla f$ is
spacelike at $q$ the argument is similar, with $N_q$  chosen to be  any timelike
vector orthogonal to $\nabla f(q)$ satisfying \eq{Ncond}.
\end{proof}

Let $N$ be any $\fourg$--timelike past directed vector field satisfying
\eq{Ncond} and \eq{Ncontr}.  Returning to \eq{eq12} we find,
\beal{eq13} i_N
  d\uee
  &=&  \sum_{i=1}^N \Big\{\underbrace{(\varphi_\eta * \ue  -\ue )}_{I}\;i_N d \varphi_i
  + \varphi_i \;\underbrace{i_N(\varphi_\eta * d\ue )} _{II}\Big\}
  \;.
 \eea
For any fixed $\epsilon$, and for any $\delta>0$  we can choose
$\eta_\delta$ so that the term $I$ is smaller than $\delta$ for all
$0<\eta<\eta_\delta$.

To obtain control of $II$, we need uniform spacelikeness of $d\ue $:

\begin{Lemma}
 \label{Lfdg2}
There exists a constant $c$ such that, for $N$ as in Lemma~\ref{Lsuee},
\bel{fourgedef2}
 N^\mu \partial_\mu \ue <-c \epsilon
% \;,
\ee
almost everywhere, for all $\epsilon>0$ sufficiently small.
\end{Lemma}

\begin{proof}
Let $\{e_\kl \nu\}$ be an $\fourg$--ON frame in which the vector field $T$ of
\eq{fourgedef} equals $T^\kl 0 e_\kl 0$.  Let $\alpha_\kl \nu$ denote the
components of $d\ue $ in a frame dual to $\{e_\kl \nu\}$. In this frame we
have
$$
 \fourg=\diag(-1,1,\ldots,1)\;,\qquad \fourge =\diag(-(1+(T^\kl 0)^2 \epsilon),1,\ldots,1)
 \;.
$$
Since $d\ue $ is  $\fourge$--null and past pointing we have
$$
 \alpha_\kl 0 = \sqrt{1+(T^\kl 0)^2\epsilon}\sqrt{ \sum \alpha_\kl i^2}
 \;.
$$
The last part of \eq{uepsco} reads
$$
 \Kz^\kl 0 \alpha_\kl 0 + \Kz^\kl i \alpha_\kl i = 1
 \;.
$$
It is straightforward to show from these two equations that there exists a
constant $c_1$ such that, for all $\epsilon$ sufficiently small,
$$
 \alpha_\kl 0 > c_1^{-1}
 \;, \quad \sqrt{ \sum \alpha_\kl i^2}>c_1^{-1}\;,\quad
 \sum |\alpha_\kl \mu| \le c_1
 %\left(\frac{\partial \ue }{\partial x^\mu}\right)^2 >c_2
 \;.
$$
Since $N$ is $\fourge$ causal past directed, \eq{Ncond} and \eq{Ncontr}
together with the construction of $N$ show that there exists a constant $c_2$ such that
$$
N^\kl 0<-c_2
\;.
$$
We then have
\beaa%l{fourgedef3}
 N^\mu \partial_\mu \ue &=& N^\kl 0 \alpha_\kl 0+ N^\kl i \alpha_\kl i
  \\
  & = &
 N^\kl 0 \sqrt{1+(T^\kl 0)^2\epsilon} \sqrt {\sum \alpha_\kl i^2}+ N^\kl i \alpha_\kl i
  \\
  & = &
 N^\kl 0   (\sqrt{1+(T^\kl 0)^2\epsilon} -1) \sqrt {\sum \alpha^2_\kl i}
  +
  \underbrace{N^\kl 0   \sqrt {\sum \alpha^2_\kl i}+ N^\kl i \alpha_\kl i}
  _{< 0 \ \scriptsize \mbox{ by Cauchy-Schwarz, as $N$ is $\fourg$--timelike}}
 \\
 & < &
  - \frac {c_2}{4 c_1}  \,\inf_{\overline\mcO}  (T^\kl 0)^2\,\epsilon =: -c\epsilon
 \;,
\eeaa
for $\epsilon$ small enough.
\end{proof}

Now, calculating  as in \eq{trnsveps}, using \eq{fourgedef2},
\beaa%l{trnsveps2} &&
%\\ \nonumber
  i_N(\varphi_\eta*d\ue  )(x) & = &     \int_{|y-x|\le \eta}
[\underbrace{(N^\mu(x)- N^\mu (y))}_{\le C\eta }\partial_\mu \ue
(y)+\underbrace{N^\mu (y)\partial_\mu \ue (y)}_{\le -c\epsilon}]
\varphi_\eta(x-y) d^{n+1}y
 \\
  & \le  & -c\epsilon +O(\eta)
  \;,
\eeaa
so that for $\eta$ small enough each such term will give a contribution to
\eq{eq13} smaller than $-c\epsilon/2$. Timelikeness of $\nabla \uee$
on $\overline\mcO$ follows now from Lemma~\ref{Lsuee2}.

Summarizing, we have shown that we
can choose $\epsilon$ and $\eta$ small enough so that the  function
$u_{\epsilon,\eta}:M\to \R $ is a time function near its zero level set. It is
rather straightforward to extend $u_{\epsilon,\eta}$ to a function on
$\doc\to \R$, with smooth spacelike zero-level-set, which coincides with
$\hyp$ at large distances. Letting $\hypo$ be this zero level set, the
function $t(p)$ is defined now as the unique value of parameter $t$ so that
$\phi_t(p)\in \hypo$; since the level sets of $t$ are smooth spacelike hypersurface,  $t$ is a smooth time function. This completes the proof  of Theorem~\ref{Tgt}.
\qed

\medskip

\subsection{Smoothness of event horizons}
 \label{Sreh}
\kk{this subsection uses asymp flatness}
The starting point to any study of event horizons in stationary space-times
is a corollary to the area theorem, essentially due to~\cite{ChDGH}, which
shows that event horizons in well-behaved stationary space-times are as
smooth as the metric allows. In order to proceed, some terminology from
that last reference is needed; we restrict ourselves to asymptotically
flat\kk{} space-times; the reader is referred to~\cite[Section~4]{ChDGH}
for the general case. Let $(\bmcM,\bfourg)$ be a $C^3$ completion of
$(\mcM,\fourg)$ obtained by adding a null conformal boundary at infinity,
denoted by $\scrip$, to $\mcM$, such that $\fourg=\Omega^{-2}\bfourg$
for a non-negative function $\Omega$ defined on $\bmcM$, vanishing
precisely on $\scrip$, and $d\Omega$ without zeros on $\scrip$. Let
$\mcE^+$ be the future event horizon in $\mcM$. We say that
$(\bmcM,\bfourg)$ is $\mcE^+$--regular if there exists a neighborhood
$\mcO$ of $\mcE^+$ such that for every compact set $C\subset \mcO$ for
which $I^+(C;\bmcM)\ne \emptyset$ there exists a generator of $\scrip$
intersecting $\overline{I^+(C;\bmcM)}$ which leaves this last set when
followed to the past.
 (Compare Remark~4.4 and Definition~4.3 in~\cite{ChDGH}).

We note the following:

\begin{Proposition}
 \label{PgoodScrip}
Consider an asymptotically flat\kk{} stationary\kk{} space-time which is vacuum at
large distances, recall that $\mcEp=\dot I^-(\Mext)\cap I^+(\Mext)$.
If $\doc$ is globally hyperbolic, then $(\mcM,\fourg)$
admits an $\mcE^+$--regular conformal completion.
\end{Proposition}

\proof Let $\bmcM$ be obtained by adding to $\Mext$ the surface $\tilde
r=0$ in the coordinate system $(u,\tilde r, \theta,\varphi)$
of~\cite[Appendix~A]{Damour:schmidt} (see also~\cite{Dain:2001kn},
where the construction of~\cite{Damour:schmidt} is corrected; those results generalize without difficulty to higher dimensions). Let $t$ be
any time function on $\doc$ which tends to infinity when $\mcE^+$ is
approached, which tends to $-\infty$ when $\dot I^+(\Mext)$ is
approached, and which coincides with the coordinate $t$ in $\Mext$ as
in~\cite[Appendix~A]{Damour:schmidt}. Let
$$
 \mcO=\{p\ |\ t(p)>0\}\cup
I^+(\mcE^+)\cup \mcE^+
 \;;
$$
%,
then $\mcO$ forms an open neighborhood of $\mcE^+$. Let $C$ be any
compact subset of $\mcO$ such that $I^+(C;\bmcM)\cap \scrip\ne
\emptyset$; then $\emptyset \ne C\cap
\doc\subset \{t>0\}$. Let $\gamma$ be any future directed causal curve from $C$ to $\scrip$,
then $\gamma$ is entirely contained in $\doc$, with $t\circ \gamma >0$. In
particular any intersection of $\gamma$ with $\partial \Mext$ belongs to
the set $\{t>0\}$, so that at each intersection point
$$
 u\circ \gamma>\inf u|_{\{t=0\}\cap \partial \Mext}=:c>-\infty\;.
$$
The coordinate $u$ of~\cite[Appendix~A]{Damour:schmidt} is null, hence
non-increasing along causal curves, so $u\circ \gamma>c$, which implies
the regularity condition.
\qed

\medskip

We are ready to prove now:

\begin{theorem}
 \label{Tsmooth}
Let $(\mcM,\fourg )$ be a smooth, asymptotically flat\kk{}, $(n+1)$--dimensional
space-time with stationary Killing vector $K_\kl 0$, the orbits of which are
complete. Suppose that $\doc$ is globally hyperbolic, vacuum at large
distances in the asymptotic region, and assume that the $\mbox{\rm null
energy condition}$ \eq{NEC} holds.
Assume that a connected component $\mcH_0$ of
$$
 \mcH:=\mcHm\cup\mcHp
$$
admits a compact cross-section satisfying $S\subset I^+(\mcMext)$. If
\begin{enumerate}
\item either
$$
 \odoc \cap I^+(\Mext) \ \textrm{is strongly causal},
$$
\item or there exists in $\doc$ a spacelike
    hypersurface $\hyp\supset\Sext$, achronal in $\doc$, so that $S$ as above
    coincides with the boundary of $\ohyp$:
    $$S= \pohyp \subset \mcEp\;,
$$
\end{enumerate}
then
$$ \cup_t \phi_t[\Kz ](S)\subset \mcH_0
$$
is a smooth null hypersurface, which is analytic if the metric is.
\end{theorem}

\begin{Remark}{\rm
The condition that the space-time is vacuum at large distances can be
replaced by the requirement of existence of an $\mcE^+$--regular
conformal completion at null infinity. }
\end{Remark}

\proof
Let $\Sigma$ be a Cauchy surface for $\doc$, and let $\bmcM$ be the
conformal completion of $\mcM$ provided by
Proposition~\ref{PgoodScrip}. By~\cite[Proposition~4.8]{ChDGH} the
hypotheses of~\cite[Proposition~4.1]{ChDGH} are satisfied, so that the
Aleksandrov divergence  $\theta_{\scriptsize{\mycal Al}}$ of $\mcE^+$, as
defined in~\cite{ChDGH}, is nonnegative. Let $S_1$ be given by
Proposition~\ref{PWald3}. Since isometries preserve area we have
$\theta_{\scriptsize{\mycal Al}}=0$ almost everywhere on
$\cup_t\phi_t(S_1)=\cup_t \phi_t(S)$. The result follows now
from~\cite[Theorem~6.18]{ChDGH}.
\qed

\subsection{Event horizons vs Killing horizons in analytic vacuum space-times}
 \label{Srt}

We have the following result, first proved by Hawking for
$n=3$~\cite{HE} (compare~\cite{FRW}
or~\cite[{Theorem~5.1}]{ChAscona}), while the result for $n\ge
4$  in the mean-non-degenerate case is due to Hollands,
Ishibashi and Wald~\cite{HIW}, see
also~\cite{VinceJimHigh,LP3,HollandsIshibashi2}:

\begin{theorem}
 \label{Torig}
Let $(\mcM,\fourg )$ be an analytic, $(n+1)$--dimensional,
vacuum space-time with complete   Killing vector $K_\kl  0$. Assume that $\mcM$ contains an analytic null hypersurface $\mcE$ with a compact cross-section $S$ transverse both to $K_\kl 0$ and to the generators of $\mcE$. Suppose that
\begin{enumerate}
\item either  $\langle \kappa \rangle_S \ne 0$, where  $\langle \kappa \rangle_S$ is defined in \eq{asg},
\item or $n=3$.
\end{enumerate}
Then there exists a
neighborhood $\mcU$ of $\mcE $  and a Killing
vector defined on $\mcU$ which is null on $\mcE$.

In fact, if $K_\kl  0$ is \emph{not} tangent to the generators of
$\mcE$, then there exist, near $\mcE$, $N$ commuting linearly independent Killing
vector fields $K_\kl 1,\ldots,K_\kl N$, $N\ge 1$, (not necessarily
complete but) with $2\pi$--periodic orbits near $\mcE$, and numbers
$\Omega_\kl 1,\ldots,\Omega_\kl N$, such that
$$
 K_\kl  0 + \Omega_\kl 1 K_\kl 1+\ldots+\Omega_\kl N K_\kl N
$$
is null on $\mcE$.
\end{theorem}

In the black hole context, Theorem~\ref{Torig} implies:

\begin{theorem}
 \label{Trig}
Let $(\mcM,\fourg )$ be an analytic, asymptotically flat\kk{}, strongly causal,
vacuum, $(n+1)$--dimensional space-time with stationary
Killing vector $K_\kl  0$, the
orbits of which are complete. Assume that $\doc$ is globally hyperbolic, that a connected component
$\mcH^+_0$ of $\mcHp$  contains a compact cross-section $S$
satisfying
$$
 S \subset  I^+(\Mext)\;,
$$
and that
\begin{enumerate}
\item either  $\langle \kappa \rangle_S \ne 0$,
\item or the flow defined by  $K_\kl 0$ on the space of the generators of  $\mcH^+_0$ is periodic.
\end{enumerate}
Suppose moreover that
\begin{enumerate}
\item[a)] either
$$
 \odoc \cap I^+(\Mext) \ \textrm{is strongly causal},
$$
\item[b)] or there exists in $\doc$ an asymptotically flat\kk{} spacelike hypersurface $\hyp$, achronal in $\doc$, so that $S$ as above coincides with the boundary of $\ohyp$:
    $$S= \pohyp \subset \mcEp\;.
$$
\end{enumerate}
If $K_\kl  0$ is \emph{\rm not} tangent to the generators of
$\mcH$, then there exist, on $\odocup$,
$N$ complete, commuting, linearly independent Killing
vector fields $K_\kl 1,\ldots,K_\kl N$, $N\ge 1$, with $2\pi$--periodic orbits, and numbers
$\Omega_\kl 1,\ldots,\Omega_\kl N$, such that the Killing vector field
$$
 K_\kl  0 + \Omega_\kl 1 K_\kl 1+\ldots+\Omega_\kl N K_\kl N
$$
is null on $\mcH_0 $.
\end{theorem}

\begin{Remark} {\em
For {\regular} four-dimensional black holes $S$ is a two-dimensional
sphere (see Corollary~\ref{Cst}), and then every Killing vector field acts
periodically on the generators of $\mcHpz$.}
\end{Remark}

\begin{proof}
Theorem~\ref{Tsmooth} shows that $\mcEpz:=\cup_t \phi_t[K_\kl 0](S) $ is
an analytic null hypersurface. By Proposition~\ref{PWald4} there exists a
smooth compact section of $\mcEpz$
which is transverse both to its generators and to the stationary\kk{} Killing vector.%
\footnote{The hypothesis of existence of such a section needs to be
added to those of~\cite[Theorem~2.1]{HIW}.}
We can thus invoke Theorem~\ref{Torig} to conclude existence of Killing
vector fields $K_\kl i$, $i=1, \ldots, N$, defined near $\mcEpz$. By
Corollary~\ref{Cdsc} and a theorem of Nomizu~\cite{Nomizu} we infer that
the $K_\kl i$'s extend globally to $\doc$. It remains to prove that the orbits
of all Killing vector fields are complete. In order to see that, we note that by
the asymptotic analysis of Killing vectors of~\cite{ChMaerten,ChBeig1}
there exists $R$ large enough so that the flows of all $K_\kl i$'s through
points in the asymptotically flat\kk{} region with $r\ge R$ are defined for all
parameter values $t\in [0,2\pi]$. The arguments in the proof of
Theorem~1.2 of~\cite{Ch:rigidity} then show that the flows $\phi_t[K_\kl
i]$'s are defined for $t\in [0,2\pi]$ throughout $\doc$. But $\phi_{2\pi}[K_\kl
i] $ is an isometry which is the identity on an open set near $\mcEpz$,
hence everywhere, and completeness of the orbits follows.
\end{proof}

\section{Stationary axisymmetric black hole space-times: the
area function}
 \label{Scosta0}
\kk{Many references to asymp flatness.
Not using simple-connectedness of doc will require a new strategy for the last part of the proof of the main thm}

As will be explained in detail below, it follows from
Theorem~\ref{Trig} together with the results on Killing vectors
in~\cite{ChBeig2,Ch:rigidity}, that
{\regular}, $3+1$ dimensional, asymptotically flat, rotating black holes have to be
axisymmetric. The next
step of the analysis of such space-times is
{the study of the area function
\index{$W$}
\bel{determinant}
 W:= -\det \Big(\fourg(K_\kl \mu, K_\kl
 \nu)\Big)_{\mu,\nu=0,1}
 \;,
\ee
with $K_\kl 0$ being the asymptotically timelike Killing
vector, and $K_\kl 1$ the axial one. Whenever $\sqrt{W}$ can be
used as a coordinate, one obtains a dramatic  simplification of
the field equations, whence the interest thereof.

The function $W$ is clearly positive in a region where $K_\kl
0$ is timelike and $K_\kl 1$ is spacelike, in particular it is
non-negative on $\Mext$. As a starting point for further considerations, one then wants to show that $W$ is
non-negative on $\doc$:}

\begin{Theorem}
 \label{Tdoc}
Let $(\mcM,\fourg )$ be a four-dimensional, analytic,
asymptotically flat, vacuum space-time with stationary Killing
vector $K_\kl 0$ and periodic Killing vector $K_\kl 1$, jointly
generating an $\R\times \Uone$ subgroup of the isometry group
of $(\mcM,\fourg )$. If $\doc$ is globally hyperbolic, then the
area function \eq{determinant} is non-negative on $\doc$,
vanishing precisely on the union of its boundary with the
(non-empty) set $\{\fourg(K_\kl 1,K_\kl 1)=0\}$.
\end{Theorem}

We also have a version of Theorem~\ref{Tdoc}, where the
hypothesis of analyticity is replaced by that of \regular ity:

\begin{Theorem}
\label{Tdoc3}   Under the remaining hypotheses of
Theorem~\ref{Tdoc}, instead of analyticity assume  that
$(\mcM,\fourg)$ is \regular. Then the conclusion of
Theorem~\ref{Tdoc} holds.
\end{Theorem}

Keeping in mind  our
discussion above, Theorem~\ref{Tdoc} follows from
Proposition~\ref{xyort} and Theorem~\ref{Tdoc2} below.
Similarly,   Theorem~\ref{Tdoc3} is a corollary of
Theorem~\ref{Tnoanalyticnew}.

\subsection{Integrability}
 \label{Sintegr}
The first key fact underlying the analysis of the area function
$W$  is the following purely local fact, observed independently
by  Kundt and Tr{\"u}mper~\cite{KundtTrumper} and by
Papapetrou~\cite{Papapetrou:ot}  in dimension four (for a
modern derivation see~\cite{Weinstein3,Heusler:book}). The
result, which does neither require $K_\kl 0$ to be stationary,
nor the $K_\kl i$'s to generate $S^1$ actions,  generalizes  to
higher dimensions as follows (compare~\cite{CarterJMP,ERWeyl}):

\begin{Proposition}
\label{xyort} Let $(\mcM,\fourg )$ be a  vacuum, possibly with
a cosmological constant, $(n+1)$--dimensional pseudo-Riemannian
manifold  with
$n-1$ linearly independent commuting Killing vector fields
$K_\kl
\mu$, $\mu=0,\ldots,n-2$. If
\bel{axcondx}
\dgt:=
\{ p\in \mcM \ | \  K_\kl{ 0} \wedge ... \wedge
K_\kl{ {n-2}}|_p =0\}\ne \emptyset
\;,
\ee
\index{${\mycal Z}_{\mbox{\scriptsize \rm dgt}}$}%
then
 \footnote{By an abuse of notation, we use the same symbols for
 vector fields and for the
associated 1-forms.}
\begin{equation}
\label{intcond} dK_\kl{\mu }\wedge K_\kl{ 0} \wedge \ldots
\wedge
K_\kl{{n-2}} =0 \;.
\end{equation}
\end{Proposition}

\begin{proof}\newcommand{\ourstar}{*}%
To fix conventions,  we use a Hodge star defined through the
formula
$$
\alpha\wedge \beta = \pm \langle \ourstar  \alpha, \beta\rangle
\mathrm{Vol}\;,
$$
where the plus sign is taken in the Riemannian case, minus  in
our
Lorentzian one, while  $\mathrm{Vol}$ is the volume form. The
following
(well {known}) identities are useful~\cite{Heusler:book};
\begin{equation}
\label{hodge1}
 **\theta=(-1)^{s(n+1-s)-1}\theta
  \;,
  \qquad \forall \theta\in\Lambda^s\;,
\end{equation}
\begin{equation}
\label{hodge2}
 i_\changedX *\theta=*(\theta\e \changedX )\;,\qquad \forall
 \theta\in\Lambda^s\;, \quad \changedX \in\Lambda^1\;.
\end{equation}
Further, for any Killing vector $K$,
\begin{equation}
\label{hodge3}
 [\Lie_K,*]=0\;.
\end{equation}
The Leibniz rule for the divergence $\delta:=*d*$ reads, for
$\theta \in
\Lambda^s$,
 \begin{align*}
\delta(\theta\e K) &=*d*(\theta\e K){\buildrel \eqref{hodge2}
                                      \over
                                      =}*d(i_K*\theta)=*(\Lie_K*\theta-i_Kd*\theta)
                                      \\
          &{\buildrel \eqref{hodge1},\eqref{hodge3}
                                      \over =}**\Lie_K\theta
                                      -*i_K(-1)^{(n+1-s+1)(n+1-(n+1-s+1))-1}**d*\theta
                                      \\
          &=(-1)^{s(n+1-s)-1}\Lie_K\theta-(-1)^{s(n+1-s)-n+1}**(\delta
         \theta\e K)\\
          &=(-1)^{s(n+1-s)-1}\Lie_K\theta+(-1)^{n+1}\delta
          \theta\e K\;.
\end{align*}
Applying this to $\theta=dK$ one obtains
 \begin{align*}
      *d*(dK\e K) &=- \Lie_K dK+(-1)^{n+1}\delta dK\e K     \\
          &= (-1)^{n+1}\delta dK\e K
          \;.
\end{align*}
As any Killing vector is divergence free, we see that
$$
 \delta d K=(-1)^{n}\Delta K=(-1)^{n+1}i_K \Ric
 \;.
$$
Assuming  that the Ricci tensor is proportional to the metric,
$\Ric=
\lambda \fourg$,  we conclude that
$$ *d*(dK\e K) =(i_K\,\lambda \fourg)\e K=0\;.$$
Let $\omega_\kl \mu$ be the $\mu$'th twist form,
$$\omega_\kl \mu :=
\ourstar (dK_\kl \mu\wedge K_\kl \mu)
 \;.
$$
%;  t
The identity
\begin{align*}
          \Lie_{K_\kl \mu}\omega_\kl \nu&=\Lie_{K_\kl
          \mu}*(dK_\kl \mu\e K_\kl \nu)                \\
          &=*(\Lie_{K_\kl \mu}dK_\kl \nu+dK_\kl
          \nu\e\Lie_{K_\kl \mu}K_\kl \nu)=0\;,
\end{align*}
together with
$$
 \Lie_{K_{\kl {\mu_1}}}(i_{K_{\kl {\mu_2}}}\ldots  i_{K_{\kl
 {\mu_{\ell}}}}\omega_{\kl {\mu_{\ell+1}}})
 =i_{K_{\kl {\mu_2}}}\ldots i_{K_{\kl {\mu_{n-1}}}}\Lie
 _{K_{\kl {\mu_{\ell}}}}\omega_{\kl {\mu_{\ell+1}}}
 =0
 \;,
$$
and with Cartan's formula for the Lie derivative, gives
\begin{equation}
\label{dii-iid}
d(i_{K_{\kl {\mu_1}}}\ldots i_{K_{\kl {\mu_{\ell}}}}\omega_{\kl
{\mu_{\ell+1}}})
 =
 (-1)^\ell i_{K_{\kl {\mu_1}}}\ldots i_{K_{\kl
 {\mu_{n-1}}}}d\omega_{\kl {\mu_{\ell+1}}}
 \; .
\end{equation}
We thus have
\begin{align*}
        d*(dK_{\kl {\mu_{1}}}\e K_{\kl {\mu_{1}}}\e\ldots \e
        K_{\kl {\mu_{{n-1}}}})
        &=d(i_{K_{\kl {\mu_{ {n-1}}}}}\ldots i_{K_{\kl
        {\mu_{2}}}}*(dK_{\kl {\mu_{1}}}\e K_{\kl {\mu_{1}}}))
        \\
        &=(-1)^{n-2}i_{K_{\kl {\mu_{n-1}}}}\ldots i_{K_{\kl
        {\mu_2}}}d\omega_{\kl {\mu_1}}               =0
        \;.
\end{align*}
So the function   $*(dK_{\kl {\mu_1}}\e K_{\kl {\mu_1}}\e
K_{\kl
{\mu_2}}\e\ldots \e K_{\kl {\mu_{n-1}}})$ is constant, and the
result follows from \eq{axcondx}.
\end{proof}

\subsection{The area function for a class of
space-times with  a commutative group of isometries}
  \label{Scosta}

The simplest non-trivial reduction of the Einstein equations by isometries,
which does \emph{not} reduce the equations to ODEs, arises when  orbits
have co-dimension two, and the isometry group is abelian. It is useful to
formulate the problem in a  general setting, with $ 1\le s\le n-1$
commuting
Killing vector fields $K_\kl \mu$, $\mu=0,\ldots,s-1$, satisfying
{the following}
orthogonal integrability  condition:%
\index{orthogonal integrability}%
\begin{equation}
\label{intcond2}
 \forall \mu=0,\ldots, s-1\qquad dK_\kl{\mu }\wedge K_\kl{ 0}
 \wedge \ldots \wedge
K_\kl{{s-1}} =0 \;.
\end{equation}
For the problem at hand,  \eq{intcond2} will hold when $s=n-1$
by Proposition~\ref{xyort}. Note further that \eq{intcond2}
with $s=1$ is the definition of staticity. So,  the analysis
that follows covers simultaneously static analytic domains of
dependence in all dimensions $n\ge 3$ (filling a gap in
previous proofs), or stationary
axisymmetric analytic four-dimensional space-times, or five
dimensional stationary analytic space-times with two further
periodic Killing vectors as in~\cite{HY}. It further covers
stationary axisymmetric \regular{} black holes in $n=3$, in
which case analyticity is not needed.

Similarly to \eq{axcondx}  we set
\index{${\mycal Z}_{\mbox{\scriptsize \rm dgt}}$}%
\bel{dgt2}
\dgt :=
\{ K_\kl{ 0} \wedge ... \wedge
K_\kl{ {s-1}} =0\}\;,
\ee
 \begin{equation}
 \label{defdeg2}
  \zh   :=\{p\in \mcM: \det\Big( \fourg(K_\kl i,K_\kl
  j)\Big)_{i,j=1,\ldots s-1}=0\}.
 \end{equation}
\index{${\widetilde{\mcZ}}$}%

In the  following result,
the proof of which builds on key ideas of
Carter~\cite{CarterJMP,CarterlesHouches}, we let $K_\kl 0$ denote the Killing vector
associated to the $\R$ factor of $\R\times \T^{s-1}$, and we let $K_\kl i$ denote
the Killing vector field associated with the    $i-th$ $S^1$ factor of $\T^{s-1}$:

\begin{Theorem}
 \label{Tdoc2}
Let $(\mcM,\fourg)$ be an $(n+1)$--dimensional, asymptotically
flat, analytic  space-time with a metric invariant under {an
action of the abelian group $G=\R\times \T^{s-1}$ with
$s$--dimensional principal orbits, $1\le s\le n-1$,
and assume
that \eq{intcond2} holds.} If  $\doc$ is globally hyperbolic,
then the function
\bel{Wdef2}
 W:= -\det\Big( \fourg(K_\kl \mu,K_\kl
 \nu)\Big)_{\mu,\nu=0,\ldots,s-1}
\ee
is non-negative on $\doc$, vanishing  on $\pdoc\cup \zh$.
\end{Theorem}

\begin{Remark}
 \label{Reman}
{\rm Here analyticity could be avoided if,  in the proof below,
one could show that one can extract out of the degenerate
$\hS_p$'s (if any) a closed embedded hypersurface.
Alternatively, the hypothesis of analyticity can be replaced by
that of non-existence of non-embedded degenerate prehorizons
within $\doc$. Moreover, one also has:
 }
\end{Remark}

\begin{Theorem}
 \label{Tnoanalyticnew}  Let $n=3$, $s=2$ and, under the
remaining conditions of Theorem~\ref{Tdoc2}, instead of
analyticity assume that $(\mcM,\fourg)$ is \regular. Then the
conclusion of Theorem~\ref{Tdoc2} holds.
\end{Theorem}

Before passing to the proof, some preliminary remarks are in order.
The fact that $\mcM \setminus \dgt  $   is open, where $\dgt$ is as
in \eq{dgt2}, together with \eq{intcond2}, establishes the
conditions of the Frobenius theorem (see, e.g.,~\cite{Hicks}).
Therefore, for every $p\notin\dgt  $ there exists a unique, maximal
submanifold (not necessarily embedded), passing through $p$ and
orthogonal to \mbox{Span}$\{K_\kl 0,...,K_\kl{s-1}\}   $, that we
denote by $\scro_p$. Carter builds his further analysis of
stationary axisymmetric black holes on the sets $\mcO_p$. This
leads to severe difficulties at the set $\zh$ of \eq{defdeg2}, which
we were not able to resolve using neither Carter's ideas, nor those
in~\cite{Weinstein1}. There is, fortunately, an alternative which we
provide below. In order to continue, some terminology is needed:

\begin{Definition}%
\index{Killing prehorizon}%
 \label{DefKH}
Let $\changedX $ be a Killing vector and set
\bel{KHd}
 \mcHN[\changedX ]:=\{ \fourg(\changedX ,\changedX )= 0\;,\ \changedX  \ne 0\}\;.
\ee
Every connected, not necessarily embedded,  null hypersurface
$\mcN_0\subset\mcHN[\changedX ]$ {to which $\changedX $ is tangent}
will be called a  $\mbox{\rm Killing prehorizon}$.
\end{Definition}

In this terminology, a {Killing horizon}
 is a  Killing prehorizon which forms an \emph{embedded} hypersurface which
\emph{coincides} with a connected component of $\mcHN[\changedX ]$.

The Minkowskian Killing vector $\partial_t-\partial_x$ provides an example
where $\mcHN$ is not a hypersurface, with every hyperplane $t+x=\const $
being a prehorizon. The Killing vector $\changedX =\partial_t +Y$ on
$\R\times \T^n$, equipped with the flat metric, where $\T^n$ is an
$n$-dimensional torus, and where $Y$ is a unit Killing vector on $\T^n$
with dense orbits, admits prehorizons which are not embedded. This last
example is globally hyperbolic, which shows that causality conditions
are not sufficient to eliminate this kind of behavior.

Our first step towards the proof of Theorem~\ref{Tdoc2} will be
Theorem~\ref{Carter69}, inspired again by some key ideas of Carter,
together with their variations by Heusler. We will assume that the $K_\kl
i$'s, $i=1,\ldots, s-1$, are spacelike (by this we mean that they are
spacelike away from their  zero sets), but no periodicity or completeness
assumptions are made concerning their orbits. This can always be
arranged locally, and therefore does not involve any loss of generality for
the local aspects of our claim; but we emphasize that our claims are global
when the
$K_\kl i$'s are spacelike everywhere.

In our analysis below we will be mainly interested in what happens in
$\doc$ where, by  Corollary~\ref{CLnozeros}, we have
 $$\zh\cap \doc = \dgt\cap \doc\;,$$
in a chronological  domain of outer communications.

{We note that $\dgt\subset\{W=0\}$, but equality does not
need to hold for  Lorentzian metrics. For example,
consider in $\R^{1,2}$, $K_\kl 0=\partial_x+\partial_t$ and
$K_\kl 1=\partial_y$; then $K_\kl 0\e\, K_\kl 1=dx\e dy-dt\e
dy\not\equiv 0$ and $W\equiv 0$.}

If the $K_\kl
i$'s generate a torus action on a stably causal manifold,%
\footnote{Let $t$ be a time-function on $(\mcM,\fourg)$; averaging $t$ over
the orbits of the torus generated by the $K_\kl i$'s we obtain a new time
function such that the $K_\kl i$'s are tangent to its level sets. This reduces
the problem to the analysis of zeros of Riemannian Killing vectors.}
it is well known
that $\zh $ is a closed, totally geodesic, timelike, stratified, embedded
submanifold of $\mcM$ with codimension of each stratum at least
two (this follows from~\cite{Kobayashi:fixed} or
\cite[Appendix~C]{ACD2}). So, under those hypotheses, within $\doc$, we will have
\index{${\mycal Z}_{\mbox{\scriptsize \rm dgt}}$}%
\beal{zond}
\mbox{
 the intersection of  $\dgt$ with any null hypersurface $\mcN$ is a \;\;\;}
 \\
\mbox{stratified submanifold of $\mcN$, with $\mcN$--codimension at least two.}
\eean
This condition will be used in our subsequent analysis. We
expect this property not to be needed, but we have
not investigated this question any further.

\begin{Theorem}
 \label{Carter69} Let $(\mcM,\fourg)$ be an
$(n+1)$--dimensional Lorentzian manifold with $s\ge 1$ linearly
independent commuting Killing vectors $K_\kl \mu$,
$\mu=0,\ldots, s-1$, satisfying the integrability conditions
\eqref{intcond2},  as well as \eq{zond}, with the $K_\kl i$'s,
$i=1,\ldots, s-1$, spacelike. Suppose that $ \{W=0\}\setminus
\dgtcp $ is not empty, and for each $p$ in this set  consider
the Killing vector field $l_p$
defined as%
\footnote{If $s=1$ then $\zh=\emptyset$ and $l_p=K_\kl 0$.}
\begin{equation}
\label{generator}
   l_p=K_\kl 0-(h^{\kl i \kl j}\fourg(K_\kl 0,K_\kl i))|_pK_\kl
   j\,,
\end{equation}
where $h^{\kl i \kl j}$ is the matrix inverse
to
    \begin{equation}
\label{h}
   h_{\kl i \kl j}:=\fourg(K_\kl{ i},K_\kl{
   j})\;,
   \quad i,j\in\{1,...,s-1\} \;.
\end{equation}
Then the distribution $  l_p^\perp \subset T\mcM$
{of vectors orthogonal to
$l_p$} is integrable over the non-empty  set
\bel{setdeff}
 \overline{\{q\in \mcM\setminus\dgt   \ | \ \fourg(
 l_p,l_p)|_{q}=0\;,\ W(q)=0\}}\setminus \{q\in \mcM\ | \
 l_p(q)=0\}
 \;.
 \ee
If we define
\index{$\hS_p$}%
$\hS_p$ to be the maximally extended over $\{W=0\}$,
connected, integral leaf of
this distribution%
\footnote{To avoid ambiguities, we emphasize that points at
which
$l_p$ vanishes do not belong to $\hSp$.}
passing through $p$, then all $\hS_p$'s are Killing
prehorizons, totally geodesic in $\mcM\setminus \{l_p=0\}$.
\end{Theorem}

In several situations of interest the $\hS_p$'s   form embedded hypersurfaces which coincide with
connected components of the set defined in \eq{setdeff}, but
this is certainly not known at this stage of the argument:

\begin{Remark}\label{RC69}
{\em Null translations in Minkowski space-time, or in
\emph{pp}-wave space-times, show that the $\hS_p$'s might be
different
from connected components of $\mcHN[l_p]$.}
\end{Remark}

\begin{Remark} {\em It follows from our analysis here that for $q\in
\hSp\setminus\dgtc $ we  have
$l_q=l_{p}$. For $q\in \hSp\cap\dgtc $ we can define $l_q$ by
setting
$l_q:=l_{p}$. We then have $l_p=l_q$ for all $q\in \hSp$.}
\end{Remark}

\proof
Let
\bel{qwdef}
 w:=K_\kl 0\e\ldots \e K_\kl {s-1}
 \;.
\ee
We need an equation of
Carter~\cite{CarterJMP}:
\begin{Lemma}[\cite{CarterJMP}] We have
 \label{LCarter}
\begin{equation}
\label{eqcarter}
 w\e dW =(-1)^sW dw\;.
\end{equation}
\end{Lemma}

\proof Let $F=\{W=0\}$. The result is trivial on the interior $\mathring F$ of
$F$, if non-empty.  By continuity, it then suffices to prove \eq{eqcarter} on
$\mcM\setminus F$. Let $\mcO$ be the set of points in $\mcM\setminus
F$ at which the Killing vectors are linearly independent. Consider any point
$p\in \mcO$, and let $(x^a,x^A)$, $a=0,\ldots,s-1$, be local coordinates
near $p$ chosen so that $K_\kl a
%\in \text{Span}\{
=\partial_a $ and
$\text{Span}\{\partial_a\} \perp\text{Span}\{ \partial_A\}$; this is possible by \eq{intcond2}. Then
$$
w = - W dx^0\wedge\ldots \wedge dx^{s-1}
\;,
$$
and \eq{eqcarter} follows near $p$.  Since $\mcO$ is open and dense,
the lemma is proved.
\qed%

\medskip

Returning to the proof of Theorem~\ref{Carter69}, as already
said, \eq{intcond2} implies that for every $p\notin\dgt  $
there exists a unique, maximal, $(n+1-s)$--dimensional
submanifold (not necessarily embedded), passing through $p$ and
orthogonal to $\mbox{\rm Span}\{K_\kl 0,\ldots ,K_\kl{s-1}\} $,
that we denote by $\scro_p$.  By definition,

\begin{equation}
  \label{propO1}
  \scro_p\cap\dgt   =\emptyset\;,
\end{equation}
and clearly
\begin{equation}
  \label{propO2}
  \scro_p\cap\scro_q\neq\emptyset \quad \Longleftrightarrow
  \quad \scro_p=\scro_q\;.
\end{equation}

Recall that $p\in \{W=0\} \setminus\dgt  $; then $K_\kl
0\e\ldots \e K_\kl {s-1}\neq
0$ in $\scro_p$ and we may choose vector fields $u_{\kl \mu}\in
TM$,
$\mu=0,\ldots ,s-1$, such that
$$K_\kl 0\e\ldots \e K_\kl {s-1}(u_\kl 0,\ldots ,u_\kl
{s-1})=1
$$
in some neighborhood of $p$. Let $\gamma$ be a $C^k$ curve,
$k\geq1$, passing through $p$ and contained
 in $\scro_p$. Since $\dot\gamma(s)\in T_{\gamma(s)}\scro_p=
 \mbox{\rm Span}\{K_\kl 0,\ldots ,K_\kl{s-1}\}^{\perp}
 |_{\gamma(s)}$, after contracting (\ref{eqcarter})
 with $(u_0,\ldots ,u_{s-1},\dot\gamma)$
 we obtain the following Cauchy problem
 \begin{equation}
 \left\{\begin{array}{cc}
 \frac{d}{ds}(W \circ\gamma)(s)\sim W \circ\gamma(s) \;,\\
 W |_{p}=0\;.
\end{array}\right.
\end{equation}
Uniqueness of solutions of this problem guarantees that
$W\circ {\gamma(s)}\equiv0$ and therefore $W$ vanishes along
the
$(n+1-s)$-dimensional submanifold ${\mycal O}_p$. Since $G$
preserves
$W$, $W$ must vanish on the sets
\bel{Spd} S_p:=G_s\cdot {\mycal
   O}_p \;.
 \ee
Here $G_s\cdot $ denotes the motion of a set using the group
generated
by the $K_\kl i$'s, $i=1,\ldots,s-1$; if the orbits of some of
the $K_\kl i$'s are
not complete, by this we mean ``the motion along the orbits of
all linear
combinations of the $K_\kl i$'s starting in the given set, as
far as those
orbits exist". Since $T_q\mcO_p$ is orthogonal to all Killing
vectors by
definition, and the $K_\kl i$'s are spacelike, the $K_\kl i$'s
are transverse
to $\mcO_p$, so that the $S_p$'s are smooth (not necessarily
embedded)
submanifolds of codimension one.

On $\{W=0\}\setminus \dgt$ the metric $\fourg$ restricted to
$\mathrm{Span}\{K_\kl 0,\ldots, K_\kl {s-1}\}$ is degenerate,
so that
$\mathrm{Span}\{K_\kl 0,\ldots, K_\kl {s-1}\}$ is a null
subspace of
$T\mcM$. It follows that for $q\in \{W=0\}\setminus \dgt$   {some} linear combination of Killing vectors is null and orthogonal to  $\mathrm{Span}\{K_\kl 0,\ldots, K_\kl {s-1}\}$, thus in
$T_q\mcO_p$. So
for $q\in
\{W=0\}\setminus \dgt$  the tangent spaces $T _q S_p$ are
orthogonal
sums of the null spaces $T_q \mcO_p$ and the  spacelike ones
$\mathrm{Span}\{K_\kl 1,\ldots, K_\kl {s-1}\}$. We conclude
that the
$S_p$'s   form smooth, null, not necessarily
embedded, hypersurfaces, with
\bel{gpsp} S_p=G\cdot {\mycal O}_p \subset\{ W=0\}\setminus
\dgt\;,
\ee
where the action of $G$ is understood as explained after
\eq{Spd}.

{Let the vector  $\lp =\Omega^\kl {\mu}K_\kl {\mu}$,
$\Omega^\kl
{\mu}\in\bbR$ be tangent to the null generators of  $S_p$,
thus
\begin{equation}
\label{l null1} \Omega^\kl {\mu}\fourg(K_\kl {\mu},K_\kl
{\nu})\Omega^\kl {\nu}=0\;.
\end{equation}
Since $\det(\fourg(K_\kl {\mu},K_\kl {\nu})) =0$ with
one-dimensional null
space on $\{W=0\}\setminus\dgt$, \eq{l null1} is equivalent
there to
\begin{equation}
\label{l null2} \fourg(K_\kl {\mu},K_\kl {\nu})\Omega^\kl {\nu}
=0\;.
\end{equation}
Since the $K_\kl i$'s are spacelike we must have $\Omega^\kl
0\ne 0$,
and it is convenient to normalize $\lp $ so that $\Omega^\kl
0=1$.
Assuming $p\not \in \zh$, from
\eq{l null2} one then immediately finds%
\begin{equation}
\label{Omega} \lp=K_\kl 0+\Omega^\kl iK_\kl i=K_\kl 0-h^{\kl
i\kl j}\fourg(K_\kl 0,K_\kl j)K_\kl i\;,
\end{equation}
where $h^{\kl i\kl j}$ is the matrix inverse to
\begin{equation}
\label{h def} h_{\kl i\kl j}=\fourg(K_\kl i,K_\kl
j)\;,\,\,i,j\in\{1,\ldots ,s-1\}\;.
\end{equation}

To continue, we show that:

\begin{Proposition}
 \label{POi} For each $j=1,\ldots,n$,  the function
$$
\Sp\ni q\mapsto  \Omega^\kl j(q):=-h^{\kl i \kl j}(q)\fourg(K_\kl
0,K_\kl i)(q)
$$
is constant over $\Sp$.
\end{Proposition}

\proof The calculations here are inspired by, and generalize
those
of~\cite[pp.~93-94]{Heusler:book}. As is well known,
\begin{equation}
\label{dh} dh^{\kl i\kl j}=-h^{\kl i\kl m}h^{\kl j\kl s}dh_{\kl
m\kl s}\,.
\end{equation}
{}From  \eqref{hodge1}-\eqref{hodge2} together with $\Lie_{K_\kl
i} K_\kl
j=0$ we have
\begin{align*}
dh_{\kl i\kl j}&=d[\fourg(K_\kl i,K_\kl j)]=di_{K_\kl i}K_\kl
j=-i_{K_\kl i}dK_\kl j\\
       &=-i_{K_\kl i}(-1)^{2(n+1-2)-1}**dK_\kl j=(-1)^{n}*(
       K_\kl i\e *dK_\kl j)\;,
\end{align*}
with a similar formula for $ d[\fourg(K_\kl 0,K_\kl j)]$.
Next,
\begin{align*}
d\Omega^\kl i &= d(-h^{\kl i\kl j}\fourg(K_\kl 0,K_\kl j))\\
          &=-[\fourg(K_\kl 0,K_\kl j)dh^{\kl i\kl j}+h^{\kl
          i\kl j}d[\fourg (K_\kl 0,K_\kl j)]]\\
          &=-[-\fourg(K_\kl 0,K_\kl j)h^{\kl i\kl m}h^{\kl j\kl
          s}dh_{\kl s\kl m}+h^{\kl i\kl m}d[\fourg (K_\kl
          0,K_\kl m)]]\\
          &=-h^{\kl i\kl m}[-(-1)^{n }\fourg(K_\kl 0,K_\kl
          j)h^{\kl j\kl s}*(K_\kl s\e *dK_\kl
          m)
          \\
          &
          +(-1)^{n}*(K_\kl 0\e*dK_\kl m)]\\
          &=(-1)^{n+1}h^{\kl i\kl m}*[(\Omega^\kl sK_\kl
          s+K_\kl 0)\e*dK_\kl m]\\
          &=(-1)^{n+1}h^{\kl i\kl m}*(\ell\e*dK_\kl m)
          \;,
\end{align*}
and
 \begin{align*}
i_{K_\kl 0}\ldots i_{K_\kl {s-1}}*d\Omega^\kl i
 &=(-1)^{n+1}i_{K_\kl 0}\ldots i_{K_\kl {s-1}}h^{\kl i\kl
 m}**(\lp\e*dK_\kl m)
  \\
                               &= h^{\kl
                               i\kl m}i_{K_\kl 0}\ldots
                               i_{K_\kl {s-1}}(\lp\e*dK_\kl m)
                               \;.
\end{align*}
Since $i_{K_\kl i}\lp|_{S_p}=\fourg(\lp,K_\kl i)|_{S_p}=0$, we
obtain
\begin{align*}
i_{K_\kl 0}\ldots i_{K_\kl {s-1}}( \lp  \e*dK_\kl
m)|_{S_p}&=i_{K_\kl 0}\ldots i_{K_\kl {s-2}}[i_{K_\kl {s-1}}
\lp  \e*dK_\kl m
 \\
 &+(-1)^1 \lp  \e
 i_{K_\kl {s-1}}*dK_\kl m]|_{S_p} \\
&=-i_{K_\kl 0}\ldots i_{K_\kl {s-2}}( \lp  \e i_{K_\kl
{s-1}}*dK_\kl m)|_{S_p}  =\ldots \\
&=(-1)^s \lp  \e i_{K_\kl 0}\ldots i_{K_\kl {s-1}}*dK_\kl
m\,|_{S_p}\\
&=(-1)^s \lp  \e*(dK_\kl m\e K_\kl {s-1}\e\ldots \e K_\kl
0)|_{S_p}\\ &{\buildrel
\eqref{intcond} \over =} 0
 \;,
\end{align*}
and therefore
\begin{equation}
\label{*dOmega} i_{K_\kl 0}\ldots i_{K_\kl {s-1}}*d\Omega^\kl
i|_{S_p}=0\;.
\end{equation}
This last result says that $d\Omega^\kl i|_{S_p}$ is a linear
combination of
the $K_\kl {\mu}$'s, so for each $i$ there exist numbers
$\alpha^\kl
{\mu}\in\bbR$ such that
\begin{equation}
\label{dOmega} d\Omega^\kl i|_{S_p}=\alpha^\kl {\mu}K_\kl
{\mu}.
\end{equation}
Now, the $\Omega^\kl i$'s are clearly invariant under the
action of the
group generated by the $K_\kl {\mu}$'s, which implies
$$
0= i_{K_\kl \mu} d\Omega^\kl i= \fourg(K_\kl \mu, \alpha^\kl
\nu K_\kl \nu)
\;.
$$
This shows that $\alpha^ \kl \mu K_\kl {\mu}$ is orthogonal to
all Killing vectors, so it must be
proportional to $ \lp  $. Since   $T_qS_p= \lp ^\perp$, we are
done.
\qed

\medskip

Returning to the proof of Theorem~\ref{Carter69}, we have shown
so far that $S_p$ is a null hypersurface in
$\{W=0\}\setminus\dgt$, with the Killing vector $l_p:=\lp $ as
in \eq{generator}   tangent to the generators of $S_p$. In
other words, $S_p$ is a prehorizon. Furthermore,
\beal{Ybel}
 \lefteqn{T_q\mcM \ni Y\in T_q S_p \ \mbox{ for some } p \quad
 \Longleftrightarrow}&&
 \\
 && \quad W(q)=0\;, \ K_\kl 0 \wedge \ldots
 \wedge K_\kl {s-1}|_q \ne 0\;, \quad Y\perp l_p \;.
\eean

For further purposes it is  necessary  to extend this result to
the hypersurface $\hSp$ defined in the statement of
Theorem~\ref{Carter69}. This  proceeds as follows:

It is  well known~\cite{Galloway:splitting} that Killing
horizons are
\emph{locally totally geodesic}, by which we mean that
geodesics
initially tangent to the horizon remain on the horizon for some
open
interval of parameters. This remains true for prehorizons:

\begin{Corollary}
 \label{Ctg0}
$\Sp$ is  locally totally geodesic. Furthermore,  if
$\gamma:[0,1)
\to \Sp$ is a geodesic such that $\gamma(1)\not \in \Sp$, then
$\gamma(1)\in \dgt$.
\end{Corollary}

\begin{proof}   Let
$\gamma:I\rightarrow \mcM$ be an affinely-parameterized
geodesic
satisfying  $\gamma(0)=q\in \Sp$ and $\dot\gamma(0)\in
T_q\Sp\Longleftrightarrow \fourg(\dot\gamma(0), l_p)=0$. Then
\bel{geodcalc}
\frac{d}{dt}\fourg(\dot\gamma(t),l_p)=
\fourg(\nabla_{\dot\gamma(t)}\dot\gamma(t),l_p)+\fourg(\dot\gamma(t),\nabla_{\dot\gamma(t)}l_p)=0
 \;,
\ee
where the first term vanishes because $\gamma$ is an affinely
parameterized geodesic, while the second is zero by  the
Killing  equation. Since $\fourg(\dot\gamma(0),l_p)=0$, we get
\begin{equation}
 \fourg(\dot\gamma(t),l_p)=0\,,\,\,\forall t\in I
 \;.
\end{equation}
We conclude that $\dot \gamma$ remains perpendicular to $l_p$,
hence remains within $\Sp$ as long as a zero of $K_\kl 0\wedge
\ldots \wedge K_\kl {s-1}$ is not
reached, compare \eq{Ybel}.
\end{proof}

Consider, now, the following set of points which can be reached by
geodesics initially tangent to $S_p$:
 \beal{tSdef}
  \tSp&:=&\{q: \ \exists \ \mbox{a geodesic segment
  $\gamma:[0,1]\to \mcM$ such }
  \\
  \nonumber
  &&
  \mbox{
  that %$\dot\gamma(0)\in l_p^\perp$,
 $\gamma(1)=q$ and
  $\gamma(s)\in S_p$ for $s\in[0,1)$}  \}\setminus\{q:l_p(q)=0\}
 \;.
\eea
Then $S_p\subset \tSp$, and if $q\in \tSp\setminus \Sp$ then
$q\in \dgt$ by Corollary~\ref{Ctg0}.
We wish to show that $\tSp$ is a smooth hypersurface, included and maximally extended in the set \eq{setdeff}; equivalently
\bel{eq}
  \tSp=\hSp
  \;.
\ee
For this, let $q\in \tSp$, let $\mcO$ be a  geodesically convex
neighborhood of $q$ not containing zeros of $l_p$, and for $r\in \mcO$ define
\bel{exp_O}
 R_r= \exp _{\mcO, r} (l_q(r)^\perp)
 \;,
\ee
here $\exp _{\mcO, r}$ is the exponential map at the point $r\in\mcO$ in the
space-time $(\mcO, \fourg|_{ \mcO})$. It is convenient to
require that $\mcO$ is included within the radius of
injectivity of all its points
(see~\cite[Theorem~8.7]{kobayashi:nomizu}).
Let $\gamma$ be as in the definition of $\tSp$. Without loss of
generality we can assume that $\gamma(0)\in \mcO$.
We have $\dot \gamma(s)\perp l_p$ for all $s\in [0,1)$, and by
continuity also at $s=1$. This shows that $\gamma([0,1])
\subset R_q$.

Now,  $R_{\gamma(0)}$ is a smooth  hypersurface in $\mcO$. It
coincides with $S_p$ near $\gamma(0)$, and every null geodesic
starting at $\gamma(0)$ and normal to $l_p$ there belongs both
to $R_{\gamma(0)}$ and $S_p$ until a point in $\dgt$ is
reached.   This  shows that  $R_{\gamma(0)}$ is null near every
such geodesic until, and including, the first point on that
geodesic at which $\dgt$ is reached (if any).  By \eq{zond} $R_{\gamma(0)}\cap S_p$ is open and dense in
 $R_{\gamma(0)}$. Thus the tangent space to $R_{\gamma(0)}$
 coincides with   $l_p^\perp$ at the open dense set of points
 $R_{\gamma(0)}\cap S_p$, with that intersection being a null,
 locally totally geodesic (not necessarily embedded)
 hypersurface. By continuity $R_{\gamma(0)}$ is a subset of~\eq{setdeff}, with $TR_{\gamma(0)}=l_p^\perp$
 everywhere. Since $R_{\gamma(0)}\subset \tSp$,  Equation~\eq{eq} follows.

The construction of the $\tSp$'s shows that every integral manifold of the distribution $l_p^{\perp}$ over the set
\bel{set0}
 \Omega:=
{\{q\in \mcM\setminus\dgt   \ | \ \fourg(
 l_p,l_p)|_{q}=0\;,\ W(q)=0\}}
 \;,
 \ee
can be extended to a maximal leaf contained in  $\overline
\Omega \setminus \{q\ | \ l_p(q)=0\}$, compare  \eq{setdeff}. To
finish the proof of  Theorem~\ref{Carter69} it thus remains  to
show that there exists a leaf through every point in $\overline
\Omega \setminus \{q\ | \ l_p(q)=0\}$.  Since this last set is
contained in the closure of $\Omega$, we} need to analyze what
happens when a sequence of null leaves $\hSpn$,   all normal to
a fixed Killing vector field $l_q$, has an accumulation point.
We show in Lemma~\ref{Lacc2} below that such sequences
accumulate to an integral leaf through the limit point, which
completes the proof of the theorem.
\qed

\bigskip

 We shall say that $S$ is an
\emph{accumulation
set} of a sequence of sets $S_n$ if $S$ is the collection of
limits,
as $i$ tends to infinity, of sequences $q_{n_i}\in S_{n_i}$.

\begin{Lemma}
 \label{Lacc2}
Let $\hSpn$ be a sequence of leaves such that $l_{p_n}=l_{q}$, for
some fixed $q$, and suppose that $ p_n\to p$. If $l_q(p)\ne 0$, then
$p$ belongs to a leaf $\hSp $ with $l_p=l_q$. Furthermore there
exists a neighborhood $\mcU$ of $p$ such that
$\exp _{\mcU,p}( l_q(p  )^\perp) \subseteq \hSp\cap\mcU$ is
the accumulation set of the sequence $\exp _{\mcU, p_n}(
l_q(p_n )^\perp)\subseteq\hSpn\cap\mcU$, $n\in\N$.
\end{Lemma}

\begin{proof}
Let $\mcU$ be a small, open, conditionally compact,
geodesically convex neighborhood of $p$
which does not contain zeros of $l_q$. Let $\hSpn$ be that
leaf,
within $\mcU$,   of the distribution $l_q^\perp$ which
contains
$p_n$. The $\hSpn $'s are totally geodesic submanifolds of
$\mcU$ by
Corollary~\ref{Ctg}, and therefore   are uniquely determined
by
prescribing $T_{p_n}\hSpn$. Now, the subspaces
$T_{p_n}\hSpn=l_q(p_n)^\perp$ obviously converge to $ l_q(p
)^\perp$
in the sense of accumulation sets.
Smooth dependence of geodesics upon
initial values implies that $\exp _{\mcU, p_n}( l_q(p_n
)^\perp)$ converges in $C^k$, for any $k$, to
$\exp _{\mcU, p}( l_q(p )^\perp)$. Since $W$ vanishes
on   $\exp _{\mcU, p_n}( l_q(p_n )^\perp)$, we obtain
that $W$ vanishes on $\exp _{\mcU, p}( l_q(p )^\perp)$.
Since $T_{q_n}\exp _{\mcU, p_n}( l_q(p_n
)^\perp)=l_p^\perp(q_n)$ for any $q_n\in \exp _{\mcU, p_n}(
l_q(p_n )^\perp)$ we conclude that  $T_{r }\exp _{\mcU, p }(
l_q(p  )^\perp)=l_p^\perp(r )$ for any $r \in \exp _{\mcU, p
}( l_q(p  )^\perp)$.
 So
$\exp _{\mcU, p }( l_q(p  )^\perp)$ is a leaf, within
$\mcU$,  through $p$ of the distribution $l_q^\perp$ over the
set \eq{setdeff}, and   $\exp _{\mcU, p }( l_q(p  )^\perp)=\hSp\cap \mcU$ is
the accumulation set   of the totally geodesic submanifolds
$\hSpn\cap \mcU$'s.
\end{proof}
\bigskip

The  remainder of the proof of Theorem~\ref{Tdoc2} consists in
showing that the $\hSp$'s cannot intersect $\doc$.
We start with an equivalent of Corollary~\ref{Ctg0}, with
identical proof:

\begin{Corollary}
 \label{Ctg} $\hSp$ is  locally totally geodesic. Furthermore,  if
$\gamma:[0,1) \to \hSp$ is a geodesic segment such that $\gamma(1)\not
\in \hSp$, then $l_p$ vanishes at $\gamma(1)$.
\qed
\end{Corollary}

Corollary~\ref{CLnozeros} shows that Killing vectors as described
there
have no zeros in $\doc$, and Corollary~\ref{Ctg} implies now:

\begin{Corollary}
 \label{Ccl}
$\hSp\cap \doc$ is totally geodesic in $\doc$ (possibly
empty).
\qed
\end{Corollary}

To continue, we want to extract, out
of the $\hS_p$'s, a closed, embedded, Killing horizon
$S^+_0$. Now, e.g. the analysis in~\cite{HIW} shows that the
gradient of $\fourg(l_p,l_p)$ is either everywhere zero on
$\hS_p$ (we then say that $\hSp$ is degenerate), or nowhere
vanishing there. One immediately concludes that non-degenerate
$\hS_p$'s, if non-empty, are embedded,  closed hypersurfaces in
$\doc$. Then, if there exists non-empty non-degenerate
$\hS_p$'s, we choose one and we set
\bel{spodef}
 S^+_0=\hS_p\;.
\ee
%.
Otherwise, all non-empty $\hS_p$'s are degenerate; to show that
such prehorizons, if non-empty, are embedded, we will invoke
analyticity (which has not been used so far).  So,
consider a degenerate component $\hS_p$, and note that $\hS_p$ does not
self-intersect, being a subset of the union of integral
manifolds of a smooth distribution of hyperplanes. Suppose
that $\hS_p$ is not embedded. Then there exists a point $q\in
\hS_p$, a conditionally compact neighborhood $\mcO$ of $q$, and
a sequence of points $p_n\in   \hS_p$ lying on pairwise
disjoint components of $\mcO\cap \hS_p$, with $p_n$ converging
to $q$. {Now, Killing vectors are solutions of the
overdetermined set of PDEs
$$
 \nabla_\mu \nabla_\nu X_\rho = R^\alpha{}_{\mu\nu\rho} X_\alpha
 \;,
$$
which imply that they are analytic if the metric is. So
$\fourg(l_p,l_p)$ is an analytic function that vanishes on an
accumulating family of hypersurfaces. Consequently
$\fourg(l_p,l_p)$ vanishes everywhere, which is not compatible
with asymptotic flatness.} Hence the $\hS_p$'s are embedded,
coinciding with connected components of the set
$\{\fourg(l_p,l_p)=0=W\}\setminus \{ l_p =0 \}$; it should be
clear now that they are closed in $\doc$. We define $S^+_0$
again using \eq{spodef}, choosing one non-empty $\hS_p$,

We can finish the proof of Theorem~\ref{Tdoc2}. Suppose that
$W$ changes sign within $\doc$. Then  $S^+_0$  is a non-empty,
closed, connected,
embedded null hypersurface within $\doc$. Now, any
embedded null hypersurface  $\Spn$ is locally two-sided, and we
can assign an intersection number one to every intersection
point of $\Spn$ with a curve that crosses $\Spn$ from its local
past to its local future, and minus one for the remaining ones
(this coincides with the oriented intersection number as
in~\cite[Chapter~3]{GuilleminPollack}). Let $p\in \Spn$, there
exists a smooth timelike future directed curve $\gamma_1$ from
some point $q\in \Mext$ to $p$. By definition there exists a
future directed null geodesic segment $\gamma_2$  from $p$ to
some point $r\in \Mext$ intersecting $S$ precisely at $p$.
Since $\Mext$ is connected there exists a curve
$\gamma_3\subset \Mext$ (which, in fact, cannot be causal
future directed, but this is irrelevant for our purposes) from
$r$ to $q$. Then the path $\gamma$ obtained by following
$\gamma_1$, then $\gamma_2$, and then $\gamma_3$ is closed.
Since $\Spn$ does not extend into $\Mext$, $\gamma$ intersects
$\Spn$ only along its timelike future directed part, where
every intersection has intersection number one, and $\gamma$
intersects $\Spn$ at least once at $p$, hence the intersection
number of $\gamma$ with $\Spn$ is strictly positive. Now,
Corollary~\ref{Cdsc} shows that $\doc$ is simply\kk{}
connected. But, by standard intersection
theory~\cite[Chapter~3]{GuilleminPollack}, the intersection
number of a closed curve with a closed, externally orientable, embedded
hypersurface in a simply\kk{} connected manifold vanishes,
which gives a contradiction  and proves that $W$ cannot change
sign on $\doc$.

It remains to show that $W$ vanishes at the boundary of $\doc$.
For this, note that, by definition of $W$, in the region
$\{W>0\}$ the subspace of $T\mcM$ spanned by the Killing
vectors $K_\kl \mu$ is timelike. Hence at every $p$ such that
$W(p)>0$ there exist vectors of the form $K_\kl 0+ \sum
\alpha_i K_\kl i$ which are timelike. But $\partial \doc\subset
\dot I^-(\Mext)\cup \dot I^+(\Mext)$, and  each of the
boundaries $\dot I^-(\Mext)$ and $\dot I^+(\Mext)$  is
invariant under the flow of any linear combination of $K_\kl
\mu$'s, and  each is achronal, hence $W\le 0$ on $\pdoc$, whence the
result.
\qed

\bigskip

In view of  what has been said, the reader will conclude:

\begin{Corollary}[Killing horizon theorem]
 \label{Cpht}
\index{Killing horizon theorem}%
Under the conditions of Theorem~\ref{Tdoc2}, away from the set $\dgt$ as
defined in \eq{dgt2},  the boundary $\odoc\setminus \doc$ is a union of
embedded Killing horizons.
\qed
\end{Corollary}

%
%\begin{proof}
%Let
%$p\in\partial\doc\setminus\dgt$, from what has been just said
%we have $W(p)=0=\fourg(l_p,l_p)$. So, by Theorem~\ref{Carter69}, $p$
%belongs to a prehorizon $\hS_p$. As explained above, in
%analytic asymptotically flat space-times all the $\hS_p
%\setminus  \dgt$'s are embedded hypersurfaces, closed in
%$\mcM\setminus \dgt$. A simple open-closed argument allows one
%to conclude
%that $S_p \setminus  \dgt$ coincides with a
%connected component of $\odoc\setminus (\doc\cup \dgt)$, as
%desired.
%\end{proof}
%
%

Let us pass now to the

\medskip

{\noindent \sc Proof of Theorem~\ref{Tnoanalyticnew}:}
Let
\newcommand{\oquosp}{\,\,\mathring{\!\! \quosp}}%
\newcommand{\quosp}{\mathcal Q}%
\newcommand{\hquosp}{\,\widehat{\!\quosp}}%
$$
 \pi:  \doc\cup \mcE^+  \to
 \underbrace{\Big(\doc\cup \mcE^+\Big) /\Big(\R\times \Uone  \Big)}_{=:\quosp}
$$
denote the quotient map. As discussed in more detail in
Sections~\ref{sSos3} and \ref{sSgcos} (keeping in mind that, by
topological censorship, $\doc$ has only one asymptotically flat
end), the orbit space $\quosp $ is diffeomorphic to the
half-plane $\{(x,y)\ |\ x\ge 0\}$ from which a finite number
$\mathring n\ge 0$ of open half-discs, centred at the axis
$\{x=0\}$, have been removed. As explained at the beginning of
Section~\ref{Sproof}, the case $\mathring n=0$ leads to
Minkowski space-time, in which case the result is clear, so
from now on we assume $\mathring n\ge 1$.

Suppose that $\{W=0\}\cap\doc$ is non-empty. Let $p_0$ be an
element of this set, with corresponding Killing vector field
$l_0:= l_{p_0}$. Let $W_0$ be the norm squared of $l_0$:
$$
 W_0:=\fourg(l_0,l_0)
  \;.
$$
In the remainder of the proof of Theorem~\ref{Tdoc3} we
consider only those $\hat S_p$'s for which $l_p=l_0$:
$$
 \hat S_p \subset \{W=0\}\cap \{W_0=0\}
\;.
$$

We denote by  $C_{\pi(p)}$
\index{$C_p$}%
the image  in $\quosp$, under the
projection map $\pi$, of $\hat S_p\cap (\doc\cup \mcE^+) $.
Define
$$
 \oquosp =  \doc /\Big(\R\times \Uone \Big)\;,
$$
\newcommand{\hmcW}{ {\mycal W}^\flat_0}%
$$
 \hmcW:= \Big(\{W_0=0\}\cap \{W=0\}\cap (\doc\cup \mcE^+) \Big)/\Big(\R\times \Uone \Big)
 \;,
$$
%%
%$$
% \ohmcW:= \Big(\{W_0=0\}\cap \{W=0\}\cap \doc \Big)/\Big(\R\times \Uone \Big)
% \;.
%$$
%%
Then $\hmcW$ is a closed subset of $\quosp$, with the following
property: through every point $q$ of $\hmcW$ there exists a
smooth maximally extended curve $C_q$, which will be called
\emph{orbit}, entirely contained in $\hmcW$.  The $C_q$'s are
pairwise disjoint, or coincide. Their union forms a closed set,
and locally they look like a subcollection of leaves of a
foliation. (Such structures are called laminations; see,
e.g.,~\cite{Gabai}.)

An orbit will be called a \emph{Jordan orbit} if $C_q$ forms a
Jordan curve.

We need to consider several possibilities; we start with the
simplest one:

\medskip

{\noindent \sc Case I:} If an orbit $C_q$ forms a Jordan curve
entirely contained in $\oquosp$, then the corresponding $\hat
S_p=\pi^{-1}(C_q)$ forms a closed embedded hypersurface in
$\doc $, and a contradiction arises as at the end of the proof
of Theorem~\ref{Tdoc2}.

\medskip
{\noindent \sc Case II:} Consider, next, an orbit $C_q$ which
meets the boundary of $\quosp$ at two  or more points which
belong to $\pi(\mcA)$, and only at such points. Let $I_q\subset
C_q$ denote that part of $C_q$ which connects any two
subsequent such points, in the sense that $I_q$ meets $\partial
\quosp$ at its end points only. Now, every $\hat S_p$ is a
smooth hypersurface in $\mcM$ invariant under $\R\times \Uone $,
and therefore meets the rotation axis $\mcA$ orthogonally. This
implies that $\pi^{-1}(I_q)$ is a closed, smooth, embedded
hypersurface in $
\doc $, providing again a contradiction.

\medskip

\begin{figure}[t]
\begin{center} {
\resizebox{3.5in}{!}{\includegraphics{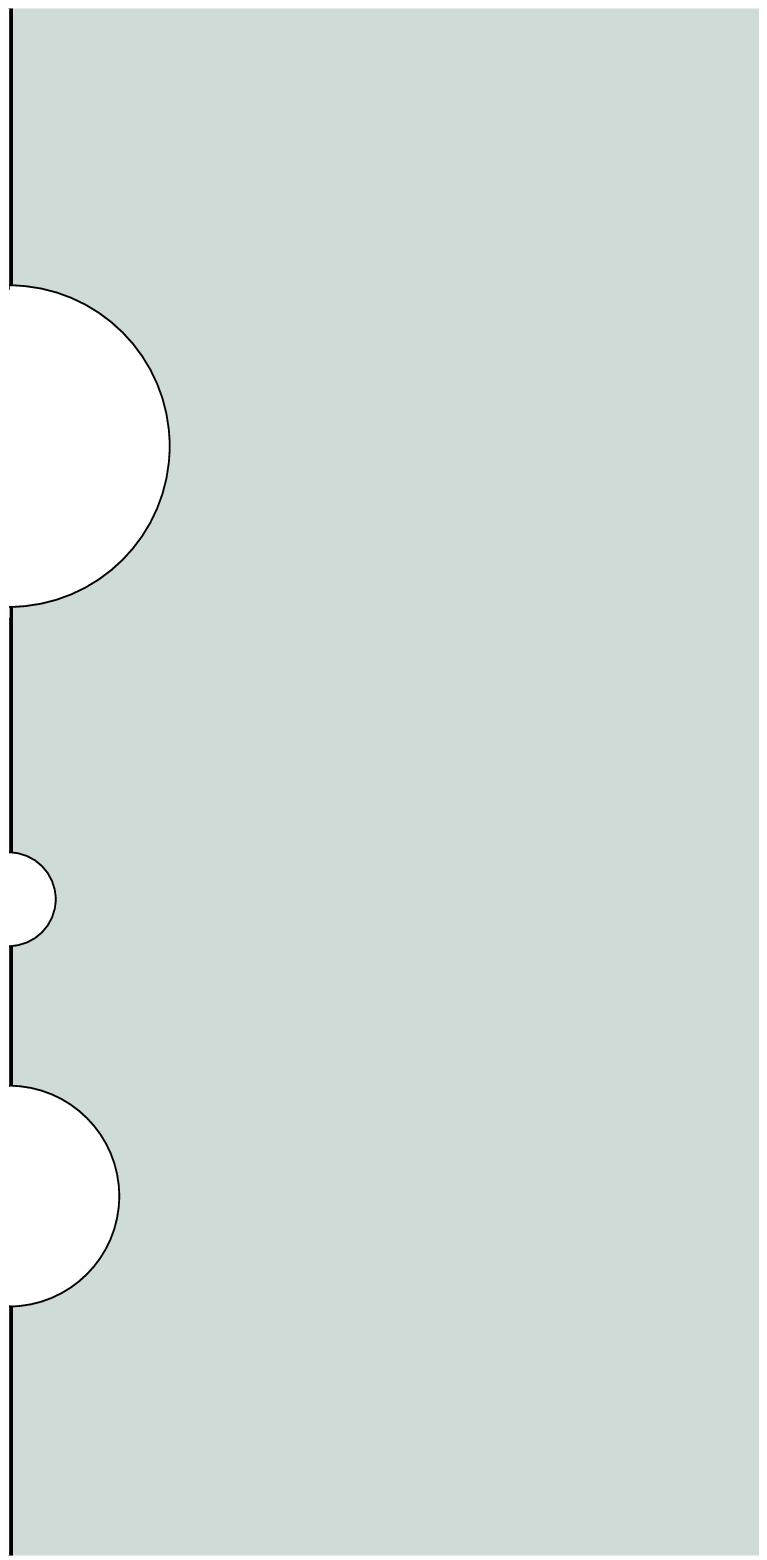}
\includegraphics{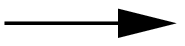}\includegraphics{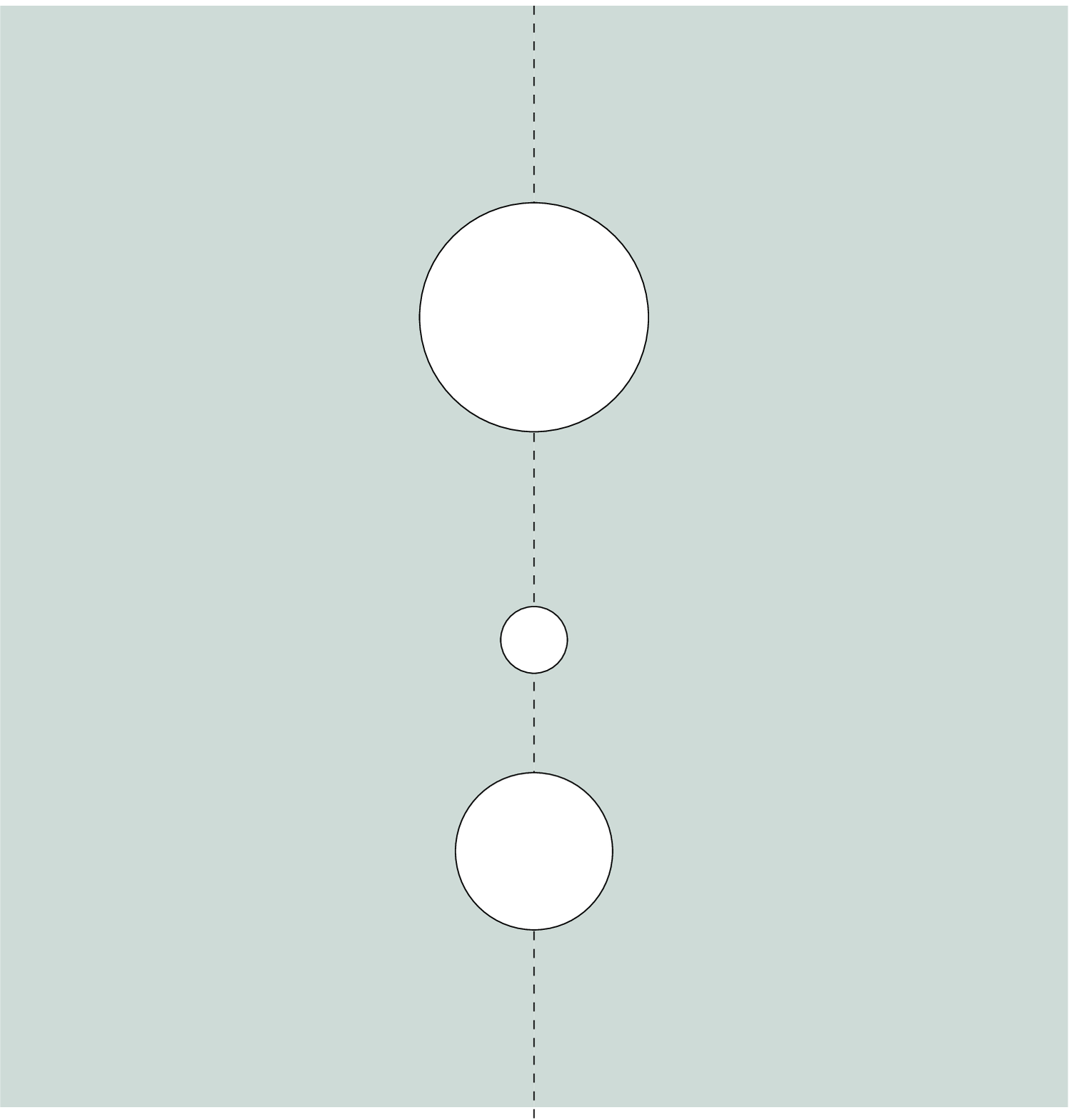}}
}
\caption{The quotient space $\quosp$ and its double $\hquosp$.
\label{Fquot}}
\end{center}
\end{figure}

To handle the remaining cases, some preliminary work is needed.
It is convenient to double $\quosp$ across $\{x=0\}$ to obtain
a manifold $\hquosp$ diffeomorphic to $\R^2$ from which a
finite number of open discs, centered at the axis $\{x=0\}$,
have been removed, see Figure~\ref{Fquot}.%
Connected components of the event horizon $\mcE^+$ correspond
to smooth circles forming the boundary of $\hquosp$, regardless
of whether or not they are degenerate. From what has been said,
every $C_q$ which has an end point at $\pi(\mcA)$ is smoothly
extended in $\hquosp$ across $\{x=0\}$ by its image under the
map $(x,y)\mapsto (-x,y)$. We will continue to denote by $C_q$
the orbits so extended in $\hquosp$.

The analysis of {\sc Cases I} and {\sc II} also shows:

\begin{Lemma}
 \label{Lcross}
An orbit $C_q$  which does not meet $\partial \hquosp$ can
cross the axis $\{x=0\}$ at most once.
\qed
\end{Lemma}

An orbit $C_q$ will be called an \emph{accumulation orbit} of
an orbit $C_r$ if there exists a sequence $q_n\in C_r$ such
that $q_n\to q$.  Every orbit is its own accumulation orbit. It
is a simple consequence of the accumulation Lemma~\ref{Lacc2}
that:

\begin{Lemma}
 \label{Lacc3}
 Let $C_q$ be an accumulation orbit of   $C_{r}$. Then for every
 $p\in C_q$ there exists a sequence $p_n\in C_r$ such that $p_n
 \to p$.
 \qed
\end{Lemma}

 We will need the following:

\begin{Lemma}
 \label{Loneint}
Let $r_n\in C_r$ be a sequence accumulating at $p\in
\pi(\mcA)\setminus \partial \hquosp$. Then $p\in C_r$, and
$C_r$ continues smoothly across $\{x=0\}$ at $p$.
\end{Lemma}

\proof By Lemma~\ref{Lacc2} there exists an orbit $C_p$
crossing the axis $\{x=0\}$ transversally at $p$. Lemma
\ref{Lacc3} shows that $C_r$ crosses the axis. But, by
Lemma~\ref{Lcross}, $C_r$ can cross the axis only once. It
follows that $C_r=C_p$ and that $p\in C_r$.
\qed

\medskip

Abusing notation, we still denote by $W$ and $W_0$ the
functions $W\circ \pi$ and $W_0 \circ \pi$. If $W$ and $W_0$
vanish at a point lying at the boundary $\partial \hquosp$,
then the corresponding circle forms a Jordan orbit. We have:

\begin{Lemma}
 \label{LnobN}
The only orbits accumulating at $\partial \hquosp$ are the
boundary circles.
\end{Lemma}

\proof Suppose that $r_n\in C_q$ accumulates at $p\in
\partial \hquosp$. Then, by continuity,
$W(p)=W_0(p)=0$, which implies that the boundary component
through $p$ is a Jordan orbit.
But it follows from Lemma~\ref{Lacc3} that any orbit
accumulating at $\partial\hquosp$ has to cross the axis more
than once, and the result follows from Lemma~\ref{Lcross}.
\qed

\medskip

The remaining possibilities will be excluded by a lamination
version of the Poincar\'e-Bendixson theorem. We will make
use of a smooth transverse orientation of all the $\hat S_p$'s;
such a structure is not available for a general lamination, but
exists in the problem at hand. More precisely, we will endow
$\doc\cup \mcE^+ $ with a smooth vector field $Z$ transverse to
all $\hat S_p$'s. The construction proceeds as follows: Choose
any decomposition of $\doc\cup \mcE^+ $ as $\R\times \ohypo$,
as in Theorem~\ref{Tgt}: thus each level set $\ohypot$ of the
time function $t$ is transverse to the stationary Killing
vector field $K_0$, with the periodic Killing vector  $K_1$
tangent to $\ohypot$. Let $q\in \hat S_p\cap \hypoz $; as the
null leaf $\hat S_p$ is transversal to $\hypoz$, the
intersection $\hypoz\cap \hat S_p$ is a hypersurface in $\mcM$
of co-dimension two. There exist precisely two null directions
at $q$ which are normal to $\hypoz\cap \hat S_p$, one of them
is spanned by $l_0(q)$; we denote by $\mathring Z_q$
the unique
future directed null vector spanning the other direction and
satisfying $\mathring Z_q= T_q + \tilde Z_q$, where $T_q$ is
the unit timelike future directed normal to $\hypoz$ at $q$,
and $\tilde Z_q$ is tangent to $\hypo$.

The above definition of $\tilde Z_q$
extends by continuity to $q\in
\hat S_p\cap\ohypoz$.

Transversality and smoothness of $l_0$  imply that there exists
a neighborhood $\mcO_q$ of $q$ and an extension $\hat Z_q$ of
$\tilde Z_q$ to $\mcO_q$ with the property that $\hat Z_q(r)$
is transverse to $\hat S_r$ for every $r\in \mcO_q$ satisfying
$W_0(r)=W (r)=0$. The neighborhood $\mcO_q$ can, and will, be
chosen to be invariant under $\R\times \Uone $;
similarly for $\hat Z_q(r)$.

Consider the covering of $\ohypoz\cap\{W_0=0\}\cap\{W =0\}$ by
sets of the form $\mcO_q\cap \ohypoz$. Asymptotic flatness
implies that $\ohypoz\cap\{W_0=0\}\cap\{W =0\}$ is compact,
which in turn implies that a  finite subcovering
$\mcO_i:=\mcO_{q_i}$ can be chosen. Let $\varphi_i$ be a
partition of unity subordinated to the covering of
$\doc\cup\mcE^+$ by the $\mcO_i$'s together with
$$
 \mcO_{0}:=\Big(\doc\cup\mcE^+\Big) \setminus \Big(\{W=0\}\cap \{W_0 =0\}\Big)
  \;.
$$
%.
The $\varphi_i$'s can, and will, be chosen to be $\R\times \Uone
$--invariant. Set
$$
 Z:= \sum_{i\ge 1}  \varphi_i \hat Z_{q_i}
 \;.
$$
Then $Z$ is smooth, tangent to $\hypoz$, and transverse to all
$\hat S_p$'s.

Choose an orientation of $\hquosp$. The vector field $Z$
projects under $\pi$ to a vector field $Z^\flat$
on $\hquosp$
transverse to each $C_q$. For each $r\in C_q$ we define a
vector $V_q(r)$ by requiring $V_q(r)$ to be tangent to $C_1$ at
$r$, with $\{V_q,Z^\flat\}$   positively oriented, and with
$V_q$ having length one with respect to some auxiliary
Riemannian metric on $\hquosp$. Then $V_q$ varies smoothly
along $C_q$, and  each $C_q$ is in fact a complete integral
curve of its own $V_q$. The vector field $V_p$ along $C_p$
defines an order, and diverging sequences, on $C_p$ in the
obvious way:  we say that a point $r'\in C_p$ is subsequent to
$r\in C_p$ if one flows from $r$ to $r'$ along $V_p$ in the
forward direction;
a sequence $r_n\in V_p$ is diverging if
$r_n=\phi(s_n)(p)$, where $\phi(s)$ is the flow of $V_p$ along
$C_p$, with $s_n \nearrow \infty$ or $s_n \searrow -\infty$.

By Lemma~\ref{Lacc2}, if a sequence $r_n\in C_{q_n}$ tends to
$r\in C_q$, then the tangent spaces $T C_{q_n}$ accumulate on
$TC_q$. This implies that there exist numbers $\epsilon_n \in
\{\pm 1\}$ such that $\epsilon_n V_{q_n}(r_n)\to V_q(r)$, and
this is the best one can say in general. However, the existence
of $Z$ guarantees that $  V_{q_n}(r_n)\to V_q(r)$.

We are ready now to pass to the analysis of

\medskip
\noindent{\sc Case III:}
In view of Lemmata~\ref{Lcross} and
\ref{LnobN}, it remains to exclude the existence of orbits
$C_q$ which are entirely contained within $\hquosp\setminus
\partial \hquosp$, and which do not intersect $\pi(\mcA)$, or which
intersect $\pi(\mcA)$ only once, and which do \emph{not} form
Jordan curves in $\oquosp$. Since $\{W=0\}\cap \ohypoz$ is
compact, there exists $p\in \hquosp$ and a diverging sequence
$q_n\in C_q$ such that $q_n\to p$. Again by
Lemmata~\ref{Lcross} and \ref{LnobN}, $p\not \in
\partial\hquosp$.
The fact that $C_p$ is a closed embedded curve follows now by
the standard arguments of the proof of the
Poincar\'e--Bendixson theorem, as e.g.\ in~\cite{HirschSmale}.
The orbit  $C_p$ does not meet $\partial \hquosp$ by
Lemma~\ref{LnobN}.  If $C_p$ met $\pi(\mcA)$, it would have an
intersection number with $\{x=0\}$ equal to one by
Lemma~\ref{Lcross}, which is impossible for a Jordan curve in
the plane. Thus $C_p$ is entirely contained in $\oquosp$, which
has already been shown to be impossible in  {\sc Case I}, and
the result is established.
\qed

\bigskip

Similarly to Corollary~\ref{Cpht}, we have the following
Corollary of Theorem~\ref{Tnoanalyticnew},
which is essentially a
rewording of Lemma~\ref{LnobN}:

\begin{Corollary}[Embedded prehorizons theorem]
 \label{Cpht2}
\index{embedded prehorizon theorem}%
Under the conditions of Theorem~\ref{Tdoc3}, away from the set
$\dgt$ as defined in \eq{dgt2},  the boundary $\odoc\setminus
\doc$ is a union of embedded Killing prehorizons.
\qed
\end{Corollary}

\subsection{The ergoset in space-time dimension four}
 \label{sSergo}
The \emph{ergoset} $E$ is defined as the set where the stationary Killing vector field $\Kz$ is spacelike or null:%
\index{ergoset}%
\bel{ergo}
 E:= \{ p\ | \ \fourg(\Kz,\Kz)|_p\ge 0\}
 \;.
\ee
In this section we wish to show that, in vacuum, the ergoset cannot
intersect the rotation axis within $\doc$, if we assume the latter to be
chronological.

The first part of the argument is purely local. For this we will assume that
the space-time dimension is four, that $\Kz\equiv X$ has no zeros near a
point $p$, that $K_\kl 1\equiv Y$ has $2\pi$--periodic orbits and vanishes
at $p$, and that $X$ and $Y$ commute.

Let $\hat T$ be any timelike vector at $p$, set
\bel{Tav}
 T:= \int_0^{2\pi} \phi_t[Y]_* \hat T dt
 \;,
 \ee
then $T$ is invariant under the flow of $Y$. Hence $T^\perp$ is also invariant under $Y$. Let $\hyp_\mcO$ denote $\exp_p(T^\perp)\cap \mcO$, where $\mcO$ is any  neighborhood of $p$ lying within the injectivity radius of $\exp_p$, sufficiently small so that $\hyp_\mcO$ is spacelike; note that $\hyp_\mcO$ is invariant under the flow of $Y$. A standard argument (see, e.g.,~\cite{ACD2} Appendix C)
shows that $Y$ vanishes on
$$
\mcA_p:=\exp_p(\Ker\, \nabla Y)
\;,
$$
%,
and that $\mcA_p$ is totally geodesic. Note that $T\in \Ker\,  \nabla Y$, which implies that $\mcA_p$ is timelike.

\renewcommand{\changedX}{X}%
We are interested in the behavior of the area function $W$ near $\mcA$, the set of points where $Y$ vanishes. We  have $\nabla W|_{\mcA}=0$ and
\beal{secWcal}
\nabla_\mu \nabla_\nu W|_\mcA &=&- \nabla_\mu \nabla_\nu \left(\fourg (\changedX,\changedX)\fourg(Y,Y)-\fourg(X,Y)^2\right)
 \\
 & = &-2\left(\fourg (\changedX,\changedX)\fourg(\nabla_\mu Y,\nabla_\nu Y)-\fourg(X,\nabla_\mu Y)\fourg(X,\nabla_\nu Y)\right)
% \\
% & = &
 \;.
\eean
The second term vanishes because $[\changedX,Y]=0$, with $Y$ vanishing on $\mcA$:
$$
 \changedX^\alpha \nabla_\nu Y_\alpha|_\mcA = - \changedX^\alpha \nabla_\alpha Y_\nu = - \changedX^\alpha \nabla_\alpha Y_\nu + \underbrace{Y^\alpha}_{=0} \nabla_\alpha \changedX_\nu=- [\changedX,Y]_\nu = 0 \;.
$$
Now, the axis $\mcA$ is timelike,
and the only non-vanishing components of the tensor $\nabla_\mu Y_\nu$ have a spacelike character on $\mcA$. This implies that the quadratic form $\nabla_\mu Y^\alpha \nabla_\nu Y_\alpha$ is semi-positive definite. We have therefore shown

\begin{Lemma}
If $\changedX$ is spacelike at $p\in \mcA$, then $W< 0$  in a
neighborhood of $p$ away from $\mcA$.
\end{Lemma}

Under the conditions of Theorem~\ref{Tdoc}, we conclude that $\changedX$ cannot be spacelike on $\mcA\cap \doc$. To exclude the possibility that  $g(\changedX,\changedX)=0$ there,%
\footnote{The analysis in Section~\ref{srhmp} shows that
$\changedX$ cannot become null on $\mcA\cap\doc$ when the
vacuum equations hold \emph{and} the axis can be identified with a smooth boundary
for the metric $q$; this can be traced to the ``boundary point
Lemma", which guarantees that the gradient of the harmonic
function $\rho$ has no zeros at the boundary $\{\rho=0\}$. But
the behavior of $q$ at those axis points which are not on a
non-degenerate horizon and on which $\changedX$ is null is not
clear.}
let $w$ be defined as in \eq{qwdef},
$$
 w = \changedX^\flat \wedge Y^\flat \;;
$$
here, and throughout this section, we explicitly distinguish between a vector $Z$ and its dual $Z^\flat:= \fourg(Z, \cdot)$.
We will further assume that $\changedX$ is causal at $p$, and that the conclusion of
Lemma~\ref{LCarter} holds:
\bel{Wweq}
dW\wedge w = W dw
\;.
\ee
Let $T$ denote the  field of vectors normal to $\hyp_\mcO$ normalized so that $\fourg(T,\changedX)=1$; note that $T_p$ is, up to a multiplicative factor, as in \eq{Tav}. Let $\gamma$ be any affinely parameterized geodesic such that  $\gamma(0)=p$, $\dot\gamma(0)\perp T_p$ and $\dot\gamma(0) \perp \changedX_p$; a calculation as in \eq{geodcalc} shows that
$$
 \fourg(Y,\dot \gamma)=\fourg(\changedX,\dot \gamma)=0
$$
along $\gamma$. As $Y$ is tangent to $\hyp_\mcO$, from \eq{Wweq} we obtain
\bel{Wweq2}
\underbrace{\frac{dW}{ds} \fourg(Y,Y)}_{=dW\wedge \changedX^\flat \wedge Y^\flat (\dot \gamma, T, Y)}= W dw(\dot \gamma, T, Y)
\;.
\ee
Now, $i_Ydw = \mcL_Yw -d(i_Y w)=-d(i_Y w)$, so that
\beaa
 dw(\dot \gamma, T, Y) &=& -d\Big(i_Y(\changedX^\flat \wedge Y^\flat)\Big) (\dot \gamma, T)
 \\&=& d\Big(-\fourg(Y,\changedX)Y^\flat+\fourg(Y,Y)\changedX^\flat\Big) (\dot \gamma, T)
 \\&=& \Big(-\fourg(Y,\changedX)d Y^\flat+\fourg(Y,Y)d\changedX^\flat\Big) (\dot \gamma, T) + \frac{d(\fourg(Y,Y))}{ds}
 \;.
\eeaa
Inserting this in \eq{Wweq2}, we conclude that
\bel{Wweq3}
\frac{d}{ds}\left(\frac W{ \fourg(Y,Y)} \right)= \underbrace{
 \Big(-\frac{\fourg(Y,\changedX) }{\fourg(Y,Y)}d Y^\flat+d\changedX^\flat\Big) (\dot \gamma, T) }_{=:f} \times \frac W{ \fourg(Y,Y)}
\;.
\ee
Let $\hthreeg$ be the metric induced on $\hyp_\mcO$ by $\fourg$. Then
$\hthreeg$ is a Riemannian metric invariant under the flow of $Y$. As is
well known (compare~\cite{ChUone}) we have $c^{-1}s^2 \le
\fourg(Y,Y)=\hthreeg(Y,Y)\le cs^2$. Since $T\in \Ker  \nabla Y$ we have
$d Y^\flat(T,\cdot)=0$ at $p$. It follows that the function $f$ defined in
\eq{Wweq3} is bounded along $\gamma$ near $p$. If
$\fourg(\changedX,\changedX)=0$ at $p$, then the limit at $p$ of  $W/{
\fourg(Y,Y)} $ along $\gamma$ vanishes by \eq{secWcal}. Using
uniqueness of solutions of ODE's, it follows from \eq{Wweq3} that $W$
vanishes along $\gamma$. But this is not possible in $\doc$ away from
$\mcA$ by Theorem~\ref{Tdoc}. We have therefore proved that the
ergoset does \emph{not} intersect the axis within $\doc$:%
\index{ergoset!theorem}%

\begin{Theorem}[Ergoset theorem]
 \label{TdocA}
In space-time dimension four, and under the conditions of
Theorem~\ref{Tdoc},  $\Kz$ is timelike on $\doc\cap \mcA$.
\qed
\end{Theorem}

A higher dimensional version of Theorem~\ref{TdocA} can
be found in~\cite{ChHighDim}.

A corollary of Theorem~\ref{TdocA} is that, under the
conditions there, the existence of an ergoset implies that of
an event horizon. Here one should keep in mind a similar result
of Haji\v  cek~\cite{HajicekErgospheres}, under conditions that
include
 the hypothesis of smoothness of $\partial E$ (which does not
hold e.g. in Kerr~\cite{LakeKerr}), and affine completeness of
those Killing orbits which are  geodesics, and non-existence of
degenerate Killing horizons. On the other hand, Haji\v cek
assumes the existence of only one Killing vector, while in our
work two Killing vectors are required.
\renewcommand{\changedX}{K}

\section{The reduction to a harmonic map problem}
 \label{srhmp}
\subsection{The orbit  space in space-time dimension four}
 \label{sSos3}
 \renewcommand{\changedX}{X}

Let $(\mcM,\fourg )$ be a chronological, four-dimensional, asymptotically flat space-time
invariant under a $\R\times \Uone $ action, with stationary Killing vector
field $K_\kl 0 \equiv \changedX $ and $2\pi$--periodic Killing vector field
$K_\kl 1 \equiv Y$.
Throughout this section we shall assume that
\bel{condgl}
  \begin{array}{l}
    \mbox{$\doc=\R\times M$, where $M$ is a three dimensional, simply  connected  manifold}\\
    \mbox{with boundary, invariant under the flow of $Y$, with the flow of $\changedX$ consisting of }\\
    \mbox{translations along the $\R$ factor. Moreover  the closure $\bar M$ of  $M$    is the union of a }\\
    \mbox{compact  set and of a finite number of asymptotically flat
ends.}
  \end{array}
\ee
Recall that \eq{condgl} follows from Corollary~\ref{Cdsc} and
Theorem~\ref{Tgt} under appropriate conditions.

Because $\changedX$ and $Y$ commute, the periodic flow of $Y$ on
$\doc$ defines naturally a periodic flow on $M$; in our context this flow
consists of rotations around an axis in the asymptotically flat regions. Now,
every asymptotic end can be compactified by adding a point, with the
action of $\Uone $ extending to the compactified manifold by fixing the
point at infinity. Similarly every boundary component has to be a
sphere~\cite[Lemma~4.9]{Hempel}, which can be filled in by a ball, with
the (unique) action of $\Uone $ on $S^2$ extending to the interior as the
associated rotation of a ball in  $\R^3$, reducing the analysis of the group
action to the boundaryless case. Existence of asymptotically flat regions,
or of boundary spheres, implies that the set of fixed points of the action is
non-empty (see, e.g.,~\cite[Proposition~2.4]{ChBeig2}). Assuming, for
notational simplicity, that there is only one asymptotically flat end, it then
follows from~\cite{Raymond} (see the italicized paragraph on p.52 there)
that, after the addition of a  ball $B_i$  to every boundary component, and
after the addition of a point $i_0$ at infinity to the asymptotic region, the
new manifold $M \cup B_i \cup \{i_0\}$ is homeomorphic to $S^3$, with
the action of $\Uone $ conjugate, by a homeomorphism, to the usual
rotations of $S^3$.  On the other hand, it is shown
in~\cite[Theorem~1.10]{Orlik} that the actions are classified, up to smooth
conjugation, by topological invariants, so that the action of $\Uone $ is
smoothly conjugate  to the usual rotations of $S^3$. It follows that the
manifold $M \cup B_i $ is diffeomorphic to $\R^3$, with the $\Uone$
action  smoothly conjugate to the usual rotations of $\R^3$.  In particular: a)
there exists a global cross-section
$\zMtwo $ for the action of $\Uone $ on $M\cup B_i$ away from the set of fixed points $\mcA$,%
\footnote{We will use the symbol $\mcA$ to denote the set of fixed points of the Killing vector $Y$  in $M$ \emph{or} in $\mcM$, as should be clear from the context.}
with $\zMtwo $ diffeomorphic to an open
half-plane;  b) all isotropy groups are trivial or equal to $\Uone $; c) $\mcA$ is diffeomorphic to $\R$.%
\footnote{We are grateful to Allen Hatcher for
clarifying comments on the classification of $\Uone $ actions.}

Somewhat more generally, the above analysis applies whenever $M$ can
be compactified by adding a finite number of points or balls. A
nontrivial example  is provided by manifolds with a finite number of
asymptotically flat and asymptotically cylindrical ends, as is the
case for the Cauchy surfaces for the domain of outer communication
of the extreme Kerr solution.

Summarizing, under \eq{condgl} there exists in $\doc$ an
embedded two-dimensional manifold $\bMtwo$, diffeomorphic to
$\hMtwo\approx [0,\infty)\times\R$ minus a finite number of
points (corresponding to the remaining asymptotic ends), and
minus a finite number of open half-discs (the boundary of each
corresponding to a connected component of the horizon). We
denote by $\Mtwo$ the manifold obtained by removing from
$\bMtwo$ all its boundaries.

\subsection{Global coordinates on the orbit space}
 \label{sSgcos}

We turn our attention now to the construction of a convenient
coordinate system on a four-dimensional, globally hyperbolic,
$\R\times \Uone $ invariant, simply connected domain of outer
communications $\doc$.   Let $\bMtwo$ and $\zMtwo$ be as in
Section~\ref{sSos3}. We will invoke the uniformization
theorem to understand the geometry of $\bMtwo$; however, some
preparatory work is useful, which will allow us to control both
the asymptotic behavior of the fields involved, as well as the
boundary conditions at various boundaries.

For simplicity we assume that $\doc$ contains only one
asymptotically flat region, which is necessarily the case under
the hypotheses of Theorem~\ref{topocensor}. On $\Mtwo$ there
is a naturally defined orbit space-metric which, away from the
rotation axis $\{Y=0\}$, is defined as follows. Let us denote
by $\fourg$ the metric on space-time, let $\changedX
_1=\changedX $, $\changedX _2=Y$, set $h_{ij}= \fourg(\changedX
_i,\changedX _j)$, let $h^{ij}$ denote the matrix inverse to
$h_{ij}$ wherever defined, and on that last set for $Z_1,Z_2\in
T_p\zMtwo$ set
\bel{OSmetst}
 q(Z_1,Z_2)=\fourg(Z_1,Z_2)- h^{ij}{\fourg(Z_1,\changedX _i)\fourg(Z_2,\changedX _j)}
 \;.
\ee
Note that if $Z_1$ and $Z_2$ are orthogonal to the Killing
vectors, then $q(Z_1,Z_2)=\fourg(Z_1.Z_2)$. This implies that
if the linear span of the Killing vectors is timelike (which,
under our hypotheses below, is the case away from the axis
$\{Y=0\}$ in the domain of outer communications),  then $q$ is
positive definite on the space orthogonal to the Killing
vectors. Also note that $q$ is independent of the choice of the
basis of the space of Killing vectors.

To take advantage of the asymptotic analysis in~\cite{ChUone}, a straightforward calculation shows that $q$ equals
\bel{OSmet}
 q(Z_1,Z_2)=\threeg(Z_1,Z_2)- \frac{\threeg(Y,Z_1)\threeg(Y,Z_2)}{\threeg(Y,Y)}
 \;,
\ee

where $\threeg$ is the (obviously $\Uone$--invariant) metric on the level sets of $t$ (where $t$ is any time function as in Section~\ref{sSos3}) obtained from the space-time metric by a formula similar
to \eq{OSmetst}:
\bel{OSmetst2}
 \threeg(Z_1,Z_2)=\fourg(Z_1,Z_2)- \frac{\fourg(Z_1,\changedX )\fourg(Z_2,\changedX )}{\fourg(\changedX ,\changedX )}
 \;.
\ee
(So $\threeg$ is \emph{not} the metric induced on the level sets of $t$ by $\fourg$.)  The right-hand-side is manifestly well-behaved  in the region where $\changedX $ is timelike; this is the case in the asymptotic region, and   near the axis on $\doc$ under the conditions of Theorem~\ref{TdocA}.

In any case, the asymptotic analysis of~\cite{ChUone} can be invoked
directly to obtain information about the metric $q$ at large distances.
Recall that if the asymptotic flatness conditions \eq{falloff1} hold with $k\ge
1$, then by the field equations \eq{falloff1} holds with $k$ arbitrarily large.
We can thus use~\cite{ChUone} to conclude that there exist coordinates
$x^A$, covering the complement of a compact set in $\R^2$ after the
quotient space has been doubled across the rotation axis, in which $q$ is
manifestly asymptotically flat as well (see Proposition~2.2 and
Remark~2.8 in~\cite{ChUone}):
\bel{twodimAFq}
 q_{AB}-\delta_{AB} = o_{k-3}(r^{-1})
 \;.
\ee
To gain insight into the geometry of $q$ near the horizons, one
can use \eq{OSmetst2} with $\changedX $ being instead the
Killing vector which is null on the horizon. It is then shown
in~\cite{Chstatic} that each non-degenerate component of the
horizon corresponds to a smooth totally geodesic boundary for
$\threeg$. (It is also shown there that every degenerate
component corresponds to a metrically complete end of infinite
extent \emph{provided} that the Killing vector tangent to the
generators of the horizon is timelike on $\doc$ near the
horizon, but it is not clear that this property holds.) Some
information on the asymptotic geometry of $\threeg$ in the
degenerate case can be obtained
from~\cite{Hajicek3Remarks,LP1}; whether or not the information
there suffices  to extend our analysis below to the
non-degenerate case remains to be seen.

\subsection{All horizons non-degenerate}
 \label{sSahnd}
Assuming that all horizons are non-degenerate, we
proceed as follows: Every non-degenerate component of the boundary
$\partial M$ is a smooth sphere $S^2$ invariant under $\Uone$. As is well
known, every isometry of $S^2$ is smoothly conjugate to the action of
rotations around the $z$ axis in a flat $\R^3$, with the rotation axis
meeting $S^2$ at exactly two points. Thus, as already mentioned in
Section~\ref{sSos3}, we can fill each component of the boundary $\partial
M$ by a smooth ball $B^3$, with a rotation-invariant metric there. We   denote by
$\gamma$ any rotation-invariant smooth Riemannian metric on
$\R^3$ which extends the original metric $\gamma$, and by $q$
the associated two-dimensional metric as in \eq{OSmet}. {}From
what has been said we conclude that every non-degenerate component of
the horizon corresponds to a smooth boundary $\partial M/\Uone$ for the
metric $ q$, consisting of a segment which meets the
rotation axis at precisely two points. The filling-in just described is
equivalent to filling in a half-disc in the quotient manifold. Since the
boundary $\partial M$  is a smooth $\Uone$ invariant surface for
$\threeg$, it meets the rotation axis orthogonally. This implies that each
one-dimensional boundary segment of $\partial M/\Uone$  meets the
rotation axis orthogonally in the metric $q$.

Consider, then, a black hole space-time which contains one
asymptotically flat end and  $N$ non-degenerate spherical
horizons. After adding $N$ half-discs as described above, the
quotient space, denoted by $\hat M^2$, is then a
two-dimensional non-compact asymptotically flat manifold
diffeomorphic to a half-plane. Recall that we are assuming
\eq{condgl}, and that there is only one asymptotically flat
region. We will also suppose that
\beal{condgl3}
&&
\mbox{$W>0$ on $\doc\setminus \mcA$, and}
 \\
 &&
 \mbox{on $\doc\cap\mcA$  the stationary Killing vector field $\changedX$ is timelike.}
\eeal{condgl2}
Note that those conditions necessarily hold under the hypotheses of Theorem~\ref{Tdoc}, compare Theorem~\ref{TdocA}.

By \eq{condgl3} the metric $q$ is positive definite away from
$\mcA$. Near $\mcA$ the metric $\threeg$ defined in
\eq{OSmetst2} is Riemannian and smooth by \eq{condgl2}, and the
analysis in~\cite{ChUone} shows that $\mcA$ is a smooth
boundary for $q$. After doubling across the boundary, one
obtains an asymptotically flat metric on $\R^2$.
By~\cite[Proposition~2.3]{ChUone}, for $k\ge 5$ in
\eq{falloff1} there exist global isothermal coordinates for
$q$:
\bel{confrep}q=e^{2u}(dx^2+dy^2)\;, \quad \mbox{with} \
u\longrightarrow_{\sqrt{x^2+y^2} \to \infty}0\;.\ee
In fact, $u=o_{k-4}(r^{-1})$. The existence of such coordinates also follows from the uniformization theorem~(see, e.g.,~\cite{Abikoff}),
but this theorem does not seem to provide the information about the asymptotic behavior in various regimes, needed here, in any obvious way.
As explained in the proof of~\cite[Theorem~2.7]{ChUone}, the coordinates $(x,y)$ can be chosen so that the rotation axis corresponds
to $x=0$, with $\hat M^2=\{x\ge 0\}$.

The next step of the construction is to modify the coordinates
$(x,y)$  of \eq{confrep} to a coordinate system $(\rho,z)$ on
the quotient manifold $\bMtwo$, covering $[0,\infty)\times \R$,
so that $\rho$ vanishes on the rotation axis \emph{and} the
event horizons. This is done by first solving the equation
$$
 \Delta_q \rho_R = 0
 \;,
$$
on $\Omega_R:= \bMtwo\cap \{x^2 + y^2 \le R^2\}$, with zero boundary value on $\partial \bMtwo$, and with $\rho_R= x$ on $\{x^2+y^2=R^2\}$.
Note that
$$
 C=\sup_{\partial  \Omega_R \setminus \mcA}x-\rho_R
 \;,
$$
is independent of $R$, for $R$ large, since $x$ and $\rho_R$ differ only on the event horizons. Since $\Delta_q x =0 $,
the maximum principle implies
$$
 x-C \le \rho_R \le x
 \ \mbox{ on $\Omega_R$}
 \;.
$$
By usual arguments there exists a subsequence $\rho_{R_i}$ which converges, as $i$ tends to infinity, to a $q$--harmonic function $\rho$ on $ \bMtwo$, satisfying the desired boundary values. By standard asymptotic expansions (see, e.g.,~\cite{ChAFT}) we find that
$\nabla \rho$ approaches $\nabla x$ as $\sqrt{x^2+y^2}\to \infty$. In fact, for any $j\in \N$ we have
\bel{rhoas}
 \rho-x =\sum_{i=0}^j \frac{\alpha_i(\varphi)}{ (x^2+y^2)^{i/2}} + O((x^2+y^2)^{-(j+1)/2})
 \; ,
\ee
where $\varphi$ denotes an angular coordinate in the $(x,y)$
plane, with $\alpha_i$ being linear combinations of
$\cos(i\varphi)$ and $\sin(i\varphi)$, with the expansion being
preserved under differentiation in the obvious way. In
particular  $\nabla \rho$ does not vanish for large $x$, so
that for $R$ sufficiently large the level sets $\{\rho=R\}$ are
smooth submanifolds. The strips $0< \rho < R$
are simply
connected so, by the uniformization theorem, there exists a
holomorphic diffeomorphism
$$
 (x,y)\mapsto (\alpha(x,y), \beta(x,y))
$$
from that strip to the set $\{0<\alpha<R\;,\ \beta\in \R\}$. By
composing with a M\"obius map we can further arrange so that
the point at infinity of the $(x,y)$--variables is mapped to
the point at infinity of the $(\alpha,\beta)$--variables. As
the map is holomorphic, the function $\alpha(x,y)$ is harmonic,
with the same boundary values and asymptotic conditions as
$\rho$,  hence $\alpha(x,y)=  \rho(x,y)$ wherever both are
defined.  If we denote by $z$ a harmonic conjugate to $\rho$,
we similarly obtain that $z-\beta$ is a constant, so that the
map
\bel{rhozmap}
 (x,y)\mapsto (\rho,z)
\ee
is a holomorphic diffeomorphism between the strips described above. Since the constant $R$ was arbitrarily large, we conclude that the map \eq{rhozmap} provides a holomorphic diffeomorphism from the interior of $\bMtwo$ to $\{\rho>0\;,\ z\in \R\}$, and provides  the desired coordinate system in which $q$ takes
the form
\bel{confrep2}
 q = e^{2\hat u}(d\rho^2+dz^2)
 \;.
\ee
{}From \eq{rhoas} and its equivalent for $z$ (which is
immediately obtained from the defining equations $\partial_x
\rho=\partial_y z$, $\partial_y \rho = - \partial_x z$) we
infer  that $\hat u\to 0$ as $\sqrt{\rho^2+z^2}$ goes to
infinity, with the decay rate $\hat u = o_{k-4}(r^{-1})$
remaining valid in the new coordinates.

In vacuum the area function $W$ satisfies $\Delta_q \sqrt{W}=0$
(see, e.g., \cite{Weinstein1}). If we assume that $W$ vanishes
on $\pdoc \cup \mcA$ (which is the case under the hypotheses of
Theorem~\ref{Tdoc}), then $W=\rho$ on $\pdoc \cup \mcA$. Since
$\Delta_q \rho=0$ as well, we have $  \Delta_q
(\sqrt{W}-\rho)=0$, with   $W-\rho$ going to zero as one tends
to infinity by \cite{ChUone}, and the maximum principle gives
\bel{Wrhoeq}
 \sqrt{W}=\rho
 \;.
\ee

\subsection{Global coordinates on $\doc$}
According to Section~\ref{sSos3} we have
  $$\doc\setminus \mcA
  \approx \bbR\times S^1 \times \bbR_+^*\times \bbR  \;,   $$
and this diffeomorphism defines a global coordinate system  $(t,\varphi,\rho,z)$ on $ \doc\setminus \mcA$, with $X=\partial_t$ and  $Y=\partial_{\varphi}$.
Letting $(x^A)=(\rho,z)$ and $(x^a)=(t,\varphi)$, we can write the metric in the form
$$
 \fourg = \fourg_{ab}(dx^a+\underbrace{\theta^a{}_A dx^A}_{=:\theta^a} ) (dx^b + \theta^b{}_B dx^B) + q_{AB}dx^A dx^B
\;,
$$
with all functions independent of $t$ and $\varphi$. The
orthogonal integrability  condition of Proposition~\ref{xyort}
gives
$$
d\theta^a=0\;,
$$
so that, by simple connectedness of $\R^*_+\times \R$, there
exist functions $f^a$ such that $\theta^a=df^a$. Redefining the
$x^a$'s to $x^a+f^a$, and keeping the same symbols for the new
coordinates, we conclude that the metric on   $\doc\setminus
\mcA$ has a global coordinate representation as
\begin{equation}
\label{space-timemetric}
\fourg=-\rho^2e^{2\llambda}dt^2+e^{-2\llambda}(d\varphi-vdt)^2+e^{2\hat u}(d\rho^2+dz^2)
\end{equation}
for some functions $v(\rho,z)$, $\llambda(\rho,z)$, with
$\rho$, $z$  and $\hat u$ as in Section~\ref{sSahnd}, see in
particular \eq{Wrhoeq}. We set
\bel{Uxrel}
 U = \llambda+ \ln \rho\;, \quad \mbox{so that} \quad
 \fourg(\partial_\varphi,\partial_\varphi) = \rho^2 e^{-2U}= e^{-2\llambda}
 \;.
\ee
{Let $\omega$ be the {\em twist potential} defined by the equation
\index{twist potential}%
\bel{twist potential} d\omega=*(dY\wedge Y)\;,
\ee
its existence follows from simple-connectedness of $\doc$ and
from $d*(dY\wedge Y)=0$ (see, e.g.,\cite{Weinstein1}). As discussed in more
detail in Section~\ref{ssCs} below
(compare~\cite[Proposition~2]{Weinstein1}),
the space-time metric is uniquely determined by  the
axisymmetric map
\bel{ harm}
 \Phi=(\llambda,\omega):\R^3\setminus \mcA \to \mathbb{H}^2
 \;,
\ee
where $\mathbb{H}^2$ is the hyperbolic space with metric
\bel{hypmetb}
 b:=d\llambda^2 + e^{4\llambda} d\omega^2\;,
\ee
and $\mcA$ is the rotation axis $\mcA:=\{(0,0,z)\;, z\in \R\}\subset \R^3$.  The metric coefficients can be determined from $\Phi$ by solving equations~\eq{metricparameters}-\eq{metricparameters3} below.
The map $\Phi$ solves the harmonic map equations~\cite{Ernst,Exactsolutions2}:
\bel{harmeq}
 |T|_b^2: = (\Delta \llambda - 2 e^{4   \llambda} |D \omega|^2)^2
  + e^{4 \llambda}(\Delta \omega +4  D  \llambda \cdot D   \omega)^2=0
 \;,
\ee
where both $D$ and $\Delta$ refer to the flat metric on $\R^3$, together with a set of asymptotic conditions depending upon the configuration at hand.

We continue with the derivation of those boundary conditions.

\subsection{Boundary conditions at non-degenerate horizons}
 \label{sSbch}

Near the points at which the boundary is analytic (so, e.g., at those points
of the axis at which $\changedX$ is timelike),  the map defined by
\eq{rhozmap} extends to a holomorphic map across the boundary (see,
e.g.,~\cite{Cohn}). This implies that $\hat u$ extends across the axis as a
smooth function of $\rho^2$ and $z$ \emph{away} from the set of points
$\{\fourg(\changedX,\changedX)=0\}$.

Let us now analyze the behavior of $\hat u$ near the points $z_i\in \mcA$
where non-degenerate horizons meet the axis. As described above, after
performing a constant shift in the $y$ coordinate, any component of a
non-degenerate horizon can locally be described by a smooth curve in the
$\zeta:=x+i y$ plane of the form
\bel{hortr}
 y= \gamma(x)\;,\ \gamma(0)=0\;, \ \gamma(x)=\gamma(-x)\;.
\ee
Near the origin, the points lying in the domain of outer communications
correspond then to the values of $x+iy$ lying in a region, say $\Omega$,
bounded by the half-axis  $\{x=0,y\ge 0\}$ and by the curve
$x+i\gamma(x)$, with $x\ge 0$.

To get rid of the right-angle-corner where the curve $x+i\gamma(x)$ meets
the axis, the obvious first attempt is to introduce a new complex coordinate
\bel{wmap}
 w:=\alpha+i\beta=-i \zeta^2
 \;.
 \ee
If we write $\gamma(x)=a_2x^2+O(x^4)$, then the image
of $\{x+i\gamma(x)\;,\ x\ge 0\}$ under \eq{wmap} becomes
\beal{atwo}
 f_1(x+i\gamma(x))&=& 2a_2 x^3 + O(x^5) -i \underbrace{(x^2-a_2^2x^4+ O(x^6))}_{=:-t}
   \\
  & = &
 i t + 2a_2 |t|^{3/2}+O(|t|^{5/2})
 \;.
 \nonumber
 \eea
The remaining part $\{iy$, $y\in \R^+\}$, of the boundary of $\Omega$, is
mapped to itself. It  follows   that the boundary of the image of  $\Omega$
by the map \eq{wmap} is a $C^{1,1/2}$ curve. Here $C^{k,\lambda}$
denotes the space of $k$-times differentiable functions, the $k$'th
derivatives of which satisfy a H\"older condition with index $\lambda$.

To improve the regularity we replace $-i\zeta^2$ by
$f_2(\zeta)=-i\zeta^2+\sigma_3 \zeta^3$ for some constant $\sigma_3$. Then \eq{atwo} becomes
\beal{atwo2}
 \phantom{xxx}
 f_2(x+i\gamma(x))&=& (2a_2 +\Re \sigma_3)x^3 + O(x^5) -i \underbrace{(x^2 + O(x^4))}_{=:-t} -\Im(\sigma_3)O(x^4)
  \\
  & = &
 i t + (2a_2 +\Re \sigma_3) |t|^{3/2}+O(|t|^{5/2})
 \;.
 \nonumber
 \eea
The remaining part of the boundary of $\Omega$ is mapped to the curve $f_2(iy)$, with $y\ge 0$:
%%
%\beal{atwo3}
%f_2(iy)&=&\Im \sigma_3 y^3 + i  {(y^2 -\Re \sigma_3 y^3)} \,,
% \eea
%
%%
\beal{atwo3}
f_2(iy)&=&\Im \sigma_3 y^3 + i \underbrace{(y^2 -\Re \sigma_3 y^3)}_{=:t}
  \\
  & = &
 i t +  \Im  \sigma_3( |t|^{3/2}+O(|t|^{2}))
 \;.
 \nonumber
 \eea
and  is thus mapped to itself if $\sigma_3$ is real. Choosing
$\sigma_3= -2a_2\in \R$     one gets rid of the offending
$|t|^{3/2}$ terms in \eq{atwo2}-\eq{atwo3}, resulting in the
boundary of $f_2(\Omega)$ of $C^{2,1/2}$ differentiability
class.

More generally,  suppose that the image of $x+i\gamma(x)$ by
the polynomial map $\zeta\mapsto
w=f_{k-1}(\zeta)=-i\zeta^2+\ldots$ has a real part equal to
$\beta_{2k-1}x^{2k-1}+O(x^{2k+1})$; then the substraction from
$f_{k-1}$ of a term $\beta_{2k-1}\zeta^{2k-1}$ leads to a new
polynomial map $\zeta\to w=f_k(\zeta)$ which has real part
$\beta_{2k+1}x^{2k+1}+O(x^{2k+3})$, and the differentiability
of the image has been improved by one. Since all the
coefficients $\beta_{2k+1}$ are real, the maps $f_k$ map the
imaginary axis to itself. One should note that this argument
wouldn't work if $\gamma$ had odd powers of $x$ in its Taylor
expansion.

 Summarizing, for any $k$ we can choose a finite polynomial $f_k(\zeta)$,
with lowest order term $-i\zeta^2$, and with the remaining coefficients real
and involving only odd powers of $\zeta$,  which maps the boundary of
$\Omega$ to a curve
\bel{bdef}
(-\epsilon,\epsilon)\ni t\mapsto (\mu(t),\nu(t)):=\left\{
        \begin{array}{ll}
         (0,t), & \hbox{$t\ge 0$;} \\
         (O(t^{k+1/2}),t), & \hbox{$t\le 0$,}

        \end{array}
      \right.
\ee
which is $C^{k,1/2}$.

Note that
\bel{psikdef}
 \psi_k(\zeta):=\sqrt{if_k(\zeta)}= \zeta\Big(1 + O (|\zeta|)\Big)
 \;,
\ee
%,
where $\sqrt{\cdot}$ denotes the principal branch of the square
root, is a holomorphic diffeomorphism near the origin. So
\bel{wdef}
 w=f_k(\zeta)=-i\psi_k^2(\zeta)
\ee
%$
and  we have
\bel{wzetm}
 dw \,d\bw = 4 |\psi_k \psi_k'|^2d\zeta \, d\bzeta = 4 |w|  |\psi_k'|^2d\zeta  \,d\bzeta
 \;.
\ee
We claim that the map
 $$
 w\mapsto \eta:=\rho+i z
 $$
extends across $\rho=0$ to a $C^{k}$ diffeomorphism near the origin.
To see this, note that we have again $\Delta \rho =0$ with respect to the metric $dw d\bw$,
with $\rho$ vanishing on a $C^{k,1/2}$ boundary.
We can straighten the boundary using the transformation
\bel{straightr}
w=(\alpha,\beta)\mapsto (\alpha-\mu(\beta),\beta)=w+(O(|\beta|^{k+1/2}),0)=w+ O(|w|^{k+1/2} )
 \;,
\ee
where $\mu $ is as \eq{bdef}, and $O(\cdot)$ is understood for small $|w|$.
Extending $\rho$ with $-\rho$ across the new boundary, one can use the standard interior Schauder
estimates on the extended function to conclude that
$w\mapsto\rho(w)$ is $C^{k,1/2}$  up-to-boundary. Now, the
condition $dz= \star d\rho$, where $\star$ is the Hodge dual of
the metric $q$, is conformally invariant and therefore holds in
the metric $dw\, d\bw$,  so $z$ is a $C^{k,1/2}$ function of
$w$.
By the boundary version of the maximum principle
we have $d\rho\ne 0$ at the boundary (when understood as a function of $w$),
and hence near the boundary, so $dz$ is non-vanishing near the boundary and orthogonal to $d\rho$.
The implicit function theorem allows us to conclude that the map $w\mapsto \eta$ is a $C^{k,1/2}$ diffeomorphism near $w=0$.

Comparing  \eq{confrep} and \eq{confrep2} we have
\bel{confrepetw}
  e^{2 \hat u} d\eta  \,d\bar \eta= q = e^{2u}d\zeta \, d\bzeta = \frac{e^{2u} }{4 |w| |\psi_k'|^2} dw  \,d\bw  \;,
\ee
in particular $dw  \,d\bw = e^{2 \tilde u_k} d\eta  \,d\bar \eta$, and from what has been said the function $\tilde u_k$ is $C^{k-1,1/2}$ up to boundary. Hence
\bel{huform}
 e^{2\hat u}=  \frac{e^{2u+2\tilde u_k} }{4 |w| |\psi_k'|^2}
\ee
where $u$ is a smooth function of $(x^2,y)$,  while $\psi_k'$ is a non-vanishing holomorphic function of $\zeta=x+iy$, $\tilde u_k$ is a $C^{k-1}$ function of $\eta=\rho+iz$, and $\eta\mapsto w$ is a $C^{k}$ diffeomorphism, with $w$ having a zero of order one where the horizon meets the axis.
Finally $x+iy$ is a holomorphic function of $\sqrt{iw}$, compare \eq{wdef}.

Choosing $k=2$ we obtain
\bel{hfin}
 \hat u = - \frac 12  \ln |w| + \hat u_1 + \hat u_2
 \;,
 \ee
where $w$ is  a smooth complex coordinate which vanishes where
the horizon meets the axis, $\hat u_2=-\ln |\psi_2'|^2/2$ is a
smooth function of $(x,y)$, and $\hat u_1$ is a $C^1$ function
of $(\rho, z)$.

Taylor expanding at the origin, from what has been said (recall
that $\eta\mapsto w$ is conformal and that, near the origin,
$\{\rho=0\}$ coincides with $\{\alpha-\mu(\beta)=0\}$) it
follows that there exists a real number $a>0$ such that
$$
(\rho,z)=\left (a^{-2}(\alpha-\mu(\beta)),a^{-2}\beta\right)+O((\alpha-\mu(\beta))^2+\beta^2)
 \;,
$$
which implies
\bel{abrzinv}
(\alpha,\beta)=  (a^2\rho,a^2z)+O(\rho^2+z^2)
 \;.
\ee
Here we have assumed that $z$ has been shifted by a constant so that it
vanishes at the chosen intersection point of the axis and of the event
horizon.

We conclude that there exists a constant $C$ such that
\bel{hfin2}
 |\hat u + \frac 12  \ln \sqrt{\rho^2+z^2}| \le C
 \quad \mbox{near $(0,0)$}
 \;.
 \ee
This is the desired equation describing the leading order behavior of $\hat
u$ near the meeting point of the axis and a non-degenerate horizon.

\subsubsection{The Ernst potential}
 \label{sSEp} We continue by deriving the boundary conditions satisfied by
the Ernst potential $(U,\omega)$
\index{Ernst potential}%
near the point where the horizon meets
the axis. Here $U$ is as in \eq{space-timemetric}-\eq{Uxrel}, and

$\omega$ is obtained from the function $v$ appearing in the metric by
solving \eq{metricparameters} below.

Our analysis so far can be summarized as:
\bel{diag}
 x+iy=\zeta \ \mapsto\  \psi_k(\zeta) =\sqrt{i f_k(\zeta)}\ \mapsto\  -i(\psi_k(\zeta)
 )^2=w
  \
\mapsto
 \
 \rho+iz
\;.
\ee
Each map is invertible on the sets under consideration; and
each is a $C^k$ diffeomorphism up-to-boundary except for the
middle one, which involves the squaring of a complex number.

Using $\zeta=\psi_k^{-1}(\sqrt{iw})$, the expansion
$$
 \psi_k^{-1}(c+id)=(c+id)\Big(1+ O(\sqrt{c^2+d^2})\Big)
\;,
$$
which follows from \eq{psikdef}, together with \eq{abrzinv}, we obtain
$$
x+iy = a \sqrt{-z+i\rho}+O(\rho^2+z^2)
\;.
$$
Equivalently,
\bel{awxy}
 x= \frac { a\rho}{\sqrt{2(z+\sqrt{z^2+\rho^2})}} +O(\rho^2+z^2)\;,
 \quad
 y =a \sqrt{\frac{z+\sqrt{z^2+\rho^2}}{2}}+O(\rho^2+z^2)
\;.
\ee

To continue, in addition to \eq{condgl}, \eq{condgl3} and \eq{condgl2} we assume that
\beal{condgl4}
 &&\mbox{The level sets of the function $t$, defined as the projection on}
 \\
 &&\mbox{ the $\R$ factor in \eq{condgl}, are spacelike, with $\partial_\varphi t=0$;}
\eean
this is justified for our purposes by Theorem~\ref{Tgt}.
Thus, the Killing vector $\partial_\varphi$  is tangent to the level sets  of $t$, so that
$$
 \fourg(\partial_\varphi,\partial_\varphi)=\hthreeg(\partial_\varphi,\partial_\varphi)
 \;,
$$
where $h$ is the Riemannian metric induced on the level sets of $t$. As shown in~\cite{ChUone},   we have
\bel{hcont}
 \hthreeg(\partial_\varphi,\partial_\varphi)= f(x,y) x^2 \;,
\ee
where the function $f(x,y)$ is uniformly bounded above and below
on compact sets.

Recall that $U$
\index{Ernst potential}%
has been defined as $-\frac 12 \ln (\fourg_{\varphi\varphi}
\rho^{-2})$, and that $(\rho,z)$ have been normalized so that
$(0,0)$ corresponds to a point where a non-degenerate horizon
meets the axis. We want to show that
\bel{Ubbhv}
 U =   \ln \sqrt{z+\sqrt{z^2+\rho^2}} +  O(1) \ \mbox{ near $(0,0)$}
 \;.
\ee
(This formula can be checked for the Kerr metrics by a direct
calculation, but we emphasize that we are considering a general
non-degenerate horizon.) To see that, we use \eq{hcont} to
obtain
\beaa
  \ln (\fourg_{\varphi\varphi} \rho^{-2})  =  \ln(x^2 \rho^{-2}) +  \ln
 (\fourg_{\varphi\varphi} x^{-2})  = 2 \ln(x \rho ^{-1}) +O(1)
 \;.
\eeaa
We assume that $\rho^2+z^2$ is sufficiently small, as required
by the calculations that follow. In the region $0\le |z|\le
2\rho$ we use \eq{awxy} as follows:
\beaa
 \ln(x  \rho  ^{-1})
 & = & \ln \left(\frac {a+  {\sqrt{2\left(\frac z \rho +\sqrt{\frac {z^2}{\rho^2}+1}\right)}}
 O(\rho^{3/2}+\frac{z^2}{\rho^{1/2}})
 }{\sqrt{2(z+\sqrt{z^2+\rho^2})}}\right)
\\
 & = &
 -\ln \left( {\sqrt{2(z+\sqrt{z^2+\rho^2})}}\right) + O(1)
 \;.
\eeaa
In the region $ z \le 0$ we note that
\beaa
 \frac 1 \rho \sqrt{2( z+\sqrt{z^2+\rho^2}) }
 &= &  \frac {\sqrt{2( z+\sqrt{z^2+\rho^2} )}\sqrt{2( -z+\sqrt{z^2+\rho^2} )}}{\rho\sqrt{2( -z+\sqrt{z^2+\rho^2} )}}
\\
 & = &
 \frac {2 }{\sqrt{2(  -z +\sqrt{z^2+\rho^2}
 )}} \le
 \frac {\sqrt{2} }{(z^2+\rho^2)^{1/4}}
 \;.
\eeaa
Hence, again by \eq{awxy},
\beaa
 \ln(x  \rho  ^{-1})
 & = & \ln \left(\frac { a  +\frac 1 \rho {\sqrt{2(z+\sqrt{z^2+\rho^2})}}O(\rho^2+z^2)
 }{\sqrt{2(z+\sqrt{z^2+\rho^2})}}\right)
\\
 & = & \ln \left(\frac { a  + O\big((\rho^2+z^2)^{3/4}\big)
 }{\sqrt{2(z+\sqrt{z^2+\rho^2})}}\right)
 =
 -\ln \left( {\sqrt{2(z+\sqrt{z^2+\rho^2})}}\right) + O(1)
 \;.
\eeaa
In the region $0\le \rho\le z/2 $ some more work is needed.
Instead of \eq{awxy}, we want to use a Taylor expansion of
$\rho$ around the axis $\alpha=0$, where $\alpha$ is as in
\eq{wmap}. To simplify the calculations, note that there is no
loss of generality in assuming that the map $\psi_k$ of
\eq{psikdef} is the identity, by redefining the original
$(x,y)$ coordinates to the new ones obtained from $\psi_k$.
Since in the region $0\le \rho\le z/2 $ we have $\beta\ge 0$,
the function $\mu(\beta)$ in \eq{straightr} vanishes, so
$$
 \alpha (\rho,z) = \underbrace{\alpha (0,z)}_{=\mu\big(\beta(0,z)\big)=0} + \partial_\rho \alpha(0, z) \rho + O(\rho^2)
  =   \partial_\rho \alpha(0, z) \rho + O(\rho^2)
 \;.
$$
Note that $\partial_\rho \alpha(0, z)$ tends to $a^2$ as $z$
tends to zero, so is strictly positive for $z$ small enough.
Instead of \eq{awxy} we now have directly
$$%\bel{awxy2}
 x= \frac { \alpha }{\sqrt{2(\beta+\sqrt{\beta^2+\alpha^2})}}
  \quad \Longrightarrow \quad
 \frac x \rho = \frac {  \partial_\rho \alpha(0, z)
  + O(\rho)}{\sqrt{2(\beta+\sqrt{\beta^2+\alpha^2})}}
 \;.
$$%\ee
In the current region $\alpha$ is equivalent to $\rho$, $\beta$
is equivalent to $z$, $\sqrt{\beta^2+\alpha^2}$ is equivalent
to $z$, and $z$ is equivalent to $ { {2(z+\sqrt{z^2+\rho^2})}}
$, which leads to the desired formula:
\beaa
 \ln(x \rho^{-1}) &=& - \ln \left(  {\sqrt{2(\beta+\sqrt{\beta^2+\alpha^2})}}\right) + O(1)
\\
 & = &
 - \ln \left(  {\sqrt{2(z+\sqrt{z^2+\rho^2})}} \frac  {{\sqrt{2(\beta+\sqrt{\beta^2+\alpha^2})}}}
   { {\sqrt{2(z+\sqrt{z^2+\rho^2})}}}\right) + O(1)
\\
 & = &
 - \ln \left(  {\sqrt{2(z+\sqrt{z^2+\rho^2})}} \right) + O(1)
 \;.
\eeaa
This finishes the proof of \eq{Ubbhv}.

Let us turn our attention now to the twist potential $\omega$:
as is well known, or from~\cite[Equation~(2.6)]{CLW} together
with the analysis in~\cite{ChUone}, $\omega$ is a smooth
function of $(x,y)$, constant on the axis $\{x=0\}$, with odd
$x$--derivatives vanishing there. So, Taylor expanding in $x$,
there exists a constant $\omega_0$ and a bounded function
$\mathring\omega$ such that
\bean
 \omega&= &\omega_0+\mathring \omega(x,y)x^2
 \\
 & = & \omega_0 + \frac{ \mathring \omega(x,y) \Big(a\rho +\sqrt{2(z+\sqrt{z^2+\rho^2})}O(\rho^2+z^2)\Big)^2  }{2(z+\sqrt{z^2+\rho^2})}
 \;.
\eeal{omex}

In our approach below, the proof of black hole uniqueness
requires a uniform bound on the distance between the relevant
harmonic maps. Now, using the coordinates $(\llambda,\omega)$
on hyperbolic space as in \eq{hypmetb}, the distance $d_b$
between two points $(x_1,\omega_1)$ and $(x_2,\omega_2)$ is
implicitly defined by the
formula~\cite[Theorem~7.2.1]{Beardon}: \beaa
    \cosh(d_b) -1 &  = &  \frac{ (e^{-2x_1}-e^{-2x_2})^2 + 4(\omega_1-\omega_2)^2}{2 e^{-2x_1-2x_2}}
 \;.
\eeaa
Using the $(U,\omega)$ parameterization of the maps, with $U$
as in \eq{Uxrel}, the distance measured in the hyperbolic plane
between   two such maps is  the supremum of the function $d_b$:
\beaa
    \cosh(d_b) -1 &  = &  \frac{\rho^4(e^{-2U_1}-e^{-2U_2})^2 + 4(\omega_1-\omega_2)^2}{2\rho^4e^{-2U_1-2U_2}}
 \\
 & = &
 \frac 12 \underbrace{(e^{2(U_1-U_2)} + e^{2(U_2-U_1)} - 2 )}_{(a)} + 2 \underbrace{\rho^{-4}e^{2(U_1+U_2)} (\omega_1-\omega_2)^2}_{(b)}
 \;.
\eeaa
Inserting \eq{Ubbhv} and the analogous expansion for the Ernst
potential of a second  metric into $(a)$ above we obviously
obtain a bounded contribution. Finally, assuming
$\omega_1(0,0)=\omega_2(0,0)$, up to a multiplicative factor
which is uniformly bounded above and bounded away from zero,
$(b)$ can be rewritten as a square of the difference of two
terms of the form
\bel{omdistf} f_i:=
\mathring \omega _i \Big(a_i + \rho^{-1}\sqrt{2(z+\sqrt{z^2+\rho^2})}O(\rho^2+z^2)\Big)^2
 \;,
\ee
with $i=1,2$. We have the following,
for all  $z^2+\rho^2\le 1$:

\begin{enumerate}
\item  The functions  $f_i$ in  \eq{omdistf} are uniformly  bounded in the sector $|z|\le  \rho$:
$$
| f_i|\le  C  \Big(a_i + \sqrt{2(z+\sqrt{z^2+\rho^2})}O(\rho +z^2/\rho )\Big) ^2 \le C'
 \;.
$$
\item
For $0\le \rho\le - z $ we write
$$
0\le
z+\sqrt{z^2+\rho^2}= |z| (\sqrt{1+\frac {\rho^2}{z^2}} - 1)\le C \frac {\rho^2}{|z|}
 \;,
$$
so that
$$
| f_i|\le  C  \Big(a_i +\frac{1}{|z|^{1/2}}O(\rho^2 +z^2 )\Big) ^2 =C(a_i+O(|z|^{3/2}))^2\le C'
 \;.
$$
\item
For $0\le \rho\le  z $   one can proceed as follows: by \eq{hcont},  together with the analysis of $\omega$ in~\cite{ChUone}, there exists a constant $C$ such that near the axis
we have
\bel{metricomomega}
 C^{-1} x^2 \le \fourg (\partial_\varphi,\partial_\varphi) =\hthreeg(\partial_\varphi,\partial_\varphi)\le C x^2\;,\quad \Big |\omega- \underbrace{\omega|_{x=0}}_{=:\omega_0} \Big| \le C x^2
% \;.
\ee
(recall that $\hthreeg$ denotes the metric induced by $\fourg$ on the slices $t=\const$, where $t$ is a time function invariant under the flow of $\partial_\varphi$).
But
\beal{omecont}
&&\\
\nonumber
\frac {(\omega_1-\omega_2)^2}{\rho^4e^{-2U_1-2U_2}}
 & = &
\frac {(\omega_1-\omega_2)^2}{ \fourg_1(\partial_\varphi,\partial_\varphi) \fourg_2(\partial_\varphi,\partial_\varphi) }
 \le 2
\frac {(\omega_1- \omega_0)^2+(\omega_2-\omega_0)^2}{ \fourg_1(\partial_\varphi,\partial_\varphi) \fourg_2(\partial_\varphi,\partial_\varphi) }
 \\
 & = &
 2
\underbrace{\left(\frac {\omega_1- \omega_0}{ \fourg_1(\partial_\varphi,\partial_\varphi) } \right)^2}_{\le C^2}
 \underbrace{\frac { \fourg_1(\partial_\varphi,\partial_\varphi)} { \fourg_2(\partial_\varphi,\partial_\varphi) }}_{=e^{2(U_2-U_1)}}+
 2
\underbrace{\left(\frac {\omega_2- \omega_0}{ \fourg_2(\partial_\varphi,\partial_\varphi) } \right)^2}_{\le C^2}
 \underbrace{\frac { \fourg_2(\partial_\varphi,\partial_\varphi)} { \fourg_1(\partial_\varphi,\partial_\varphi) }}_{=e^{2(U_1-U_2)}}
\;,
\eean
where $\fourg_i$ denotes the respective  space-time metric,
while $x_i$ denotes the respective $x$ coordinate.
Uniform boundedness of this expression, in a neighborhood of the intersection point, follows now from \eq{Ubbhv}.

%The reader should note that this argument does \emph{not} apply for $z<0$,
%because there is no reason for $\omega_1$ to coincide with $\omega_2$ on the axis  for $z< 0$ near zero.
\end{enumerate}

We are ready now to prove one of the significant missing elements of all previous uniqueness claims for the Kerr metric:

\begin{Theorem}
 \label{Tbdist} Suppose that \eq{condgl}, \eq{condgl3}-\eq{condgl2} and
\eq{condgl4} hold. Let $(U_i, \omega_i)$, $i=1,2$, be the Ernst potentials
\index{Ernst potential}%
associated with two vacuum, stationary, asymptotically flat axisymmetric
metrics with smooth non-degenerate event horizons. If
$\omega_1=\omega_2$ on  the rotation axis,  then the hyperbolic-space
distance between $(U_1,\omega_1)$ and $(U_2,\omega_2)$ is bounded,
going to zero as $r$ tends to infinity in the asymptotic region.
\end{Theorem}

\proof We have just proved that the distance between two
different Ernst potentials is bounded near the intersection
points of the horizon and of the axis. In view of \eq{condgl2},
the distance is bounded  on bounded subsets of the axis away
from the horizon intersection points by the analysis
in~\cite{ChUone}. Next,  both  $ \omega_a$'s  are bounded on
the horizon, and both functions $\rho^2 e^{-2U_a}$'s are
bounded on the horizon away from its end points. Finally, both
$\omega_a$'s approach the Kerr twist potential at infinity by
the results in~\cite{SimonBeig} (the asymptotic
Poincar\'e Lemma~8.7 in~\cite{ChDelay} is useful in this
context), so the distance   approaches zero as one recedes to
infinity by a calculation as in \eq{omecont}, together with the
asymptotic analysis of~\cite{ChUone}; a more detailed
exposition can be found in~\cite{CostaPhD}.
\qed

\subsection{The harmonic map problem: existence and uniqueness}
 \label{sShmp}

 In this section we consider Ernst maps
\index{Ernst potential}%
satisfying  the following conditions, modeled on the local behavior of the Kerr solutions:

\begin{enumerate}

\item
There exists $N_{\mbox{\scriptsize\rm  dh}}\ge 0$ degenerate event horizons,
which are represented by punctures
$(\rho=0,z=b_i)$, together with a mass parameter $m_i>0$
and angular momentum parameter  $a_i=\pm m_i$, with the following behavior for small $r_i:=\sqrt{\rho^2+(z-b_i)^2}$,
\begin{equation}
  \label{eq:20e}
U= \ln\Big(\frac{ r_i} {2m_i}\Big)+\frac 12 \ln\left( {1+ \frac{(z-b_i)^2}{r_i^2}
}\right)+ O(r_i).
\end{equation}
The twist potential $\omega$ is a bounded, angle-dependent function
which  jumps by $-4J_i=-4a_im_i$ when crossing $b_i$ from $z<b_i$ to $z>b_i$,
where $J_i$ is the ``angular momentum of the puncture".

\item There exists $N_{\mbox{\scriptsize\rm  ndh}} \ge 0$
    non-degenerate horizons, which are represented by bounded open
    intervals $(c_i^-,c_i^+)=I_i\subset \mcA$, with none of the previous $b_j$'s
    belonging to the union of the closures of the $I_i$. The functions
    $U-2\ln \rho$ and $\omega$ extend smoothly across each interval
    $I_i$, with the following behavior near the end points, for some
    constant $C$, as derived in \eq{Ubbhv}:
\bel{hfin3} |U- \frac 12  \ln (\sqrt{\rho^2+(z-c_i^\pm)^2}+z-c_i^\pm )| \le C \quad
 \mbox{near $(0,c_i^\pm )$} \;. \ee
The function $\omega$ is
assumed to be locally constant on $\mcA\setminus (\cup _i\{b_i\}\cup_j I_j)$, with  expansions as in \eq{omex} nearby.

\item The functions $U$ and $\omega$ are smooth across  $\mcA\setminus (\cup _i\{b_i\}\cup_j I_j)$.
\end{enumerate}

A collection $\{b_i,m_i\}_{i=1}^{\Ndh}$, $I_j$, $j=1,\ldots,\Nndh$, and
$\{\omega_k\}$, where the $\omega_k$'s are the values of $\omega_i$ on
the connected components of  $\mcA\setminus (\cup _i\{b_i\}\cup_j I_j)$,
will be called \emph{``axis data"}.
\index{axis data}%

We have the following~\cite[Appendix~C]{CLW}
(compare~\cite{Weinstein:Hadamard,Dain:2006} and references
therein for previous related results):

\begin{Theorem}
 \label{Tuniquehm}
For any set of axis data  there exists a unique harmonic map $\Phi:\R^3\setminus \mcA\to \mathbb{H}^2$
which lies a finite distance from a solution with the properties 1.--3. above, and such that $\omega=0$ on $\mcA$ for large positive $z$.
\qed
\end{Theorem}

Here the distance between two maps $\Phi_1$ and $\Phi_2$ is defined as
$$
d(\Phi_1,\Phi_2)= \sup_{p\in\R^3\setminus \mcA}
d_b(\Phi_1(p),\Phi_2(p))
 \;,
$$
where the distance $d_b$ is taken with respect to the hyperbolic metric \eq{hypmetb}.

We emphasize the following corollary, first established by Robinson~\cite{RobinsonKerr}
using different methods (and assuming $|a|<m$, which Weinstein~\cite{Weinstein1} does not);
the approach presented here is due to Weinstein~\cite{Weinstein1}:%
\footnote{Yet another  approach can be found
in~\cite{Neugebauer:2003qe}; compare~\cite[Section~2.4]{MAKNP}.
In order to become complete, the proof there needs to be
complemented by a justification of the assumed behavior of
their potential $\Phi$ (not to be confused with the map $\Phi$
here) on the set $\{\rho=0\}$. More precisely, one needs to
justify differentiability of $\Phi$ on $\{\rho=0\}$ away from
the horizons,  continuity of $\Phi$ and $\Phi'$ at the points
where the horizon meets the rotation axis, as well as the
detailed differentiability properties of $\Phi$  near
degenerate horizons as implicitly assumed
in~\cite[Section~2.4]{MAKNP}.}

\begin{Corollary}
\label{CKu}
For each mass parameter $m$ and angular momentum parameter $a\in (-m,m)$
there exists only one map $\Phi$ with the behavior at the axis corresponding to {\aregular} axisymmetric vacuum black hole with a connected non-degenerate
horizon centered at the origin and with $\omega$ vanishing on $\mcA$ for large positive $z$. Furthermore, no {\regular} non-degenerate axisymmetric vacuum black holes with $|a|\ge m$ exist.
\end{Corollary}

\begin{proof}
Theorem~\ref{Tgt} shows that \eq{condgl} and \eq{condgl4} hold,
\eq{condgl3} follows from Theorem~\ref{Tdoc}, while \eq{condgl2} holds by
the Ergoset Theorem~\ref{TdocA}. One can thus introduce $(\rho,z)$
coordinates on the orbit space as in Section~\ref{sSgcos}, then the event
horizon corresponds to a connected interval of the axis of length $\ell$, for
some $\ell>0$. Let $(U,\omega)$ be the Ernst potential corresponding to
the black hole under consideration, with $\omega$ normalized to vanish
on $\mcA$ for large positive $z$. Let $J$ be the total angular momentum
of  the black hole, there exists a Kerr solution $(U_K,\omega_K)$, with
$\omega_K $ normalized to vanish on $\mcA$ for large positive $z$, and
such that the corresponding ``horizon interval" has the same length $\ell$.
We can adjust the $z$ coordinate so that the horizon intervals coincide.
The value of $\omega$ on the axis for large negative $z$ equals $4J$,
similarly for $\omega_K$, hence $\omega=\omega_K$ on the axis except
possibly on the horizon interval.  Theorem~\ref{Tbdist} shows  that
$(U,\omega)$  lies at a finite distance from $(U_K,\omega_K)$. By the
uniqueness part of Theorem~\ref{Tuniquehm} we find
$(U,\omega)=(U_K,\omega_K)$, thus the ADM mass of the black hole
equals   the mass of the comparison Kerr solution, and  $|a|< m$ follows.
\end{proof}

\subsection{Candidate solutions}
 \label{ssCs}

Each harmonic map $(\llambda,\omega)$  of Theorem~\ref{Tuniquehm} with $ N_{\mbox{\scriptsize\rm  dh}} + N_{\mbox{\scriptsize\rm  ndh}} \ge 1$ provides a  candidate for a solution with $ N_{\mbox{\scriptsize\rm  dh}} + N_{\mbox{\scriptsize\rm  ndh}} $ components of the event horizon, as follows: let the functions $v$ and $\hat u$ be the unique solutions of the set of equations
\begin{eqnarray}
\label{metricparameters}
 &
 \partial_\rho v = -e^{4\llambda}\rho\:\partial_z\omega
 \;, \qquad
 \partial_z v = e^{4\llambda}\rho\:\partial_{\rho}\omega
 \;,
 &
 \\
&
\partial_{\rho} \hat u = \rho\left[ (\partial_{\rho}\llambda)^2 -(\partial_{z}\llambda)^2
+\frac{1}{4}e^{4\llambda}((\partial_{\rho}\omega)^2-(\partial_{z}\omega)^2)\right]
+\partial_{\rho}\llambda
 \;,
 &
\\
&
\partial_{z} \hat u = 2\:\rho\left[ \partial_{\rho}\llambda\: \partial_{z}\llambda
+\frac{1}{4}e^{4\llambda}\partial_{\rho}\omega\:\partial_{z}\omega\right]
+\partial_{z}\llambda
\;,
\label{metricparameters3}
&
\end{eqnarray}
which go to zero at infinity. (Those equations are compatible whenever $(\llambda,\omega)$ satisfy the harmonic map equations.)
Then the metric \eq{space-timemetric}
satisfies the vacuum Einstein equations
(see, e.g.,~\cite[Eqs.~(2.19)-(2.22)]{Weinstein3}). Every such solution provides a candidate for a regular, vacuum, stationary, axisymmetric black hole with several components of the event horizon. If $ N_{\mbox{\scriptsize\rm  dh}} + N_{\mbox{\scriptsize\rm  ndh}}=1$ the resulting metrics are of course the Kerr ones.

At the time of writing of this work, it is not known whether any such candidate solution other than Kerr itself describes {\aregular} black hole.
It should be emphasized that there are two separate issues here: The first is that of uniqueness, which is settled by the uniqueness part of Theorem~\ref{Tuniquehm} together with the remaining analysis in this section:
\emph{if} there exist stationary axisymmetric multi-black hole solutions, with all components of the horizon non-degenerate, \emph{then} they belong to the family described by   the harmonic maps of Theorem~\ref{Tuniquehm}.  Note that Theorem~\ref{Tuniquehm} extends to those solutions with degenerate horizons with the behavior described in \eq{eq:20e}. Conceivably this covers all degenerate horizons, but this remains to be established.

Another question is that of the global properties of the candidate solutions:
for this  one needs, first, to study the behavior of the harmonic maps of
 Theorem~\ref{Tuniquehm} near the singular set  in much more detail in
 order  to establish e.g. existence of a smooth event horizon;  an analysis
 of this issue has only been done so far~\cite{Weinstein1,LiTian} if
 $N_{\mbox{\scriptsize\rm  dh}}=0$ away from the points where the axis
 meets the horizon, and the question of space-time regularity at those points is wide open.
Regardless of this, one expects that for all such solutions the integration
of the remaining equations \eq{metricparameters}-\eq{metricparameters3}
will lead to singular ``struts" in the space-time metric \eq{space-timemetric} somewhere on $\mcA$.

\section{Proof of Theorem~\ref{Tubh}}
 \label{Sproof}

If $\mcEp$ is empty, the conclusion follows from the Komar identity and the rigid positive energy theorem (see, e.g.~\cite[Section~4]{Chstatic}).
Otherwise the proof splits  into two cases, according to whether or not $\changedX $ is tangent to the generators of $\mcEp$, to be covered separately in Sections~\ref{sSRh} and \ref{sSnrc}.

\subsection{Rotating horizons}
  \label{sSRh}
Suppose, first, that the Killing vector  is not tangent to the generators of
some connected component $\mcE^+_0$  of $\mcE^+=\mcH^+\cap I^+(\Mext)$.
Theorem~\ref{Trig} shows  that the isometry group of $(\mcM,\fourg)$ contains
$\R\times \Uone $. By Corollary \ref{Cdsc} $\doc$ is simply connected so that,
in view of Theorem~\ref{Tgt}, the analysis of Section~\ref{srhmp} applies,
leading to the global representation \eq{space-timemetric} of the metric.
The analysis of the behavior near the symmetry axis of the harmonic map $\Phi$
of Section~\ref{sSbch} shows that $\Phi$ lies a finite distance
from one of the solutions of Theorem~\ref{Tuniquehm}, and the uniqueness part
of that last theorem allows us to conclude; compare Corollary~\ref{CKu}  in the connected case.

\subsection{Non-rotating case}
\label{sSnrc}

The case where the stationary Killing vector $\changedX $ is
tangent to the generators of every component of $\mcH^+$ will
be referred to as the \emph{non-rotating one}. By hypothesis
$\nabla (\fourg(\changedX ,\changedX ))$ has no zeros on
$\mcEp$, so all components of the future event horizon are
non-degenerate.

Deforming $\hyp$ near $\partial \hyp$ if necessary, we may
without loss of generality assume that $\hyp$ can be extended
across $\mcEp$ to a smooth spacelike hypersurface there.

For the proof we need a new hypersurface  $\hyp''$ which is
maximal, Cauchy for $\doc$, with $\changedX $ vanishing on
$\partial \hyp''$. Under our hypotheses such a hypersurface
will not exist in general, so we start by replacing
$(\mcM,\fourg)$ by a new space-time $(\mcM',\fourg')$  with the
following properties:

\begin{enumerate}
\item $(\mcM',\fourg')$ contains a region $\doc'$ isometric  to $(\doc,\fourg)$;
\item $(\mcM',\fourg')$ is invariant under the flow of a
    Killing vector $\changedX '$ which coincides with
    $\changedX $ on $\doc$;
\item    Each connected component of the horizon $\mcEpz'$ is
    contained in a bifurcate Killing horizon, which contains a
    ``bifurcation surface" where $\changedX '$ vanishes. We will
    denote by $S$ the union of these bifurcation surfaces.
 \end{enumerate}

This is done by attaching to $\doc$ a bifurcate horizon   near each connected component of $\mcEp$ as in~\cite{RaczWald2}, invoking Corollary~\ref{Cpht}.

We wish, now to construct a Cauchy surface $\hyp'$ for $\doc'$   such that  $\partial \hyp'=S$. To do
that, for $\epsilon>0$ let $\fourg_\epsilon$ denote a family of metrics such that $\fourg_\epsilon $ tends  to $\fourg$, as $\epsilon$ goes to zero, uniformly on compact sets, with the property that null directions for $\fourg_\epsilon$ are spacelike for $\fourg$. Consider the family of $\fourg_\epsilon$--null Lipschitz hypersurfaces
$$
 \mcN_\epsilon:= \dot J^+_\epsilon(S)\cap \mcM
 \;,
$$
where $\dot J^+_\epsilon$ denotes the boundary of the causal future with respect to the metric $\fourg_\epsilon$. The $\mcN_\epsilon$'s are threaded with $\fourg_\epsilon$--null geodesics, with initial points on $S$, which converge uniformly to $\fourg$-null geodesics starting from $S$, hence to the generators of $\mcEp$ (within $\mcM'$).  It follows that, for all $\epsilon$ small enough,  $\mcN_\epsilon$ intersects $\hyp$  transversally. Furthermore, since $\mcEp$ is smooth, decreasing $\epsilon$ if necessary, continuity of Jacobi fields with respect to $\epsilon$ implies that the $\mcN_\epsilon$'s remain smooth in the portion between $S$ and their intersection with $\hyp$. Choosing $\epsilon$ small enough, one obtains a smooth $\fourg$-spacelike hypersurface $\hyp'$, with boundary at $S$, by taking the union of  the portion of $\mcN_\epsilon$ between $S$ and where it meets $\hyp$, with that portion of  $\hyp$  which extends to infinity and which is bounded by the intersection with $\mcN_\epsilon$, and smoothing out the intersection. The hypersurface $\hyp'$ can be shown to be Cauchy by the usual arguments~\cite{BILY,Galloway:cauchy}.

By~\cite{ChWald1} there exists an asymptotically flat  Cauchy hypersurface $\hyp''$ for $\doc$, with boundary on $S$, which is maximal.

We wish to show, now, that $\doc'$ , and hence $\doc$, are
static; this has been first proved in~\cite{Sudarsky:wald}, but
a rather simple proof proceeds as follows: Let us decompose
$\changedX' $ as $Nn+Z$, where $n$ is the future-directed
normal to $\hyp''$, while $Z$ is tangent.  The space-time
Killing equations imply
\bel{Yeq} D_i Z_j + D_j Z_i = -2 N K_{ij}\;,
\ee
where $g_{ij}$ is the metric induced on $\hyp''$, $K_{ij}$ is its extrinsic curvature tensor, and $D$ is the
covariant derivative operator of $g_{ij}$. Since $\hyp''$ is
maximal, the (vacuum) momentum constraint reads
\bel{MomC} D_iK^{ij} = 0\;.
\ee
{}From \eq{Yeq}-\eq{MomC} one obtains
\bel{fi1}
D_i(K^{ij}Z_j) = -N K^{ij}K_{ij}\;. \ee
Integrating \eq{fi1} over $\hyp''$, the boundary integral in the
asymptotically flat regions gives no contribution because $K_{ij}$
approaches zero there as $O(1/r^{n-1})$, while  $ Z$
approaches zero there as $O(1/r^{n-2})$~\cite{ChMaerten}.
The boundary integral at the horizons
vanishes since $Z$ and $N$ vanish on $S=\partial \hyp''$ by
construction. Hence
\bel{Vanint}
\int _{\hyp''} NK^{ij}K_{ij}=0\;.
 \ee
On a maximal hypersurface the normal component $N$ of
a Killing vector satisfies the equation
\bel{Neq1}
\Delta N = K^{ij}K_{ij} N\;,
\ee
and the maximum principle shows that $N$ is strictly positive
except at $\partial \hyp''$. Staticity of $\doc'$ along
$\hyp''$ follows now from \eq{Vanint}. Moving the $\hyp''$'s
with the isometry group one covers $\doc'$~\cite{ChWald1}, and
staticity of $\doc'$ follows. Hence $\doc$ is static as well,
and Theorem~\ref{Tubhs} allows us to conclude that $\doc$ is
Schwarzschildian. This achieves the proof of
Theorem~\ref{Tubh}.
\qed

\section{Concluding remarks}
 \label{sCr}

To obtain a satisfactory uniqueness theory in four dimensions,
 the following issues remain to be addressed:

\begin{enumerate}
\item The previous versions of the uniqueness theorem required
    analyticity of \emph{both} the metric \emph{and} the
    horizon. As shown in Theorem~\ref{Tsmooth}, the latter
    follows from the former. This is a worthwhile improvement,
    as even $C^1$--\emph{differentiability  of the horizon} is not clear a
    priori. But the hypothesis of analyticity of the metric remains to be removed.

In this context one should keep in mind the Curzon solution,
where analyticity of the metric fails precisely at the horizon.
We further note an interesting recent uniqueness theorem for Kerr
without analyticity conditions~\cite{IonescuKlainerman1}.
However, the examples constructed at the end of
Section~\ref{ssnhg} show that further insights are needed to be
able to conclude along the lines envisaged there.

The hypothesis of analyticity is particularly annoying in the
static context, being needed there only to exclude
non-embedded Killing prehorizons. The nature of that
problem seems to be rather different from Hawking's rigidity,
with presumably a simpler solution, yet to be found.

\item The question of uniqueness of black holes with degenerate
    components of the Killing horizon requires further
    investigations. Recall that non-existence of stationary,
    vacuum, {\regular}  black holes with \emph{all } components
    of the event horizon non-rotating and degenerate, follows
    immediately from the Komar identity and the positive energy
    theorem~\cite{Horowitz}
    (compare~\cite[Section~4]{Chstatic}).  Furthermore, the
    results here go a long way to prove uniqueness of
    degenerate, stationary, axisymmetric, rotating
    configurations: the only element missing is an equivalent
    of Theorem~\ref{Tbdist}. We expect that  Theorem~\ref{TLP}
    can be useful for solving this problem, and we hope to
    return to that question in the near future.

In any case, the above would not cover solutions with
degenerate non-rotating components. One could exclude such
solutions by proving existence of maximal hypersurfaces within
$\doc$ with an appropriate asymptotic behavior at the
cylindrical ends. The argument presented in Section~\ref{sSnrc}
would then apply to give staticity, and non-existence would
then follow from~\cite{CRT}, or from Theorem~\ref{Tubhs}.

\item The question of existence of multi-component solutions needs to be settled.
\end{enumerate}

And, of course, the question of classification of higher dimensional stationary black holes is largely unchartered territory.

\bigskip

\noindent {\sc Acknowledgements:} We are grateful to J.~Isenberg for
numerous comments on a previous version of the paper. PTC is grateful to
R.~Wald and G.~Weinstein for useful discussions. Similarly JLC wishes to
thank J.~Nat\'ario for many useful discussions.

%...
%...
%...
\backmatter
%\appendix
\bibliographystyle {smfplain}
%\bibliographystyle{amsplain}
%\bibliographystyle{/usr/share/texmf/tex/revtex/prsty}
%\bibliography{%ptjjk,
%../../references/newbiblio,%
%../../references/reffile,%
%../../references/bibl,%
%../../references/Energy,%
%../../references/hip_bib,%
%../../references/netbiblio}

\bibliography{%ptjjk,
../references/newbiblio,%
../references/newbib,%
../references/reffile,%
../references/bibl,%
../references/Energy,%
../references/hip_bib,%
../references/netbiblio,../references/addon}
%\texttt{\input{READMEl}}
%\bibliography {}
\printindex
\end {document}

%% file: bhmacros.tex
%This mnote stuff should  be on the top of the file
%
%
\newcommand{\mnote}[1]%{}
{\protect{\stepcounter{mnotecount}}$^{\mbox{\footnotesize $%
\!\!\!\!\!\!\,\bullet$\themnotecount}}$ \marginpar{%\color{red}
\raggedright\tiny\em $\!\!\!\!\!\!\,\bullet$\themnotecount: #1} }

\newcommand{\Spn}{S^+_0}

\newcommand{\mcEp}{{\mcE^+}}
\newcommand{\mcEpz}{{\mcE^+_0}}

\newcommand{\kk}[1]{}%{\mnote{{\bf If we consider the KK case:} #1}}

\newcommand{\zR}{\mathring R}
\newcommand{\zD}{\mathring D}
\newcommand{\zA}{\mathring A}
\newcommand{\mzh}{\mathring h}

\newcommand{\odoc}{\overline{\doc}}
\newcommand{\ohyp}{\,\,\overline{\!\!\hyp}}
\newcommand{\pohyp}{\partial\ohyp}

\newcommand{\aregular}{{an {\regular}}}
\newcommand{\regular}{$I^+$--regular}%\newchange{terminology changed (macro, can be restored by resetting to ``regular")}}

\newcommand{\hthreeg}{h}

\newcommand{\odocup}{{\doc}\cup \mcH_0^+}
\newcommand{\llambda}{\lambda}

\newcommand{\ue}{u_{\epsilon}}
\newcommand{\uee}{u_{\epsilon,\eta}}

\newcommand{\bmcM}{\,\,\,\,\widetilde{\!\!\!\!\mcM}}
\newcommand{\bfourg}{\widetilde{\fourg}}

\newcommand{\eean}{\nonumber\end{eqnarray}}

\newcommand{\lp}{\ell}

\newcommand{\tSp}{\tilde S_p}
\newcommand{\Sp}{S_p}

\newcommand{\dgtcp}{\dgt}%{(\dgt\cup \zh)}
\newcommand{\dgtc}{\dgt}%{\dgt\cup \zh}

\newcommand{\mcHp}{{\mcH^+}}
\newcommand{\mcHpz}{{\mcH^+_0}}
\newcommand{\mcHm}{{\mcH^-}}

\newcommand{\Nndh}{N_{\mbox{\scriptsize\rm  ndh}}}
\newcommand{\Ndh}{N_{\mbox{\scriptsize\rm  dh}}}

\newcommand{\nic}{}

\newcommand{\hypo}{\,\,\mathring{\!\! \hyp}}
\newcommand{\ohypo}{\overline{\hypo}}
\newcommand{\hypot}{\,\,\mathring{\!\! \hyp_t}}
\newcommand{\hypoz}{\,\,\mathring{\!\! \hyp_0}}
\newcommand{\ohypoz}{\overline{\hypoz}}
\newcommand{\ohypot}{\overline{\hypot}}

\newcommand{\Kz}{K_\kl 0}

\newcommand{\fourge}{\fourg_\epsilon}

\newcommand{\Cp}{ \mcC^+ }
\newcommand{\Cpe}{ \mcC^+ _\epsilon}
\newcommand{\Ct}{ \mcC^+_t}
\newcommand{\Ctm}{ \mcC^+_{t_-}}
\newcommand{\hStmR}{\hat S_{t_-,R}}
\newcommand{\hStR}{\hat S_{t,R}}
\newcommand{\hSzR}{\hat S_{0,R}}
\newcommand{\hSts}{\hat S_{\tau,\sigma}}

\newcommand{\Sone}{\Sz}
\newcommand{\Sonep}{\Sz}%{S_{0,p}}

\newcommand{\Sz}{S_0}

\newcommand{\bw}{\bar w}
\newcommand{\bzeta}{\bar \zeta}

\newcommand{\zMtwo}{\mathring{M}{}^2}
\newcommand{\Mtwo}{{M}{}^2}
\newcommand{\bMtwo}{{\bar M}{}^2}
\newcommand{\hMtwo}{{\hat M}{}^2}

\newcommand{\Mtext}{\Sext}%{M_{\mbox{\scriptsize \rm ext}}}
\newcommand{\Mint}{\mcM_{\mbox{\scriptsize \rm int}}}
\newcommand{\mcMext}{\Mext}

\newcommand{\mcHN}{\mcN}

\newcommand{\hS }{{\hat S }}

\newcommand{\hSp}{{\hat S_p}}
\newcommand{\hSpn}{{\hat S_{p_n}}}

\newcommand{\mcA}{\mycal A}
\newcommand{\mcZ}{\mycal Z}

\newcommand{\zh}{{\,\,\widetilde{\!\!\mcZ}}}
\newcommand{\dgt}{{\mycal Z}_{\mbox{\scriptsize \rm dgt}}}

\newcommand{\kl}[1]{{(#1)}}

\newcommand{\Uone}{{\mathrm{U(1)}}}

 % The next ones are HR

\newcommand{\bbR}{\mathbb{R}}

\def\scro{{\mycal O}}
%{{\mycal Doc}}
\def\scrip{\scri^{+}}%
\def\e{\wedge}

\def\K0{\phi^{K_0}}

\def\X.{\phi^{X}\cdot}
%{|K_0\e...\e K_{D-3}|^2}

%\newcommand{\Ric}{\operatorname{Ric}}

{\catcode `\@=11 \global\let\AddToReset=\@addtoreset}
\AddToReset{equation}{section}
\renewcommand{\theequation}{\thesection.\arabic{equation}}

\newcommand{\fourg}{{\mathfrak g }}

\newcommand{\mcN}{{\mycal N}}

\newcommand{\nopcite}[1]{}

\newcommand{\Ker}{\mbox{\rm Ker}}

%{{\widehat \riemg}}
%{{\widehat \riemgz}}

%\newcommand{\<}{\langle}
%\renewcommand{\>}{\rangle}

%\renewcommand{\hbar}{{\overline \riemgz}}

%\newcommand{\letters}
% {\renewcommand{\theenumi}{\alph{enumi}}
%   \renewcommand{\labelenumi}{(\theenumi)}}
%\newcommand{\romanletters}
  {\renewcommand{\theenumi}{\roman{enumi}}
   \renewcommand{\labelenumi}{(\theenumi)}}

\newcommand{\const}{\mathrm{const}}

\newcommand{\mcE}{{\mycal E}}
\newcommand{\mcC}{{\mycal C}}

\newcommand{\mcW}{{\mycal W}}

\newcommand{\Lie}{\EuScript L}
\newcommand{\nablash}{\nabla{\kern -.75 em
     \raise 1.5 true pt\hbox{{\bf/}}}\kern +.1 em}
\newcommand{\Deltash}{\Delta{\kern -.69 em
     \raise .2 true pt\hbox{{\bf/}}}\kern +.1 em}
\newcommand{\Rslash}{R{\kern -.60 em
     \raise 1.5 true pt\hbox{{\bf/}}}\kern +.1 em}

\newcommand{\Ric}{\operatorname{Ric}}

\newcommand{\mcO}{{\mycal O}}

\newcommand{\mcU}{{\mycal U}}
\newcommand{\mcV}{{\mycal V}}

 % exterior differential

\newcommand{\hyp}{{\mycal S}}%{\Sigma}{{\mycal S}}

%{\,\,\,\overline{\!\!\!\mycal S}}

 % koneksja tla
 % metryka tla

 %{\pi}} % kontrawariantna gestosc metryki
                            %na czasoprzestrzeni
 % ped sprzezony do kontrawariantnej gestosci
                    % metryki (A duze w innych pracach JK i JJ)
%\newcommand{\E}[1]{{\rm e}^{#1}}
%\newcommand{\kolo}[1]{\stackrel{\circ}{#1}}
%\newcommand{\ve}{\varepsilon}
 %three dimensional metric in
                                %space-time, used for the inverse

\newcommand{\threeg}{\gamma}

 % determinant of the
                                             % three dimensional
                                             % metric pulled back
                               %to the model space
 % the lapse function on the model space
 % the shift vector field on the model
                             % space
 %three dimensional ADM momentum pulled back
                               %to the model space
 % conformally rescaled metric
 % standard round metric on the two
                              % sphere

 % pochodna metryki tla
 % pochodna metryki tla
 %{{}^{n-1}M} %the n-1 dimensional manifold

\newcommand{\mcM}{{\mycal M}}

\newcommand{\mcH}{{\mycal H}}

\newcommand{\bea}{\begin{eqnarray}}
\newcommand{\beaa}{\begin{eqnarray*}}
\newcommand{\bean}{\begin{eqnarray}\nonumber}

\newcommand{\bel}[1]{\begin{equation}\label{#1}}
\newcommand{\beal}[1]{\begin{eqnarray}\label{#1}}
\newcommand{\beadl}[1]{\begin{deqarr}\label{#1}}
\newcommand{\eeadl}[1]{\arrlabel{#1}\end{deqarr}}
\newcommand{\eeal}[1]{\label{#1}\end{eqnarray}}
\newcommand{\eead}[1]{\end{deqarr}}
\newcommand{\eea}{\end{eqnarray}}
\newcommand{\eeaa}{\end{eqnarray*}}

\newcommand{\be}{\begin{equation}}
\newcommand{\ee}{\end{equation}}

%{\mbox{\rm \scriptsize ext}\,}

\newcommand{\eq}[1]{\eqref{#1}}
\newtheorem{defi}{\sc Coco\rm}[section]
\newtheorem{theorem}[defi]{\sc Theorem\rm}
\newtheorem{Theorem}[defi]{\sc Theorem\rm}

\newtheorem{Conjecture}[defi]{\sc Conjecture\rm}

\newtheorem{Definition}[defi]{\sc Definition\rm}

\newtheorem{Proposition}[defi]{\sc Proposition\rm}

\newtheorem{Lemma}[defi]{\sc Lemma\!\rm}

\newtheorem{Corollary}[defi]{\sc Corollary\!\rm}
%\theoremstyle{remark}
%\theorembodyfont{\upshape}

\newtheorem{Remark}[defi]{{\sc Remark}\rm}

%\newtheorem{example}[defi]{{\sc Example}\rm}
%\newtheorem{remark}[defi]{{\sc Remark}\rm}

%%%%%%%%%%%%%%%%%%%%%%%%%%%%%%%%%%%%%%%%%%%%%%%%%%%%%

%\newcommand{\proof}{\noindent {\sc Proof:\ }}

\def \R {\Reel}

\newcommand{\mcL}{{\mycal L}}
\def \Nat{\mathbb{N}}

\def \N {\Nat}

%\newcommand{\T}{\Bbb{T}}

%\newcommand{\Reel}[0]{\mbox{$\mathbb{R} $}}

%\newcommand{\Hyp}[0]{\mbox{$\mathbb{H} $}}

%\newcommand{\Nat}[0]{\mbox{$\mathbb{N} $}}

%\newcommand{\Sphere}[0]{\mbox{$\mathbb{S} $}}

%\newcommand{\Reel}[0]{\mbox{$\mathit{R} $}}

%\newcommand{\Hyp}[0]{\mbox{$\mathit{H} $}}

%\newcommand{\Nat}[0]{\mbox{$\mathit{N} $}}

%\newcommand{\Sphere}[0]{\mbox{$\mathit{S} $}}

%%%%%%%%%%%%%%%%%%%%%%% notes en marge numerotees %%%%%%%%%%%%%%%%%%%

\newcounter{mnotecount}[section]

\renewcommand{\themnotecount}{\thesection.\arabic{mnotecount}}

\newcommand{\ednote}[1]{}%{\mnote{#1}}

 % The next ones are HR%
%\newcommand{\Sm}{\ensuremath{\Sigma_{-}}}
%\newcommand{\No}{\ensuremath{N_{1}}}
%\newcommand{\Nt}{\ensuremath{N_{2}}}
%\newcommand{\Nth}{\ensuremath{N_{3}}}

%\definecolor{turquoise}{rgb}{0.25,0.8,0.7}
\definecolor{bluem}{rgb}{0,0,0.5}

%Il existe deux repères pour cela :
%+ cyan, magent, yellow, black
\definecolor{mycolor}{cmyk}{0.5,0.1,0.5,0}
\definecolor{michel}{rgb}{0.5,0.9,0.9}
%\newcmykcolor{le_nom_de_la_couleur}{w x y z}
%avec w,x,y,z entre 0.0 et 1.0

%+ red,green, blue et la commande :
\definecolor{turquoise}{rgb}{0.25,0.8,0.7}
\definecolor{bluem}{rgb}{0,0,0.5}

\definecolor{MDB}{rgb}{0,0.08,0.45}
\definecolor{MyDarkBlue}{rgb}{0,0.08,0.45}

\definecolor{MLM}{cmyk}{0.1,0.8,0,0.1}
\definecolor{MyLightMagenta}{cmyk}{0.1,0.8,0,0.1}

\definecolor{HP}{rgb}{1,0.09,0.58}

\newcommand{\opp}[1]{}%{\label{{\mnote{{\color{HP} open problem}}}

\newcommand{\pdoc}{\partial \doc}

\newcommand{\Sext}{\hyp_{\mathrm{ext}}}
\newcommand{\Mext}{\mcM_{\mathrm{ext}}}

\newcommand{\doc}{\langle\langle \mcMext\rangle\rangle}

%{\,\,\,\,\widetilde{\!\!\!\!\cM}}

%\newcommand{\tg}{{\tilde g}}

%{\hskip10pt {\overline{\phantom{I}}\hskip-15pt{\mycal M}}}

\newcommand{\mcB}{{\mycal B}}
%\newcommand{\cL}{{\mycal  L}}

 %background Riemannian metric
 %identity matrix
 %constants

%\def\ba{\begin{eqnarray}}
%\def\ea{\end{eqnarray}}

%\def\half {{1\over 2}}

\def\exp{\,\mbox{exp}}

\def\emph#1{{\it #1}}
\def\textbf#1{{\bf #1}}

\def\AA{{\mycal  A}}

\def\Lie{{\mycal  L}}

\def\L{{\bf L}}
\def\O{{\bf O}}

\def\R{{\mathbb R}}

\def\K{{\bf K}}

\def\T{{\Bbb T}}

\newcommand{\changedX}{K}

\newcommand{\hahyp}{\,\,\widehat{\!\!\hyp}}%

\def\diag{{\mbox{diag}}}

\def\2{{\overline 2}}

\newcommand{\beqa}{\begin{eqnarray}}
\newcommand{\eeqa}{\end{eqnarray}}

{\catcode `\@=11 \global\let\AddToReset=\@addtoreset}
\AddToReset{figure}{section}
\renewcommand{\thefigure}{\thesection.\arabic{figure}}